\documentclass{aa}  

\usepackage{graphicx}
\usepackage{txfonts}
\usepackage[colorlinks,linktocpage=true]{hyperref}
\usepackage{ulem}
\newcommand{\kms}{km~s$^{-1}$}

\begin{document} 

  \title{Probing the shape of the Milky Way dark matter halo with hypervelocity stars: a new method}
  

   \author{Arianna Gallo\thanks{a.gallo@unito.it}\inst{1,2}
  	\and
  	Luisa Ostorero\inst{1,2}
  	\and
  	Sankha Subhra Chakrabarty\inst{1,2}
  	\and
  	Stefano Ebagezio\inst{3,1}
  	\and
  	Antonaldo Diaferio\inst{1,2}}
  
  \institute{Dipartimento di Fisica, Universit\`a di Torino,  Via P. Giuria 1, I-10125 Torino, Italy
  	\and
  	Istituto Nazionale di Fisica Nucleare (INFN), Sezione di Torino, Via P. Giuria 1, I-10125 Torino, Italy 
  	 \and I.~Physikalisches Institut, Universit\"at zu K\"oln , Z\"ulpicher Str.\ 77, D-50937 K\"oln, Germany}

   \date{Received date; accepted date}


   \abstract
    {We propose a new method to determine the shape of the gravitational potential of the dark matter (DM) halo of the Milky Way (MW) with the galactocentric tangential velocities of a sample of hypervelocity stars (HVSs). We compute the trajectories of different samples of HVSs in a MW where the baryon distribution is axisymmetric and the DM potential either is spherical or is spheroidal or triaxial with radial-dependent axis ratios. We create ideal observed  samples of HVSs with known latitudinal component of the tangential velocity, $v_{\vartheta}$, and azimuthal component of the tangential velocity, $v_{\varphi}$. We determine the shape of the DM potential with the distribution of $|v_{\vartheta}|$, when the Galactic potential is axisymmetric, or with the distribution of $|v_{\vartheta}|$ and of a function, $\bar v_{\varphi}$, of $v_{\varphi}$ when the Galactic potential is non-axisymmetric. We recover the correct shape of the DM potential by comparing the distribution of $|v_{\vartheta}|$ and $\bar v_{\varphi}$ of the ideal observed  sample against the corresponding distributions of mock samples of HVSs that traveled in DM halos of different shapes. We use ideal observed optimal samples of $\sim 800$ HVSs, which are the largest samples of $4~M_\sun$ HVSs ejected with the Hills mechanism at a rate $\sim 10^{-4}$~yr$^{-1}$, currently outgoing, and located at more than 10~kpc from the Galactic center. In our ideal case of galactocentric velocities with null uncertainties and no observational limitations, the method recovers the correct shape of the DM potential with a success rate $S\gtrsim 89\%$ when the Galactic potential is axisymmetric, and $S > 96\%$ in the explored non-axisymmetric cases. The unsuccessful cases yield axis ratios of the DM potential that are off by $\pm 0.1$. The success rate decreases with decreasing size of the HVS sample: for example, for a spherical DM halo, $S$ drops from $\sim 98\%$ to $\sim 38\%$ when the sample size decreases from $\sim 800$ to $\sim 40$ HVSs.
    Accurate estimates of the success rate of our method applied to real data require more realistic samples of mock observed HVSs. Nevertheless, our analysis suggests that a robust determination of the shape of the DM potential requires the measure of the galactocentric velocity of a few hundred HVSs with  robustly confirmed galactocentric origin.
}

%


  \maketitle
  
%

\section{Introduction}
\label{sec:introduction} 
    The Lambda Cold Dark Matter ($\Lambda$CDM) cosmological model predicts the existence of dark matter (DM) halos surrounding the Milky Way (MW) and the external galaxies. Although at first order the spherically symmetric Navarro–Frenk–White (NFW) profile 
	\citep{navarro1997} can provide a good approximation to the shape of the DM halos, the first numerical N-body simulations \citep{Frenk1988,Dubinski1991,Warren1992,Cole1996} found the shape of the DM halos to be triaxial, and subsequent works \citep[e.g., ][]{Jing2002,Bailin2005,Hayashi2007,Vera-Ciro2011} confirmed this results. 
	In these simulations, the DM halos are triaxial with a tendency to prolateness in the center, while they become more and more triaxial/oblate in the outer part, as a consequence of an accretion history that is more anisotropic at early times and becomes more isotropic at later times.

	The results obtained from these DM-only simulations are affected by the absence of the baryonic component, whose inclusion is of fundamental importance to properly describe the formation of small scale systems like the MW and the other galaxies. The inclusion of the gas dynamics in the simulations was the first attempt to account for the baryonic effects on the shape of the DM halo in the 1990's \citep[e.g.,][]{Katz1991,Katz1993,Dubinski1994}. In the following years, cosmological hydrodynamic simulations were used to investigate the impact of radiative cooling, star formation, and supernova feedback on the final shape of the DM halos \citep[e.g.,][]{Gnedin2004,Kazantzidis2004,Gustafsson2006,Tissera2010,Abadi2010,Zemp2012,Bryan2013,Butsky2016}. The inclusion of these effects leads to a larger sphericity of the DM halos, that are now predicted to be, in the central regions of galaxies, rounder or more oblate than previously thought.

    In the last decades, the predictions on the shape of the DM halos have been largely tested on the MW, with the aim of constraining the shape of its DM halo; those tests used different tracers of the Galactic gravitational potential, such as the distribution and kinematics of halo stars \citep[e.g.,][]{smith2009,loebman2014}, the kinematics of disk stars in the Solar neighborhood \citep{olling2000}, the tidal streams from satellite galaxies and from globular clusters \citep[e.g.,][]{helmi2004,Johnston2005,fellhauer2006,ruzicka2007,law2009,koposov2010,law2010,veraciro2013,kupper2015,bovy2016,Malhan2019}, the distribution of both globular clusters \citep[e.g.,][]{posti2019} and the MW satellite galaxies \citep{zentner2005}, and the flaring of the HI distribution \citep[e.g.,][]{olling2000,banerjee2011}. 

    Because of the use of different tracers, the results of these studies hold on different spatial scales. However, even on comparable scales, these results are not all consistent with one another, partly because of the use of different techniques and partly because of the different working assumptions.
  
   The Galactic DM halo is suggested to be overall close to spherical on the basis of the tilt of the velocity ellipsoid of a sample of halo subdwarf stars located at galactocentric cylindrical radii of 7-10~kpc and depth $\lesssim 4.5$ kpc below the Galactic plane, in the 250 deg$^{2}$ sky area covered by SDSS Stripe 82 \citep{smith2009}. 
   Under the assumptions that the MW DM halo is a spheroid and the full Galactic gravitational potential is axisymmetric, a variety of results are found.
   The GD-1 stellar stream excludes a significantly oblate DM halo at the GD-1 location, $r \sim 14$~kpc, where the vertical-to-planar axis ratio (hereafter referred to as ``flattening'') of the gravitational equipotential surfaces is constrained to $q_{\Phi} > 0.89$, according to  \citet{koposov2010}; the same stellar stream is found to provide a stronger constraint, yielding a prolate DM halo with mass density flattening $q_\rho = 1.27$, by \citet{bovy2016}.
   On the other hand, the stellar stream Pal 5 constrains the DM halo to be mildly oblate at $r \sim 19$ kpc, with either  potential flattening $q_{\Phi}=0.95$ \citep{kupper2015} or  density flattening $q_\rho = 0.9$  \citep{bovy2016},
   suggesting a radial-dependent flattening for the DM halo.
   Finally, the combination of Pal 5 and GD-1, together with constraints on the force field near the Galactic disk, return a nearly spherical DM halo, with $q_\rho=1.05$, \citep{bovy2016} within the inner 20 kpc.
   
   These results are at odds with those found with different probes on similar scales: the combination of the kinematics of disk stars in the vicinity of the Sun with the flaring of the HI disk is found to consistently constrain the DM halo to be oblate, with density flattening $q_\rho \sim 0.8$ \citep{olling2000}; the kinematics of halo stars also  constrains the dark halo to be significantly oblate, with flattening $q_\rho=0.4$ (or, equivalently, $q_{\Phi}=0.7$; \citeauthor{loebman2014} \citeyear{loebman2014}).
   Conversely, the distribution of globular clusters suggests a prolate DM halo with density flattening $q_\rho=1.3$ \citep{posti2019}. 

    At larger galactocentric distances, $20 \lesssim r \lesssim 60$~kpc, the tidal streams of the Sagittarius dwarf spheroidal (Sgr dSph) galaxy lead to conclude that the DM halo potential has to be mildly oblate ($q_{\Phi}=0.90-0.95$;
    \citeauthor{Johnston2005} \citeyear{Johnston2005}) or nearly spherical ($q_{\Phi}=0.92-0.97$; \citeauthor{fellhauer2006} \citeyear{fellhauer2006}) to explain the precession of Sgr dSph's orbit, while extremely oblate halos with density flattening $q_{\rho}<0.7$ are ruled out \citep{ibata2001}; on the other hand, the potential of the dark halo has to be prolate, with $q_{\Phi}=1.25-1.5$ to explain the kinematics  of Sgr dSph's older, leading stream \citep{helmi2004}. Finally, on scales  $r \lesssim 200$ kpc, the modeling of the Magellanic stellar streams generated by the interaction of the MW with the Magellanic system favors a DM halo that has a globally oblate potential with $q_{\Phi}<1$ \citep{ruzicka2007}. The uncertainties on the quoted flattening range from $\sim 5\%$ to $\sim 20\%$.

    Some of the apparent inconsistencies among the results on the shape of the dark halo can be solved by assuming that the DM halo is triaxial, so that the Galactic potential is globally non-axisymmetric.
    For instance, a triaxial DM potential with intermediate-to-major axis ratio $(b/a)_{\Phi}=0.99$ and a minor-to-major axis ratio $(c/a)_{\Phi}=0.72$ on scales $20 \lesssim r \lesssim 60$ kpc enables \citet{law2010} to explain, at the same time, both the angular precession and the radial velocities of the stars in the Sgr dSph leading stream \citep[see also][]{law2009}. 
  
    Other inconsistencies on the shape of the dark halo, especially those on different spatial scales, may in principle be relieved by discarding the simplifying assumption of a radial-independent shape for the halo: although common to all the above-mentioned models, this assumption is not supported by the results of N-body simulations.
    In this context, by assuming an axisymmetric dark halo to model the flaring of the HI disk, \citet{banerjee2011} find the halo to be prolate, with a density flattening $q_{\rho}$ increasing from 1 to 2 in the range $9 \lesssim r \lesssim 24$~kpc.  Conversely, with a non-axisymmetric model applied to the Sgr dSph's streams, and accounting for the effects of the Large Magellanic Cloud (LMC), \citet{veraciro2013} constrain the DM halo potential to be mildly oblate at $r \lesssim 10$~kpc, where $q_{\Phi}=0.9$, and smoothly translating to a triaxial shape at larger radii, where the intermediate-to-major axis ratio is $(b/a)_{\Phi}=0.9$ and the minor-to-major axis ratio is $(c/a)_{\Phi}=0.8$, in agreement with cosmological simulations. All these results clearly show that, despite the large efforts, the shape of the MW DM halo remains uncertain and that further work is necessary.

	In this paper, we use the hypervelocity stars (HVSs) as tracers of the shape of the MW DM halo.
	The existence of HVSs was postulated by \citet{hills1988}, who defined as an HVS a star ejected from the Galactic center after a close encounter between a stellar binary and the supermassive black hole (SMBH) associated with SgrA$^{\star}$: the ejected star is characterized by a present-day speed exceeding the Galactic escape velocity, while the companion star becomes gravitationally bound to the SMBH.  
	Alternative mechanisms for the generation of the HVSs were subsequently proposed, both before and after the first observational evidence for HVSs \citep{brown2005}: a three-body interaction between a single star and a binary BH \citep{yu2003}, the scattering of a star off a stellar-mass BH \citep{oleary2008}, the star formation along the path of an outflow driven by an active galactic nucleus \citep{silk2012}, the interaction between a globular cluster and a BH binary \citep{fragione2016}, and a four-body interaction between a binary star and a binary BH \citep{wang2018}.	
	
	On the observational side, it was W.~R.~Brown who serendipitously discovered the first HVS candidate: a B-type star escaping the MW with a galactocentric velocity of $\sim 700$~\kms~ \citep{brown2005}. Many HVS candidates were later found in both targeted and not targeted surveys  \citep[e.g.,][]{hirsch2005,edelmann2005,brown2006a,brown2006b,brown2007b,brown2007c,brown2009a,brown2012a,brown2014,Tillich2011,Li2012,Li2015,Pereira2013,zheng2014,huang2017,Neugent2018,Du2019,Luna2019,koposov2020,li2021}. 
	Some of these observed stars have speeds smaller than the Galactic escape velocity: they are referred to as ``bound HVSs''.
	As noted by \cite{brown2015}, among the mechanisms proposed for the HVS production, the Hills mechanism has a unique ability in generating a large number of unbound main-sequence stars and in explaining the presence of the so-called S-stars in close orbit around the central SMBH \citep[e.g.,][]{ghez2003,ghez2005,gillessen2009,gillessen2017,genzel2010}.
	
	We live in the Gaia era \citep[e.g.,][]{gaiaDR1summ2016,gaiamission2016,gaiaDR1cat2016,gaiaDR2summ2018,gaiaDR2cat2018,gaiaED3summ2021} and for the last years several studies have been conducted to forecast the number of HVSs that Gaia can detect by the end of the mission \citep{marchetti2018}, search for new HVSs candidates \citep[e.g.,][]{marchetti2017, bromley2018, marchetti2019, marchetti2021}, and revalue the classification of a star as HVS candidate based on the improved accuracy on proper motions provided by Gaia \citep[e.g.,][]{boubert2018,brown2018,irrgang2018a,Erkal2019,kreuzer2020,irrgang2021}. The birth place of many stars is indeed still uncertain and some of the HVS candidates may actually be: (i) runaway stars ejected from the Galactic disk \citep[e.g.,][]{blaauw1961,poveda1967,Leonard1991}; (ii) halo stars outliers whose velocities can reach $450$~\kms~ \citep{smith2009b} or fast halo stars, that can be unbound to the MW, produced by a tidal interaction between a dwarf galaxy and the MW near the Galactic center \citep[e.g.,][]{abadi2009,huang2021}; (iii) stars ejected as either HVSs or runaway stars from the nearest satellite galaxies of the MW, as the LMC \citep{boubert2016,boubert2017} and the Sgr dSph galaxy \citep{huang2021}.
    From the available literature, we estimate a sample of $\sim 70$ HVS candidates whose galactocentric origin has not been unambiguously ruled out. This sample includes both unbound and bound HVS candidates, that make up $\sim 40\%$ and $\sim 60\%$ of the full sample, respectively.
    These HVS candidates are the result of heterogeneous classification methods, and the number of true HVSs remains uncertain until the galactocentric origin of these stars is unambiguously confirmed.

    Since the discovery of the very first HVS candidate, the HVSs have been recognized as a powerful tool to probe either the shape of the Galactic DM halo \citep{gnedin2005,yu2007}, or its mass \citep{rossi2017,fragione2017}, or both \citep{contigiani2019}. 
    They were also used to discriminate among different models of the Galactic gravitational potential in Newtonian gravity \citep{perets2009} and between different theories of gravity \citep{perets2009,chakrabarty2022}.
    In this paper, we assume that Newtonian gravity holds on Galactic scales, we fix the mass of the DM halo, and we use the kinematical properties of the HVSs to constrain the shape of the dark halo. We account for neither the gravitational effects of the LMC on the HVS trajectories \citep{kenyon2018} nor the time dependence of the gravitational potential of the MW due to its  interaction with the LMC \citep{boubert2020}.
   Because the HVSs are ejected radially with high speed and may cross the entire Galaxy before dying out, in an isolated MW, the small, typically a few per cent, deviations from straight lines of their trajectories are determined by the asphericity of the MW gravitational potential well, dominated by the DM halo at large galactocentric distances. 
   
   Previous studies aimed at constraining the shape of the gravitational potential of the dark halo made use of (i) sufficiently accurate ($\sigma_{\mu}\lesssim 10~ \mu$as yr$^{-1}$) proper motion measurements of either one observed HVS with known distance or a set of two or more HVSs with unknown distance, with the constraints becoming tighter for larger samples of observed HVSs \citep{gnedin2005}; (ii) a triaxiality estimator that is a function  of the components of the specific angular momentum of HVSs located at Galactocentric distances $r \gtrsim 50$ kpc \citep{yu2007}; (iii) a likelihood function constructed by back-propagating the phase-space position of each HVS to the Galactic center, in order to reproduce its observed phase-space coordinates and mass \citep{contigiani2019}.
   
   The above methods have been applied to triaxial \citep{gnedin2005,yu2007} or axisymmetric \citep{contigiani2019} DM halos. 
   The techniques used by \citet{gnedin2005} and \citet{contigiani2019} do not depend on the model of the potential assumed for the DM halo, but require the integration of the HVS trajectories; conversely, the angular-momentum technique designed by \citet{yu2007} does depend on the dark halo potential, but does not require the trajectory integration. In all these models, the shape of the DM halo potential is constant with radius.
   Furthermore, all the techniques require intermediate steps that involve each HVS of the sample individually, even though the final result depends on the contribution of all the HVSs of the sample.
   
   In this work, we propose a statistical method to constrain the axis ratios of a radial-dependent, triaxial gravitational potential of the DM halo from the distributions of the HVS phase space coordinates that are mostly affected by the asphericity of this potential, namely the components of the galactocentric tangential velocities. 
   Unlike the techniques illustrated above, our method does not require intermediate steps that involve each star of the sample, such as the trajectory integration or the evaluation of the angular momenta of the HVSs.
   Our method can be applied to different models of the MW gravitational potential, as it is the case for the methods by \cite{gnedin2005} and \cite{contigiani2019}.
    
	The paper is organized as follows. In Sect.~\ref{sec:model_simulations}, we describe our numerical simulations of the initial velocity distribution of a sample of HVSs ejected according to the Hills mechanism, and the simulations of the HVS trajectories in a Galactic gravitational potential generated by DM halos with different shapes; we also illustrate the construction of our HVS phase space mock catalogs.
	In Sect.~\ref{sec:triaxiality_indicators}, we show how the asphericity of the DM halo mostly affects the HVS tangential velocity: we identify this velocity as the key variable to statistically discriminate between different shapes of the DM halo, and we select the appropriate HVS sample to pursue this goal.
	In Sect.~\ref{sec:method}, we present our statistical method to recover the shape of the DM halo from a distribution of HVS tangential velocities. 
	In Sects.~\ref{sec:axisymmetric_Galactic_potential} and \ref{sec:non-axisymmetric_Galactic_potential}, we show the results of the application of our method to an ideal sample of mock observed HVSs with null uncertainties and no observational limitations that traveled in an axisymmetric and non-axisymmetric Galactic gravitational potential, respectively.
	In Sect.~\ref{sec:sample_size}, we investigate the effect of the size of the ideal sample of mock observed HVS on the success rate of our method. We finally discuss our results and conclude in Sects.~\ref{sec:Discussion}~and~\ref{sec:conclusions}, respectively.

\section{Numerical simulations and mock catalogs} 
\label{sec:model_simulations}

	In Sect.~\ref{sec:ejection_velocities}, we illustrate the generation of the distribution of the initial velocities of HVSs ejected with the Hills mechanism. In Sect.~\ref{sec:MW_potential}, we describe our model for the gravitational potential of the MW, generated by the baryonic components and a DM halo with different shapes, and in Sect.~\ref{sec:orbit_integration} we illustrate the simulation of the trajectories of the ejected HVSs across the Galaxy.
	Finally, in Sect.~\ref{sec:mock_catalogs}, we describe the mock catalogs that we built from the HVS phase space distribution.

\subsection{Ejected stars: velocity distribution} \label{sec:ejection_velocities}

Following \citet{hills1988} \citep[see also][]{bromley2006}, we simulated the ejection of stars from the Galactic Center with a 3-body numerical code which reproduces the gravitational interaction of a binary star system with the SMBH associated with SgrA$^{\star}$.  
For the ejected stars, our code provides the distribution of the ejection velocities, $v_{\rm ej}$.
A detailed description of the code will be provided in a separate paper.

We set the mass of the SMBH to $4 \times 10^6 M_{\sun}$, consistent with different estimates \citep[e.g.,][]{boehle2016, gillessen2017}. For simplicity, we restricted our simulations to equal-mass stellar binaries of $4 + 4~M_{\sun}$ on hyperbolic orbits. 
The mass of 4~$M_{\sun}$ is representative of the upper end of the mass distribution of the HVS candidates observed in currently available surveys (see references in Sect.~\ref{sec:introduction}).
We will further discuss our mass choice in Sect.~\ref{sec:Discussion}.

For a fixed mass of the binary members, the velocity distribution of the ejected stars depends upon a series of parameters: 
(i) the stellar binary semi-major axis, $a$; (ii) the minimum approach distance, $R_{\rm min}$, between the center of mass of the binary and the SMBH; (iii) the inclination angle, $i$, between the orbital plane of the binary star and the orbital plane of the binary's center of mass and the SMBH; (iv) the initial phase, $\phi$, of the binary star. We randomly sample $i$ and $\phi$ in the interval $[0, 2\pi]$ with uniform probability density function. 
As for $a$ and $R_{\rm min}$, which determine the probability of ejection of the primary star, we randomly sampled $a$ from the interval $0.05-4$~AU with probability density function $p(a)\propto 1/a$, and we randomly drew $R_{\rm min}$ from the interval $1-700$~AU with probability density function $p(R_{\rm min})\propto 1/R_{\rm min}$ \citep[][and references therein]{bromley2006}. 

\begin{figure}[htb!]
	\includegraphics[width=8.8cm,height=6.0cm]{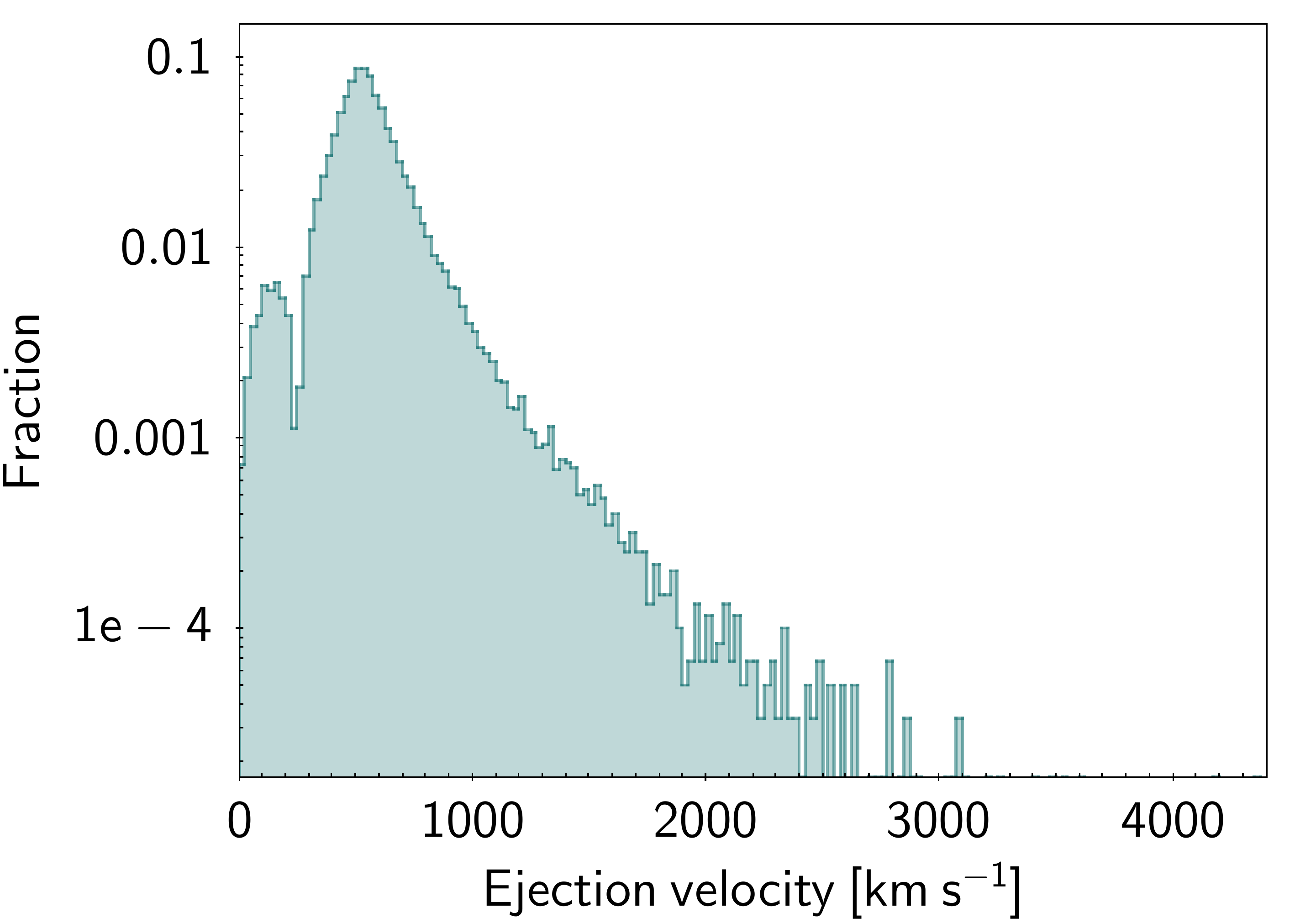}
	\caption{Distribution of ejection velocities for the ejected star(s) of a $4+4$~$M_{\sun}$ binary star system, after a close encounter with a $4\times 10^6\:M_{\sun}$ SMBH.
	The distribution consists of $N_{\rm ej} = 60,000$ stars ejected in $N_{\rm int} \sim 240,000$ 3-body interactions.
\label{fig:v_ej}}
\end{figure}

Figure~\ref{fig:v_ej} shows the ejection velocity distribution for a simulation of $N_{\rm int} \sim 240,000$ 3-body interactions that produces 
$N_{\rm ej}=60,000$ ejection events. 
Most of the ejected stars possess ejection velocities in the range $\sim 250-4,000$~\kms; the velocity distribution has a major peak at $v_{\rm ej} \sim 510$~\kms~ and has positive skewness.
A minority of stars are ejected with speeds lower than $\sim 250$~\kms: these stars are the result of rare, double-ejection events, where neither of the binary members becomes gravitationally bound to the SMBH, and both stars are ejected. 
Because we assumed that the binary system starts its approach trajectory towards the SMBH with a velocity at infinity $v_{\rm inf}=250$~\kms~ \citep{hills1988}, energy conservation requires that one of these stars is ejected with velocity lower than 250~\kms. 
Our distribution of ejection velocities is comparable to the ejection speed distribution obtained from the analytical prescription by \citet{bromley2006}.

We note that the ejection velocity $v_{\rm ej}$ is defined as the speed that an ejected star would have at infinite distance from the SMBH, in absence of other gravitational sources \citep{hills1988,bromley2006}, as in our 3-body simulation.
However, the ejection velocity distribution shown in Fig.~\ref{fig:v_ej} can be used as a distribution of initial velocities when simulating the trajectories of the HVSs from the Galactic center to the outer halo, as we explain in Sect.~\ref{sec:orbit_integration}.

\subsection{Milky Way gravitational potential} 
\label{sec:MW_potential}

We modeled the Milky Way gravitational potential $\Phi$ as the superposition of the potentials generated by three distributions of baryonic matter and one distribution of dark matter:
\begin{equation}
\Phi = \Phi_{\rm BH} + \Phi_{\rm b} + \Phi_{\rm d} + \Phi_{\rm h} ,  
\label{eq:phi_grav}
\end{equation}
where $\Phi_{\rm BH}$ is the potential generated by the SMBH located in the Galactic Center, $\Phi_{\rm b}$ is generated by the Galactic bulge, $\Phi_{\rm d}$ is the disk potential, and $\Phi_{\rm h}$ is the potential of the dark matter halo embedding the Galaxy.
We consider the Galaxy as isolated, neglecting both the presence of the LMC \citep{kenyon2018} and any time dependence of the Galactic potential deriving from the interaction of the MW with the LMC \citep{boubert2020}.

In a reference frame with the origin at the Galactic Center, we used spherical coordinates $(r, \vartheta , \varphi)$ for the spherically symmetric components of the potential (with $ -90^\circ \le \vartheta \le 90^\circ$), cylindrical coordinates $(R, \varphi, z)$ for the axisymmetric components, and Cartesian coordinates $(x, y, z)$ for the triaxial component. 
We took the $x$-$y$ plane as the equatorial plane of the disk, with the $x$-axis corresponding to the direction from the Sun to the Galactic center, and we took the $z$-axis as the vertical axis.

We included the contribution of the SMBH to the gravitational potential as:
\begin{equation}
\Phi_{\rm BH}(r) = - \frac{G\:M_{\rm BH}}{r} ,
\label{eq:smbh}
\end{equation}
where $M_{\rm BH}$ is the mass of the SMBH.

For the bulge component, we adopted the \cite{hernquist1990} potential:
\begin{equation}
\Phi_{\rm b}(r) = - \frac{G\:M_{\rm b}}{r + r_{\rm b}} ,
\label{eq:bulge}
\end{equation}
where $M_{\rm b} = 3.4 \times 10^{10} \ M_{\sun}$ and $r_{\rm b} = 0.7 \ {\rm kpc}$ \citep{kafle2014,price-whelan2014,rossi2017,contigiani2019} are the scale mass and the scale radius of the model, respectively.

For the disk component, we adopted the axisymmetric potential by \cite{miyamoto1975}:
\begin{equation}
\Phi_{\rm d}(R,z) = - \frac{G\:M_{\rm d}}{\sqrt {R^2 + \left( a_{\rm d} + \sqrt{z^2 + b_{\rm d}^2} \right)^2}}\, ,
\label{eq:disk}
\end{equation}
where $M_{\rm d} = 1.0 \times 10^{11} \ M_{\sun}$, $a_{\rm d} = 6.5 \ {\rm kpc}$ and $b_{\rm d} =  0.26 \ {\rm kpc}$ \citep{kafle2014,price-whelan2014,rossi2017,contigiani2019} are the scale mass and the scale lengths of the model, respectively. 

Finally, we modeled the contribution of the DM halo with the triaxial generalization of the spherically symmetric NFW \citep{navarro1997} potential proposed by \cite{vogelsberger2008}:

\begin{equation}
\Phi_{\rm h}(\tilde{r}) = -\frac{G\:M_{200}}{f(C_{200})}\:\frac{\mathrm{ln}\left( 1 + \frac{\tilde{r}}{r_{\rm s}}\right)}{\tilde{r}}\, ,
\label{eq:halo}
\end{equation}
where $f(w) = {\rm ln} (1 + w) - w/( 1 + w)$, $M_{200} = 8.35 \times 10^{11} M_{\sun}$ is the DM halo mass enclosed within $r_{200}$,\footnote{$r_{200}$ is the radius of a spherical volume whose mean mass density is 200 times the critical density of the Universe} $C_{200} = r_{200}/r_s = 10.82$ is the halo concentration parameter, and $r_{\rm s} = 18$~kpc is a generalized scale radius. For the above parameters, we adopted the values used by \cite{hesp2018}, 
that are consistent with the values derived from halo stars \citep{Xue2008}, blue horizontal branch stars \citep{deason2012}, the massive satellite population of the MW \citep{Cautun2014}, and Cepheids \citep{Ablimit2020}. 
The coordinate $\tilde r$ is a generalized radius that replaces the radius $r$ of the NFW spherical potential,
\begin{equation}
\tilde{r} = \frac{\left( r_{\rm a} + r\right) r_{\rm E}}{r_{\rm a} + r_{\rm E}} \, .
\end{equation}
Here, 
$r_{\rm E}$ is an ``ellipsoidal radius'', 
\begin{equation}
r_{\rm E} = \sqrt{\frac{x^2}{a^2} + \frac{y^2}{b^2} + \frac{z^2}{c^2}}\, ,
\end{equation}
where the three ellipsoid parameters, $a$, $b$, and $c$ have to fulfill the condition $a^2 + b^2 + c^2 = 3$, and their combination defines the degree of triaxiality of the potential well. Specifically, the axis ratio of the equipotential surfaces on the $x$-$y$ plane is $q_y=b/a$, whereas the axis ratio of the equipotential surfaces on the $x$-$z$ plane is $q_z=c/a$.
The parameter $r_{\rm a}$ is the scale length where the smooth transition from a triaxial potential to a nearly spherical potential occurs; we took it to be $1.2 r_{\rm s}$, as in  \citet{hesp2018}: the halo is triaxial ($\tilde{r} \approx r_E$) in the inner region ($r \ll r_a$), whereas it is approximately spherical ($\tilde{r} \approx r$) in the outer region ($r \gg r_a$). 

In general, the potential $\Phi_{\rm h}({\tilde r})$ given by Eq.~\ref{eq:halo} is triaxial with  $q_y \ne 1$, $q_z \ne 1$, and $q_y \ne q_z$. However, this potential becomes spheroidal when either $q_y$ or $q_z$ are equal to 1 or when $q_y = q_z \ne 1$: when $q_y=1$, the potential is axisymmetric about the $z$-axis; when  $q_z=1$, the potential is axisymmetric about the $y$-axis; when $q_y = q_z \ne 1$ the potential is axisymmetric about the $x$-axis. When both $q_y$ and $q_z$ are equal to 1, the DM halo potential is spherically symmetric.

Because $ \Phi_{\rm BH}$ and $\Phi_{\rm b}$ are spherically symmetric, and $\Phi_{\rm d}$ is axisymmetric about the $z$-axis, our gravitational potential $\Phi$ (Eq.~\ref{eq:phi_grav}) is globally axisymmetric about the $z$-axis when the DM halo is either spherical or spheroidal and axisymmetric about the $z$-axis. On the other hand, the Galactic gravitational potential is non-axisymmetric when the DM halo is either triaxial or spheroidal with a symmetry axis misaligned with respect to the $z$-axis. 

In all of our simulations, we adopted the same parameters for the SMBH, the bulge, and the disk potentials (Eqs.~\ref{eq:smbh}-\ref{eq:disk}). We also adopted the same parameters for the potential of the DM halo (Eq.~\ref{eq:halo}), with the exception of (i) the triaxiality parameter $q_z$, which was set to a different value in each of the simulations of the axisymmetric Galactic potential, while $q_y$ was kept fixed to 1, and (ii) both triaxiality parameters, $q_y$ and $q_z$, in simulations of a non-axisymmetric Galactic potential. 
By varying the triaxiality parameter of the DM halo $q_y$ and $q_z$ in appropriate ranges of values, we explored the effect of the halo shape on the HVS observables. 

\begin{figure}[ht!]
	\includegraphics[width=8cm,height=5.5cm]{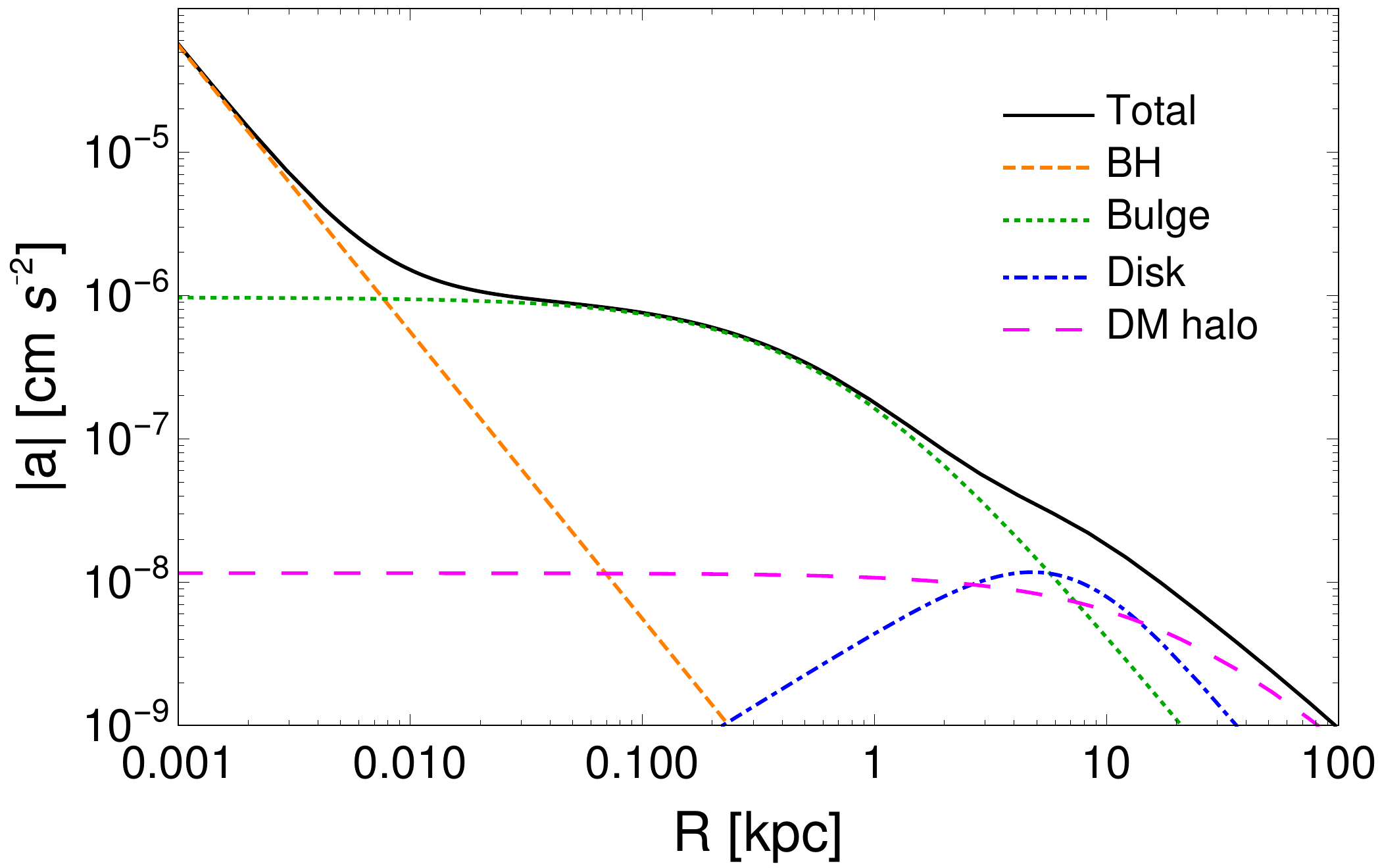}
	\caption{Magnitude of the radial gravitational acceleration in the plane of the disk due to our MW potential $\Phi$ (Eq.~\ref{eq:phi_grav}), as a function of the radial cylindrical coordinate $R$. The black solid line represents the total gravitational acceleration. The dashed and dotted lines represent the contributions of the SMBH (dashed orange line), the bulge (dotted green line), the disk (blue dot-dashed line), and a spherical DM halo (magenta long-dashed line).
	\label{fig:pot}}
\end{figure}

Figure~\ref{fig:pot} shows the magnitude of the radial gravitational acceleration in the plane of the disk associated with our MW potential $\Phi$ as a function of the cylindrical coordinate $R$, for a spherical DM halo ($q_y = 1, q_z = 1$).
The chosen gravitational potential generates masses enclosed within $120$~pc and within $100$~kpc that agree with the observed values derived by \citet{Launhardt2002} and reported in \citet{Dehnen1998}, respectively. It also reproduces a circular velocity of 235~\kms~ at the solar neighborhood, in agreement with the observational values of $231 \pm 6$~\kms~  by \citet{Bobylev2017} and of $230\pm12$~\kms~ by  \citet{BobylevBajkova2016}.

\subsection{Orbit integration} 
\label{sec:orbit_integration}

In the Galactic gravitational potential illustrated in Sect.~\ref{sec:MW_potential}, we simulated the time evolution of the position, velocity, and acceleration of a sample of stars ejected from the Galactic Center according to the Hills mechanism (see Sect.~\ref{sec:ejection_velocities}).
For each star, we numerically integrated Newton's equation of motion in a galactocentric reference frame and in Cartesian coordinates, using the Velocity Verlet algorithm \citep[e.g.,][]{frenkel2001}. We traced the star  trajectory until the star death.
In our simulations, the total energy of a star is conserved with a relative accuracy of $\sim 10^{-8}$.

In Sect.~\ref{sec:ejection_velocities}, we showed that our sample of ejected star consists of $N_{\rm ej} = 60,000$ HVSs; we adopted this sample as our full sample of initial velocity magnitudes. This choice is legitimate: even though $v_{\rm ej}$ is defined as the speed that an ejected star would have at infinite distance from the SMBH in a 3-body interaction (see Sect.~\ref{sec:ejection_velocities}), our SMBH is embedded in the Galactic mass distribution; thus, the ejected stars can be assumed to move at $v_{\rm ej}$ at the outer edge of the sphere of influence of the SMBH, just before the Galactic gravitational potential starts to overcome the SMBH potential.
The Hills ejection mechanism yields an isotropic distribution of ejected stars. We thus assigned each of the $N_{\rm ej}$ stars an initial position $(r_0, \vartheta, \varphi)$, with $r_0=3$~pc the radius of the sphere of influence of the SMBH \citep{genzel2010}, and $(\vartheta, \varphi)$ randomly drawn from a uniform  distribution over the surface of a sphere; we assigned each star a velocity direction $\hat{n}(\vartheta, \varphi)$. We end up with a sample of $N_{\rm ej}$ initial conditions.

\begin{figure}[ht!]
	\includegraphics[width=8.6cm,height=6.cm]{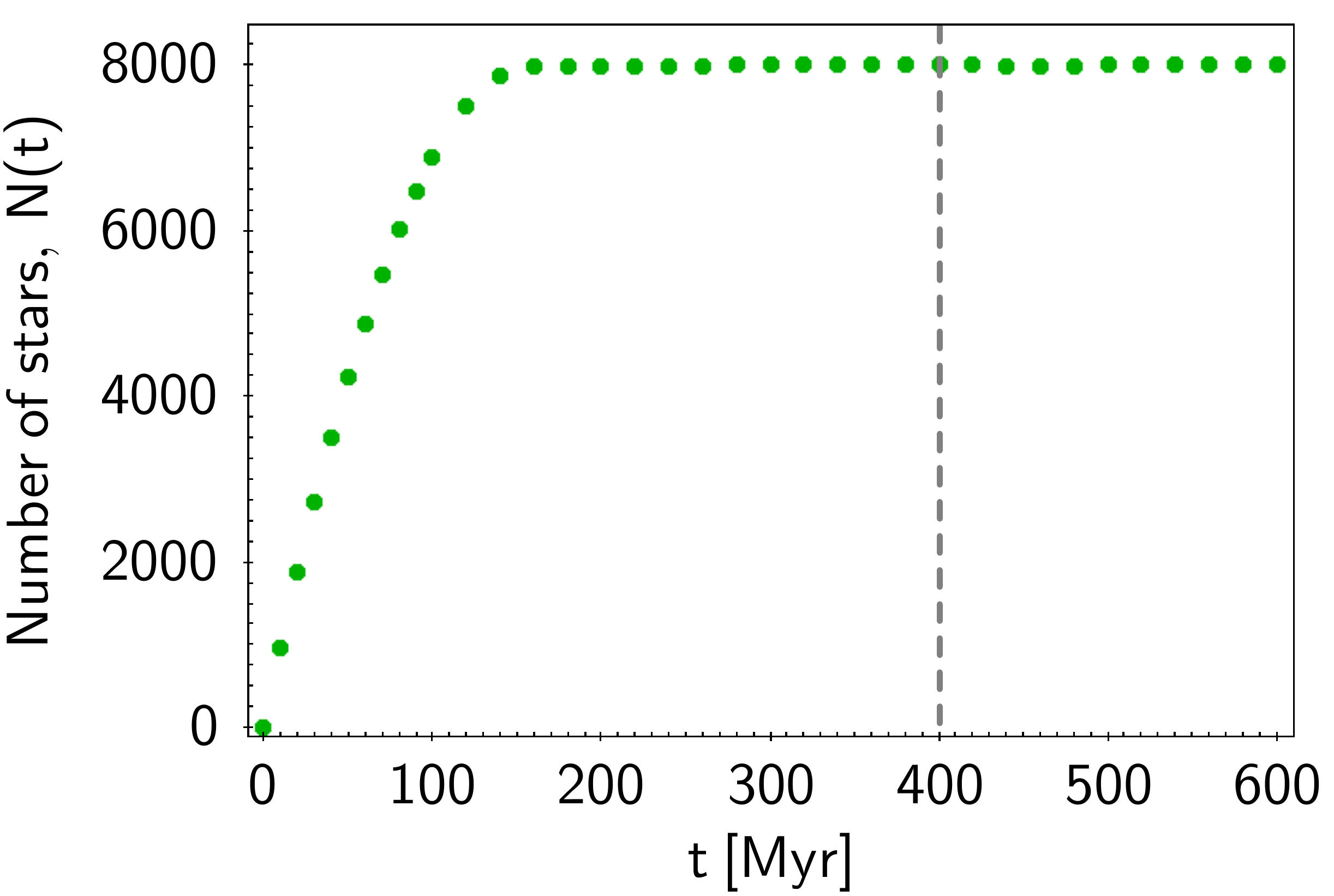}
	\caption{
    The number of 4~$M_{\sun}$ observable HVSs, that started to be ejected at $t=0$ at a rate $10^{-4}$~yr$^{-1}$, stops increasing beyond  $t \simeq \tau_{\rm L} = 160$ Myr. The vertical, dashed line marks the time of the steady state when we chose to observe the system, $t_{\rm obs}=400$ Myr.
	}
	\label{fig:star_number}
\end{figure}

In our simulations below, we computed the trajectories of the HVSs in a large number of different subsamples of size smaller than $N_{\rm ej}$. Each subsample was generated by adopting a random subsample of the $N_{\rm ej}$ initial conditions. Each subsample was selected as follows. 

At an average rate $R = 10^{-4}$ yr$^{-1}$, roughly consistent with the estimate by  \citet{bromley2012} and \citet{zhang2013} \citep[see also][] {hills1988,yu2003}, all the $N_{\rm ej}$ HVSs would be ejected in $600$~Myr. We thus assigned the $i$-th HVS of each subsample an ejection time $t_{\rm ej,i}$ uniformly sampled in the range $0-600$ Myr. The stars have a finite lifetime and each star can thus have a different residual lifetime at $t_{\rm ej,i}$.
A 4~$M_{\sun}$ star with Solar metallicity has a main sequence lifetime $\tau_{\rm ms} \simeq 160$ Myr \citep{schaller1992,brown2006b}: we took this lifetime as the total lifetime of the star, $\tau_{\rm L}$. Thus, at the time of ejection, the $i$-th star is also assigned an age $\tau_{\rm ej,i}$ randomly sampled from a uniform distribution between 0 and $\tau_{\rm L}$.

Figure~\ref{fig:star_number} shows that the number $N(t)$ of observable stars, whose lifetime is $\tau_{\rm L}$, reaches a steady state $N(t) \simeq N_{\rm s}\simeq 8000$ at $t \simeq \tau_{\rm L}$. 
The steady state is the result of the combination of (i) a continuous ejection of stars, with average rate $R$, and (ii) the finite lifetime of the stars, $\tau_{\rm L}$.
We chose to observe our star sample at the observation time, $t_{\rm obs}= 400$ Myr.
All the stars whose ejection time was larger than $t_{\rm obs}$ were discarded from the sample.
Among the remaining HVSs, we selected the $N_{\rm obs}$ stars that are alive at the observation time $t_{\rm obs}$, namely the stars that satisfy the condition $t_{\rm obs} - t_{\rm ej,i} + \tau_{\rm ej,i} < \tau_{\rm L}$. For these stars, we computed the trajectory through the Galaxy for a travel time $t_{\rm travel,i} = t_{\rm obs}-t_{\rm ej,i}$.

We computed the orbits of the sample of $N_{\rm obs}\simeq N_{\rm s} \simeq 8000$ HVSs in a series of 64 simulations, each of them corresponding to a different combination ($q_y, q_z$) of the triaxiality parameters of the DM halo (see Sect.~\ref{sec:MW_potential}).
The different combinations were obtained by varying both $q_y$ and $q_z$ in the range $0.7-1.4$ in steps of $0.1$.
A summary of the simulated shapes of the DM halo is reported in Table~\ref{tab:overall_cases}.

\begin{table*}[ht]
	\caption{Summary of the shapes of the DM halo gravitational potential used to simulate the HVS trajectories.}
    \centering
	\begin{tabular}{cccc}
		\hline \hline
		Shape of the DM halo & Axis of symmetry & Axial ratios & Number of combinations \\
		\hline
		Spherical  & $-$  & $q_y = q_z = 1$         & 1 \\
		Spheroidal & $z$      &  $q_y = 1$ \& $q_z \neq 1$ & 7 \\
		Spheroidal & $y$      &  $q_y \neq 1$ \& $q_z = 1$ & 7 \\
		Spheroidal & $x$      &  $q_y = q_z \neq 1$     & 7\\
	    Triaxial   & none     &  $q_y \neq q_z$ \& $q_y \neq 1$ \& $q_z \neq 1$ & 42 \\
        \hline
	\end{tabular}
	\label{tab:overall_cases}
\end{table*}

\subsection{Mock catalogs A and B}
\label{sec:mock_catalogs}

With the simulations described in Sect.~\ref{sec:orbit_integration}, we created two series of mock catalogs, hereafter referred to as mock catalogs A and B. In both series, each mock catalog includes the phase space distribution of  $N_{\rm obs}$ HVSs at the observation time $t_{\rm obs}$ in a DM halo with a given shape. However, the two series differ in the initial conditions of the HVSs. 

Each mock catalog A was generated with the same set of $N_{\rm obs}$ initial conditions.
In other words, all the stars with the same identification index $i$ in all  catalogs A are given the same combination $S^i=$\{$v_{\rm ej}$, $\hat{n}(\vartheta, \varphi)$, $t_{\rm ej}$, $\tau_{\rm ej}$\}$^i$ of initial conditions, namely the same ejection velocity, ejection time $t_{\rm ej}$, and  age $\tau_{\rm ej}$ at ejection.
The only difference among these mock catalogs is the set of triaxiality parameters of the DM halo, $(q_y, q_z)$. 
We used mock catalogs A to highlight the effect of the degree of triaxiality of the DM halo on the distribution of the phase space coordinates of the HVSs, and to identify the phase space coordinates that can serve as shape indicators  (Sect.~\ref{sec:halo_impact}).

On the other hand, mock catalogs B were generated with  different
sets of $N_{\rm obs}$ initial conditions. In other words, the $i$-th stars in different catalogs B are given a different combination $S^i$ of initial conditions.
Therefore, mock catalogs B differ from one another not only for the shape of the DM halo, but also for the sample of ejected stars. 

We used mock catalogs B to explore the effect of the variation of the initial conditions on the detection of deviations from spherical symmetry of the shape of the DM halo potential (Sect.~\ref{sec:IC_impact}). We also used catalogs B to implement our method to recover the shape of the DM halo of the MW from real HVS samples (Sect.~\ref{sec:method}).

Each mock catalog A or B includes the positions and the velocities of the $N_{\rm obs}$ bound and unbound HVSs in the galactocentric, the Galactic heliocentric, and the equatorial reference frames, at the observation time $t=t_{\rm obs}$.
In our simulations, the star position in the galactocentric reference frame is  $(r,\vartheta,\varphi)$ in spherical coordinates and $(x,y,z)$ in Cartesian coordinates, with $\varphi=0^{\circ}$ on the $x$-axis, and $\vartheta=0^{\circ}$ on the $x$-$y$ plane;
the star galactocentric velocity is $(v_r, v_\vartheta,v_\varphi)$, with $v_r$ the radial velocity, $v_\vartheta$ the latitudinal velocity, and $v_\varphi$ the azimuthal velocity.
We defined as tangential velocity the vector whose components are the latitudinal and the azimuthal velocities: $\vec v_{\rm t}=(v_\vartheta, v_\varphi)$; its magnitude is $v_{\rm t} \equiv |\vec v_{\rm t}| = (v_\vartheta^2 + v_\varphi^2)^{\frac{1}{2}}$.
We show the distribution of the galactocentric distance, as well as the distributions of the galactocentric radial, latitudinal, and azimuthal velocities of the HVSs for one of our mock catalogs in Appendix \ref{sec:appendixA}.

As will be shown at the end of Sect.~\ref{sec:halo_impact}, we evaluated the possible use of galactocentric, heliocentric, and equatorial phase space coordinates to infer the shape of the dark halo. To this aim, we converted the star phase space coordinates from the galactocentric to the Galactic heliocentric reference frame, by choosing the Sun to be located at $(x,y,z)=(-R_{\sun},0,0)$ and to have velocity $(U_{\sun},V_{\sun}+\Theta_0,W_{\sun})$ in the galactocentric reference frame.
We used $R_{\sun} = 8.277$~kpc \citep{Abuter2021}, and our model's rotational velocity at $R_{\sun}$, $\Theta_0 = 235$~\kms, in agreement with \citet{Bobylev2017} and \citet{BobylevBajkova2016} (see Sect.~\ref{sec:MW_potential}). For the velocity of the Sun with respect to the local standard of rest, we used $U_{\sun} = 11.1$~\kms, $V_{\sun} = 12.24$~\kms, and $W_{\sun} = 7.25$~\kms~\citep{schonrich2010}.
The equations for the transformation of coordinates and velocities from the galactocentric to the Galactic heliocentric reference frame, and from  the Galactic heliocentric to the equatorial reference frame, are reported in Appendix \ref{sec:appendixB}. 


\section{Indicators of the shape of the DM halo}
\label{sec:triaxiality_indicators}

We used mock catalogs A described in Sect.~\ref{sec:mock_catalogs} to investigate the effect of the triaxiality parameters of the gravitational potential of the DM halo on the HVS phase space distribution. Our final goal was to identify the HVS phase space coordinates (Sect.~\ref{sec:halo_impact}) and define the HVS samples (Sects.~\ref{sec:halo_impact} and \ref{sec:sample_selection})  
that are best suited to discriminate between different shapes of the DM halo.

\subsection{Effect of the shape of the DM halo on the HVS phase space distribution}
\label{sec:halo_impact}

We explored the effect of the triaxiality parameters of the gravitational potential of the DM halo on the phase space distribution of the simulated HVSs, with the goal of identifying the phase space coordinates that are most sensitive to the shape of the DM halo, and can thus be identified as triaxiality indicators.
For this purpose, we resorted to mock catalogs A: because these catalogs are all generated with a specific sample of stars, characterized by fixed distributions of initial conditions (Sect.~\ref{sec:mock_catalogs}), any significant difference among two of these mock catalogs can only be ascribed to the different triaxiality parameters of the gravitational potential of the DM halo.

For our investigation, from each  mock catalog A of $N_{\rm obs}$ stars (see Sect.~\ref{sec:mock_catalogs}),  we selected a subsample of stars that, at the time of observation $t_{\rm obs}$, are located at galactocentric distances $r>10$~kpc, where the gravitational effect of the DM halo starts to be relevant (see Fig.~\ref{fig:pot}). We also required the stars to possess positive galactocentric radial velocity, $v_{\rm r} >0$, to match the observational HVS selection criterion. Applying these selection criteria returns, for each mock catalog, a subsample of $N \simeq N_{\rm obs}/10 \simeq 800$ stars.
In Sect.~\ref{sec:sample_selection} we demonstrate that this reasonable sample selection turns out to be the best selection on the basis of kinematic arguments.

We performed our analysis by means of a statistical approach. 
Specifically, we made use of the two sample Kolmogorov-Smirnov test \citep{press2007}, hereafter referred to as ``KS test'', to check whether the null hypothesis $H_0$ is true, namely whether the distribution of a given phase space coordinate in a spherical DM halo and the distribution of the same coordinate in a non-spherical DM halo are consistent with being drawn from the same parent population, and are thus indistinguishable.
If the $p$-value of the KS test is $p\leq \alpha$, then $H_0$ is rejected at the adopted significance level $\alpha$: in this case, we considered the two distributions as significantly different from each other, and we investigated the use of that phase space coordinate as an indicator of the shape of the DM halo. We adopted a significance level $\alpha=5\%$.

For our KS tests, we considered the distributions of the components of the star position and velocity in the galactocentric reference frame, in both spherical and Cartesian coordinates, for a series of pairs of DM halos; each halo pair is composed of a spherical DM halo and a non-spherical DM halo with different triaxiality parameters.
For the sake of simplicity, we illustrate here the details of our investigation for spheroidal DM halos with $q_y=1$, namely for spheroidal halos that are axisymmetric about the $z$-axis and yield a global axisymmetric Galactic potential; we only comment on the case of a non-axisymmetric Galactic potential. This restriction will not imply a loss of generality of our main result.
Both the axisymmetric and the non-axisymmetric cases are carefully explored in Sects.~\ref{sec:axisymmetric_Galactic_potential} and \ref{sec:non-axisymmetric_Galactic_potential}.

We created 50 series of mock catalogs A. Each series consisted of a spherical DM halo and six spheroidal DM halos axisymmetric about the $z$-axis and with $q_z$ ranging from $q_z=0.7$ (extremely oblate halo) to $q_z=1.4$ (extremely prolate halo). The set of initial conditions is the same for each halo in the same series, and differs from one series to the other. We computed the $p$-values of the KS tests between the distributions of the phase-space coordinates in the spherical halo and those in the spheroidal halos of the same series. Table~\ref{tab:ks} shows the range of these $p$-values for the 50 series of mock catalogs.

The distributions of the magnitude of the latitudinal velocity, $|v_\vartheta|$,\footnote{$|v_\vartheta|$ is equivalent to the magnitude of the tangential velocity, $v_{\rm t}$, in a spheroidal DM halo axisymmetric about the $z$-axis.} are the only distributions to be significantly different in a spherical DM halo and in a spheroidal DM halo axisymmetric about the $z$-axis: for DM halos whose gravitational potential well displays a deviation from the spherical shape $|q_z-1| \ge 0.1$, we always got $p<\alpha$, implying that the null-hypothesis of the KS test could be rejected at the significance level $\alpha$. In particular, for the largest deviations from the spherical DM halo considered in this work, the $p$-value is $< 10^{-10}$.
On the other hand, for a very mild deviation of the potential well of the DM halo from the spherical shape (i.e., for $|q_z-1| \le 0.05$) the distributions of  $|v_\vartheta|$ are never significantly different from the spherical case. 
Finally, for DM halos whose potential well displays a deviation from the spherical shape $|q_z-1|$ in the range $(0.05-0.1)$, the distributions of  $|v_\vartheta|$ can either be or not be significantly different from the spherical case, depending on the set of stars' initial conditions used for the generation of mock catalogs A. For example, when $|q_z-1| = 0.075$, we found $p$ in the range $ 0.01-0.14$ for the case of spherical vs. oblate (with $q_z = 0.925$) DM halo, and $p$ in the range $0.01-0.11$ for the case of spherical vs. prolate (with $q_z = 1.075$) DM halo.

For all the distributions of phase space coordinates different from $|v_\vartheta|$, we found $p > 0.7$: we could not reject the null-hypothesis of the KS test at the significance level $\alpha=5\%$, regardless of the shape of the DM halo.
Therefore, we conclude that the magnitude of the HVS latitudinal velocity, $|v_\vartheta|$, is the only phase space coordinate whose distribution can be used to discriminate between a spherical and a spheroidal DM halo axisymmetric about the $z$-axis.

This result is not surprising. HVSs are ejected radially outward from the Galactic center, but they attain non-zero tangential velocities, $\vec v_{\rm t}$, as they travel through a non-spherically symmetric potential.
In our model (see Eq.~\ref{eq:phi_grav}), the potentials of both the SMBH and the bulge are spherically symmetric (see Eqs.~\ref{eq:smbh} - \ref{eq:bulge}). However, the disk potential (see Eq.~\ref{eq:disk}) is axially symmetric about the $z$-axis. Hence, it contributes to the component of the tangential velocity along the polar angle, that is the latitudinal velocity $v_\vartheta$.
When the DM halo is either spherical or spheroidal with axial symmetry about the $z$-axis, $v_\vartheta$ is still the only non-null component of the tangential velocity. In particular, when the DM halo is spherical, only the gravitational pull of the disk induces non-null $v_\vartheta$; on the other hand, when the DM halo is spheroidal with axial symmetry about the $z$-axis, the gravitational pull of the disk combines with that of the DM halo: in the case of prolate spheroidal DM halo ($q_z > 1$) the pull of the disk is opposite to that exerted by the DM halo, while in the case of oblate spheroidal DM halo ($q_z < 1$) both the disk and the halo attract the stars towards the Galactic plane, leading to higher tangential velocities.

\begin{table*} 
	\caption{Ranges of the $p$-values of the KS tests between the distribution of the HVS phase space coordinates in a spherical DM halo and those in different spheroidal halos. Each entry shows the ranges of $p$-values  obtained from a sample of 50 series of mock catalogs A. Each series is made of a spherical halo and spheroidal halos with axial symmetry about the $z$-axis with six different values of the triaxiality parameter $q_z$.
	Column 1: HVS phase space coordinate in the galactocentric reference frame. Columns 2-4: $p$-value of the KS test for a spherical halo compared against a prolate halo (with $q_z = 1.4$, $q_z = 1.1$, $q_z = 1.05$). Columns 5-7: $p$-value of the KS test for a spherical halo compared against an oblate halo (with $q_z = 0.7$, $q_z = 0.9$, $q_z = 0.95$). 
	A single $p$-value is shown instead of a $p$-value range when the relative difference between the limits of the $p$-value interval is smaller than $10^{-2}$.
	With the adopted 5\% significance level, $p$-values smaller than 5\% indicate that the compared distributions are not drawn from the same parent distribution, implying that the corresponding phase space variable can be an indicator of the shape of the DM halo.}
	\label{tab:ks}
	\begin{center}
	\setlength{\tabcolsep}{3pt}
		\begin{tabular}{ccccccc}
			\hline \hline
			Phase space &  & $p$-value                 &  &  & $p$-value & \\
			coordinate  &  & (spherical vs. prolate) &  &  & (spherical vs. oblate) & \\
			\hline
			& $q_z=1.4$ & $q_z = 1.1$ & $q_z = 1.05$ & $q_z = 0.7$ & $q_z = 0.9$ & $q_z = 0.95$\\
			\hline
			$x$	&1.00		&1.00	&1.00	&1.00		&1.00	&1.00\\
			$y$	&1.00		&1.00	&1.00	&1.00		&1.00	&1.00\\
			$z$	&[0.99 - 1.00]	&1.00	&1.00	&[0.88 - 1.00]	&1.00	&1.00\\
			$r$	&1.00		&1.00	&1.00	&1.00		&1.00	&1.00\\
			$\vartheta$&[0.94 - 1.00]	&1.00	&1.00	&[0.71 - 0.99]	&1.00	&1.00\\
			$v_x$	&[0.98 - 1.00]	&1.00	&1.00	&1.00		&1.00	&1.00\\
			$v_y$	&[0.98 - 1.00]	&1.00	&1.00	&1.00		&1.00	&1.00\\
			$v_z$	&[0.93 - 1.00]	&1.00	&1.00	&[0.76 - 0.99]	&1.00	&1.00\\
			$v_r$	&1.00		&1.00	&1.00	&1.00		&1.00	&1.00\\
			$v_{\rm t}=|v_\vartheta|$	&[6$\times 10^{-22}$ - 5$\times 10^{-14}$]	&[2$\times 10^{-4}$ - 0.01] &[0.06 - 0.44] &[7$\times 10^{-21}$ - 6$\times 10^{-11}$] &[8$\times 10^{-4}$ - 0.04] &[0.11 - 0.47]\\
			\hline			
		\end{tabular}
	\end{center}
\end{table*}

\begin{figure}[ht]
	\includegraphics[width=7.6cm,height=6.2cm]{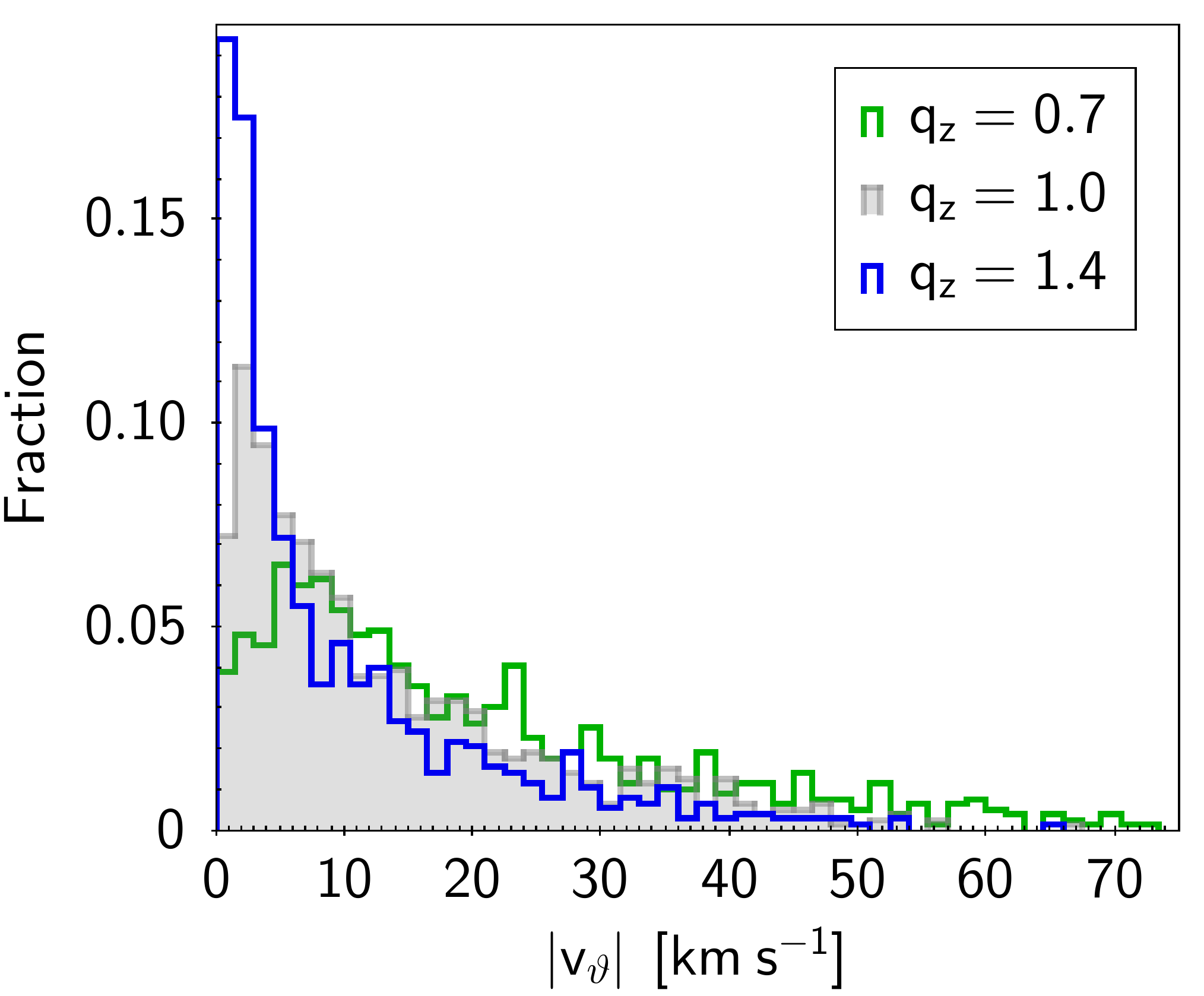}
	\caption{Distribution of the magnitude of the HVS galactocentric latitudinal velocity, $|v_\vartheta|$, in a Galaxy with an extremely prolate DM halo ($q_z = 1.4$; blue histogram), a spherical DM halo  ($q_z = 1$; gray, shaded histogram), and an extremely oblate DM halo ($q_z = 0.7$; green histogram). The distributions were generated with the same initial conditions (mock catalogs A), to highlight the effect of the different geometries of the DM halo. \label{fig:v_theta_axisymmetric}}
\end{figure}

This situation is illustrated in Fig.~\ref{fig:v_theta_axisymmetric}, where we show a comparison of the distributions of the magnitude of the latitudinal velocity, $|v_{\vartheta}|$, for simulated HVSs in a Galaxy with a spheroidal DM halo which is axisymmetric about the $z$-axis and characterized by an extremely prolate ($q_z = 1.4$), spherical ($q_z = 1$), and extremely oblate ($q_z = 0.7$) shape.
The fraction of HVSs with higher $|v_\vartheta|$ is larger in the case of the oblate DM halo (green histogram) than in the case of the spherical DM halo (gray, shaded histogram), because of the concordant gravitational pull of disk and DM halo. Conversely, the fraction of stars with lower $|v_\vartheta|$ is larger in the case of the prolate DM halo (blue histogram) than in the case of the spherical DM halo, because of the opposite pull of disk and DM halo.

The results reported in Table~\ref{tab:ks} show that,  
while the latitudinal velocity $v_\vartheta$ is an indicator of the shape of the DM halo, the polar angle $\vartheta$ is not, even though any difference in the distributions of $\vartheta$ for stars that have traveled in different DM halos is only determined by the halo shape, as $v_\vartheta$ is. This higher sensitivity of $v_\vartheta$ to the changes of the shape of the DM halo has two explanations: (i) $v_\vartheta$ is the time derivative of $\vartheta$; any variation in $v_\vartheta$ implies a variation $\Delta \vartheta$ of the coordinate $\vartheta$ over a finite time interval $\Delta t$; however, significant $\Delta \vartheta$ can be achieved only over $\Delta t$ that, on average, are larger than the HVS travel time; (ii) the $v_\vartheta$'s are null at the ejection, and the final distribution of $v_\vartheta$ is determined by the gravitational potential alone; instead, the distribution of $\vartheta$ is uniform in ${\rm cos}\vartheta$ at the start: hence, its final distribution is the result of the combination of the initial randomness and of the action of the gravitational potential.

Table~\ref{tab:ks} also shows that neither the Cartesian spatial coordinates nor the Cartesian components of the HVS velocity are significantly affected by the deviation from spherical of the shape of the DM halo; therefore, they are not useful indicators of the triaxiality of the DM halo.
This result is expected for the Cartesian spatial coordinates $x$ and $y$, as well as for the velocity components $v_x$ and $v_y$, because the gravitational potential is symmetric about the $z$-axis. On the other hand, for the same reason, the result is not fully expected for $z$ and $v_z$.
However, our simulations show that the projection, on the $z$-axis, of any non-null $v_\vartheta$ induced by a non-spherical gravitational potential in the star motion is overwhelmed by the projection, on the same axis, of the radial velocity component of the star. 
If $v_z$ is not a good indicator of the triaxiality of the DM halo, the spatial coordinate $z$ cannot be a good indicator either: as $\vartheta$ is less sensitive than $v_\vartheta$ to the shape of the DM halo, $z$ is less sensitive to the halo shape than $v_z$. 

The results that we obtained for a spheroidal DM halo symmetric about the $z$-axis (i.e., for a DM halo with $q_y=1$ and $q_z \ne 1$) can be extended to the more complex cases of a fully triaxial DM halo and of a spheroidal DM halo with a symmetry axis misaligned with respect to the $z$-axis (i.e., for a DM halo with $q_y \ne 1$).
In those cases, the stars acquire both a non-null latitudinal velocity, $v_\vartheta$, and a non-null azimuthal velocity, $v_\varphi$. Therefore, both components of $\vec v_{\rm t}=(v_\vartheta,v_\varphi)$ can be used as indicators of the triaxiality parameters of the DM halo. 

We note that significant differences between the distributions of HVS tangential velocity components in a spherical and in a non-spherical DM halo only emerge when those velocity components are computed in the galactocentric reference frame.
In the Galactic heliocentric reference frame, no phase space coordinate is characterized by distributions that significantly differ in the cases of spherical and non-spherical DM halos, according to the KS test. Specifically, this is the case for each of the components of the star velocity. Indeed, all the components of the heliocentric velocity $(v_d,v_l,v_b)$ are a composition of $v_{\rm r}$ and $\vec v_{\rm t}=(v_\vartheta,v_\varphi)$ (see Eqs.~\ref{eq:vd}-\ref{eq:vb}), and the information on the triaxiality parameters stored in the galactocentric $\vec v_{\rm t}$'s is diluted in the velocity transformation from the galactocentric to the heliocentric system. The same results hold for the phase space coordinates in the equatorial reference frame.

\subsection{Star kinematics and sample selection} 
\label{sec:sample_selection}

As illustrated in Sect.~\ref{sec:halo_impact}, the components $v_\vartheta$ and $v_\varphi$ of the tangential velocity of the HVSs can effectively probe the non-spherical components of the gravitational potential. 
This result was demonstrated for a subsample of mock HVSs characterized by $r>10$~kpc and $v_{\rm r}>0$.
Here, we show that these sample selection criteria turn out to be the most suitable selection criteria also on the basis of the star kinematics, for HVSs of $4\, M_{\sun}$. Indeed, these criteria enable us to select those stars whose tangential velocity is not affected by effects other than the shape of the gravitational potential well.
We adopted an analogous HVS sample selection based on stellar kinematics in \citet{chakrabarty2022}.

\begin{figure*}[ht]
	\centering
	\includegraphics[width=9.1cm,height=5.8cm]{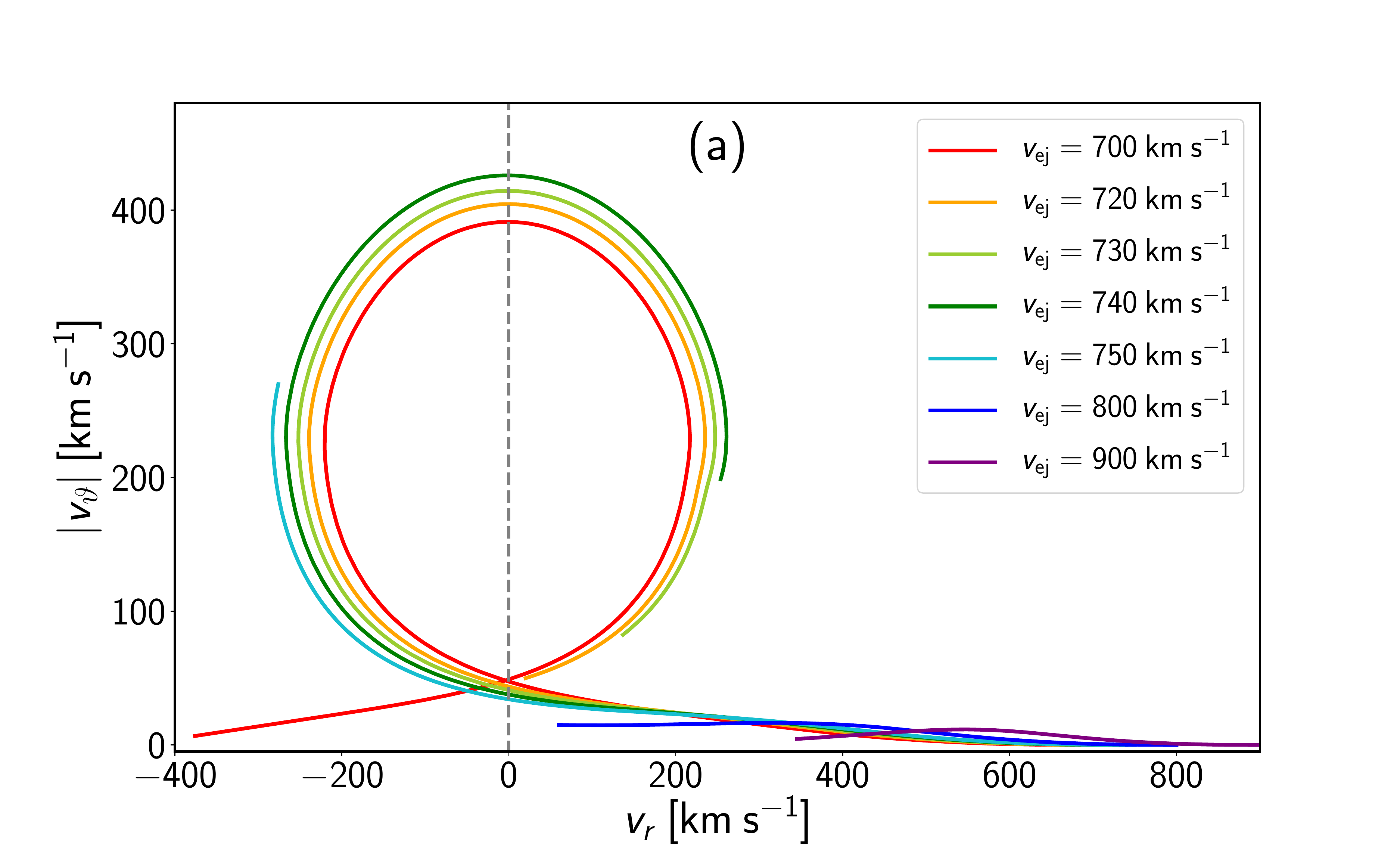}
	\includegraphics[width=8.9cm,height=5.8cm]{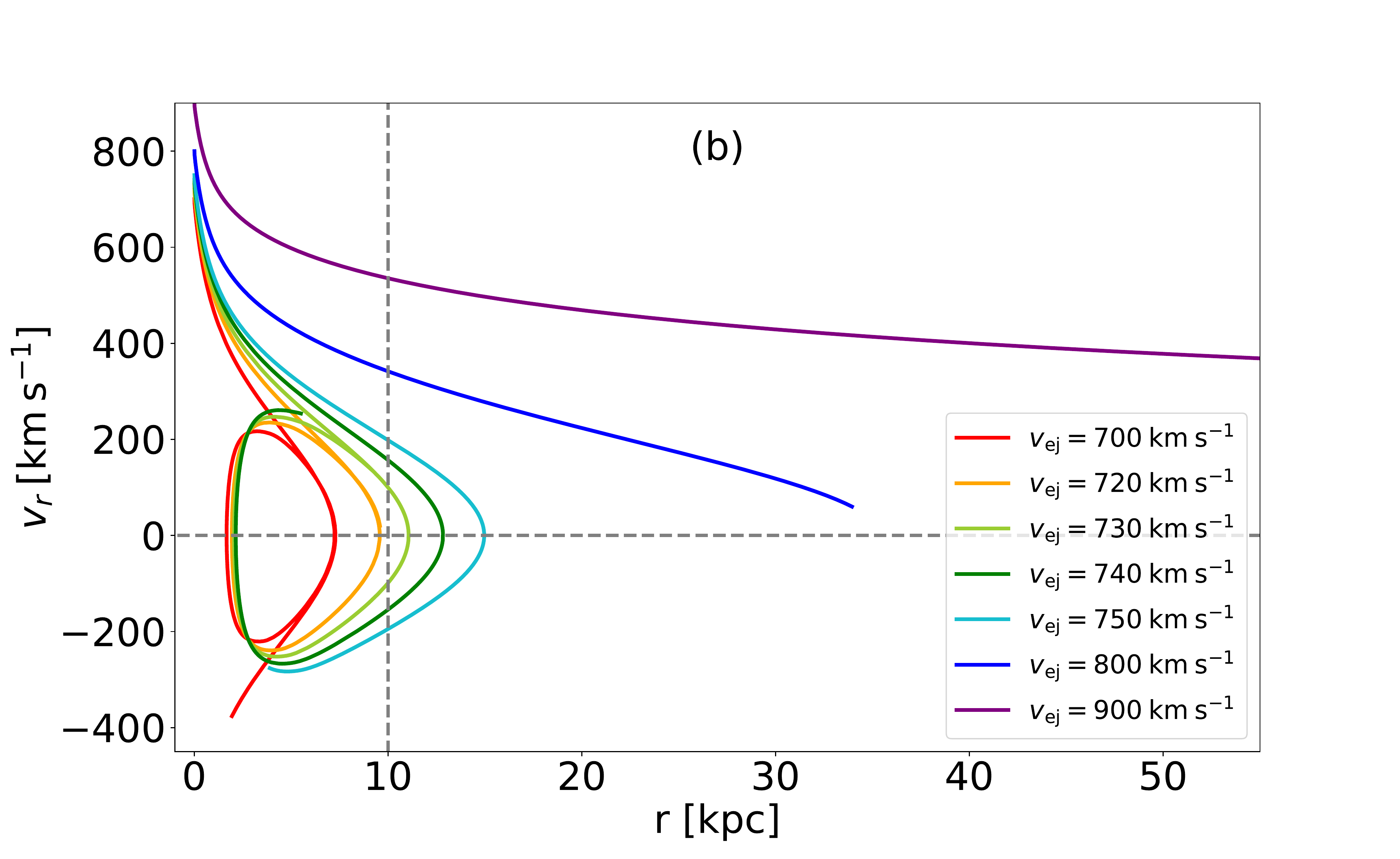}
	\caption{{\it Panel a}: Magnitude of the galactocentric latitudinal velocity, $|v_\vartheta|$, as a function of the galactocentric radial velocity $v_r$, for HVSs ejected with different speed $v_{\rm ej}$ and traveling in a spherical DM halo. Initially, the star velocity is purely radial, and $|v_\vartheta| = 0$; as time goes on, $|v_\vartheta|$  becomes non-null due to the non-spherical potential of the Galactic disk (see Eq.~\ref{eq:disk}). For stars with smaller ejection speed ($v_{\rm ej} \lesssim 800$~\kms), radial and angular dynamics are coupled: this coupling results into a fast growth of $|v_\vartheta|$ after the first turnaround of the star. However, such couplings are not manifested for stars with larger ejection speed. {\it Panel b}: galactocentric radial velocity $v_r$ as a function of the galactocentric distance $r$, for HVSs ejected radially outward with different ejection velocities $v_{\rm ej}$. The vertical dashed line of panel $a$ and the horizontal dashed line of panel $b$ corresponds to $v_r = 0$, while the vertical dashed line of panel $b$ corresponds to $r = 10$~kpc. In both panels, HVSs are ejected in the direction $\hat n (\vartheta, \varphi) = \hat n (45^\circ , 45^\circ)$ with a travel time of $160$~Myr, which is the largest possible travel time for a 4~$M_{\sun}$ star. 
	\label{fig:vtheta_vr_vr_r_different_vej}}
\end{figure*}

In our model of the Galactic potential (see Sect.~\ref{sec:MW_potential}), stars with ejection speed $v_{\rm ej} \gtrsim 800$~\kms~ always possess tangential velocities that are independent of their radial velocities, and are induced only by the non-spherical shape of the gravitational potential well. Those stars are robust indicators of the shape of the DM halo in any stage of their trajectory.
Conversely, for stars with ejection speed $v_{\rm ej} \lesssim 800$~\kms, the tangential velocity, $v_{\rm t}$, can be strongly coupled with the radial velocity, $v_{\rm r}$, for a significant fraction of their trajectory, and $v_{\rm t}$ can be very high regardless of the shape of the DM halo.
Indeed, if any of these stars is sufficiently young at ejection, it may experience the outer turnaround before dying out. At the outer turnaround, the star starts falling back toward the Galactic center, its radial velocity becomes negative, and its tangential velocity starts growing; this growth continues until the star experiences the inner turnaround, and then starts a new outward trajectory, with positive radial velocity and a tangential velocity that decreases with time. Therefore, those stars may serve as indicators of the halo shape only in specific parts of their outward trajectory, namely only during specific periods of their lifetime.

The situation is illustrated in Fig.~\ref{fig:vtheta_vr_vr_r_different_vej} for the case of a spherical DM halo, and for a set of HVSs ejected radially outward in the direction $\hat n (\vartheta, \varphi) = \hat n (45^\circ , 45^\circ)$ and with representative values of the ejection speed, $|v_{\rm ej}| = \{700, 720, 740, 800, 900 \}$~\kms.
All the stars are ejected with null age, $\tau_{\rm ej}=0$, to illustrate the evolution of the stars' observables during the largest possible travel time (i.e., 160 Myr, for the 4~$M_{\sun}$ stars considered in this work). Stars not ejected with null age would experience only part of the evolution shown in the figure. 
We note that the choice of the ejection direction of the HVSs does not affect any of the results that are presented in the following.

Panel $a$ of Fig.~\ref{fig:vtheta_vr_vr_r_different_vej} shows the relation between the magnitude of the tangential velocity, $v_{\rm t} = |v_\vartheta|$, and the radial velocity, $v_{\rm r}$, for the above set of stars. We note that the magnitude of the tangential velocity, $v_{\rm t} = (v_\vartheta^2 + v_\varphi^2)^{\frac{1}{2}}$ is equivalent to the magnitude of the latitudinal velocity $|v_{\vartheta}|$ because the DM halo is spherical: since the axisymmetric disk potential is the only source of non-zero $v_{\rm t}$, the azimuthal velocity $v_\varphi$ is null.

Panel $b$ of Fig.~\ref{fig:vtheta_vr_vr_r_different_vej} shows the relation between the radial velocity, $v_{\rm r}$, and the distance to the Galactic center, $r$, for the same set of HVSs.
Both the vertical dashed line in panel $a$, and the horizontal dashed line in panel $b$ correspond to a null radial velocity, $v_{\rm r}=0$, that the star possesses at its outer and inner turnaround radii.

For the HVSs generated with ejection speeds $v_{\rm ej} \gtrsim 800$~\kms~ (blue and purple lines), both panels $a$ and $b$ show that these stars can never reach their outer turnaround radius ($r=r_{\rm out}$) before dying out. Indeed, they never cross the line $v_{\rm r}=0$ from right to left in panel $a$, and the line $v_{\rm r}=0$ from top to bottom in panel $b$.

For these stars, panel $a$ shows that $|v_\vartheta|$ is always $\lesssim$ few 10~\kms, regardless of $v_{\rm r}$: these values of $v_\vartheta$ are entirely determined by the deviation from the spherical symmetry of the gravitational potential.
This deviation is due to the axisymmetric disk alone, in the case of the spherical DM halo considered here; it is due to a combination of disk and triaxial DM halo when the DM halo is non-spherical, as discussed in Sect.~\ref{sec:halo_impact}. 
At any stage of the stars' trajectories, the distributions of $|v_\vartheta|$ for these stars can be used as an indicator of the shape of the DM halo.

On the contrary, for the HVSs with ejection speed $v_{\rm ej}\lesssim 800$~\kms~ (cyan, dark green, green, orange, and red lines), both panels $a$ and $b$ show that these stars may undergo at least the first, outer turnaround before dying out.
Indeed, these stars may cross both the line $v_{\rm r}=0$ from right to left in panel $a$, and the line $v_{\rm r}=0$ from top to bottom in panel $b$ at the outer turnaround radius, $r=r_{\rm out}$, if they are ejected with sufficiently low age. 
When $v_{\rm ej} \lesssim 740$~\kms, the stars may also experience the second, inner turnaround. In this case, they cross again the line $v_{\rm r}=0$ from left to right in panel $a$, and correspondingly cross the line $v_{\rm r}=0$ from bottom to top in panel $b$, at the inner turnaround radius, $r=r_{\rm in}$.
Finally, when $v_{\rm ej} \lesssim 720$~\kms, the stars may also undergo subsequent turnarounds.

For the HVSs with $v_{\rm ej} \lesssim 800$~\kms,
panel $a$ shows that, after an initial phase where the star velocity is almost purely radial, with $|v_\vartheta| \lesssim$ few 10~\kms~ regardless of $v_{\rm r}$,  $|v_\vartheta|$ quickly increases after the outer turnaround and becomes very large for those stars that undergo the inner turnaround.  
These large values of $|v_\vartheta|$ are determined by the
exchange of kinetic energy between radial and angular degrees of freedom, rather than by the shape of the gravitational potential well, especially close to the inner turnaround.
Therefore, the tangential velocity of these stars could in principle be used as an indicator of the shape of the DM halo only in specific parts of the star's trajectory. 

To perform our investigation on the shape of the DM halo by means of the HVS tangential velocities, we needed to select only the stars that, at $t=t_{\rm obs}$, were on their first outward trajectory from the Galactic center, namely the stars that had not experienced an inner turnaround yet. 
Therefore, we first excluded the stars that, at $t=t_{\rm obs}$, were on an inward trajectory ($v_{\rm r} < 0$) toward the Galactic center: we thus required $v_{\rm r} > 0$ for the stars of our sample. 
From all the outgoing stars, we then excluded those that had already experienced the inner turnaround and may thus had uninterestingly large $|v_\vartheta|$. To do so, we note that panel $b$ of Fig.~\ref{fig:vtheta_vr_vr_r_different_vej} shows that stars characterized by ejection velocities $v_{\rm ej} \lesssim 740$~\kms~ may live long enough to experience at least one inner turnaround.  However, panel $b$ of Fig.~\ref{fig:vtheta_vr_vr_r_different_vej} also shows that these stars can never go back to galactocentric distances $r > 10$~kpc, after having undergone the inner turnaround.
We thus required a galactocentric distance $r > 10$~kpc for the stars of our sample. This detailed analysis of the star kinematics thus supports our choice of the two preliminary selection criteria, $v_{\rm r} > 0$ and $r > 10$~kpc, adopted in Sect.~\ref{sec:halo_impact}.

\begin{figure}[ht]
	\includegraphics[width=9.1cm,height=6.cm]{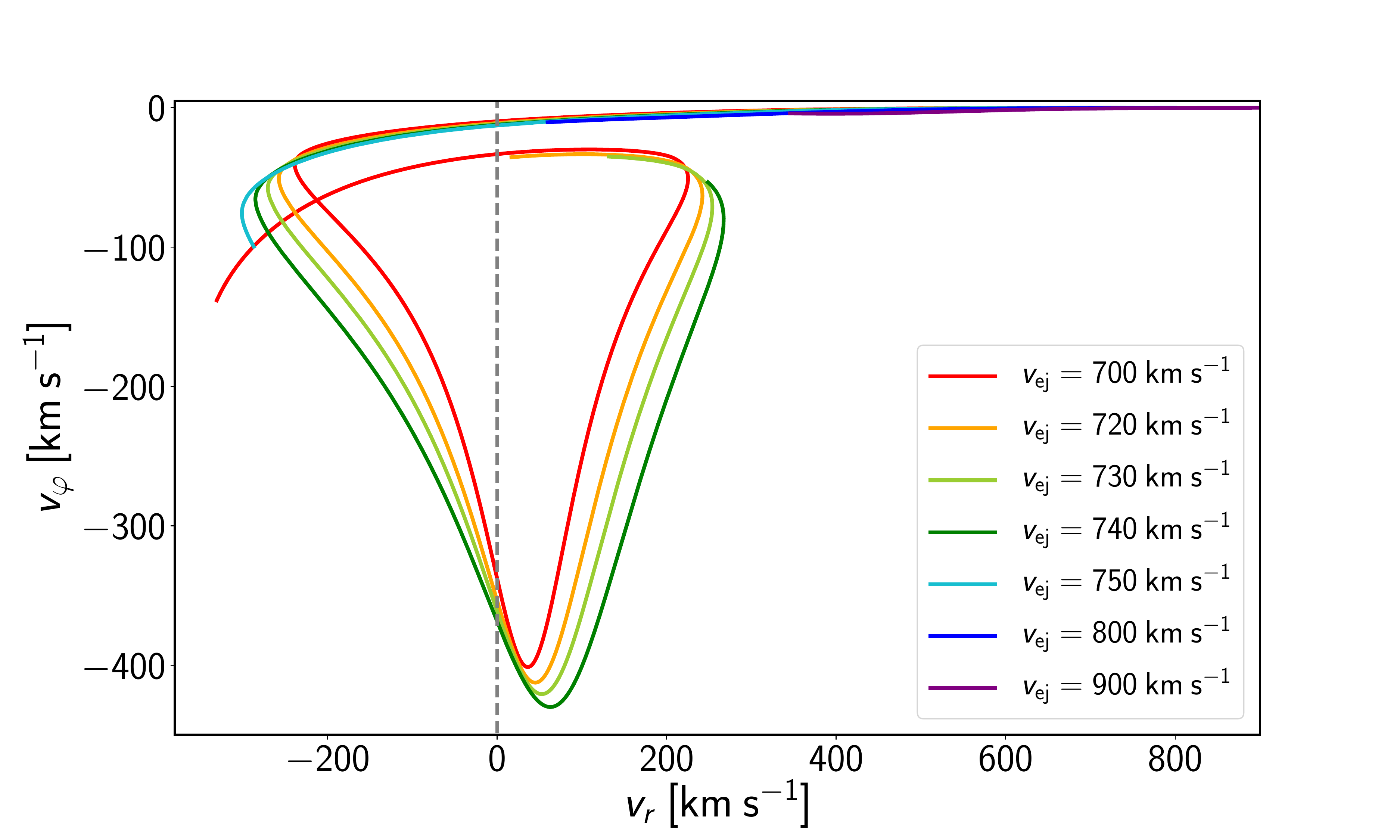}
	\caption{The HVS galactocentric	azimuthal velocity $v_\varphi$ as a function of the radial velocity $v_{\rm r}$, for HVSs ejected with different speed $v_{\rm ej}$ and traveling for 160 Myr.
	Here, the gravitational potential of the MW is non-axisymmetric, with a $q_y = 0.8$ and $q_z = 1.0$ DM halo potential. Initially, the star velocity is purely radial, and $v_\varphi = 0$; as time goes on,  $v_\varphi$ becomes non-null due to the DM halo asymmetry with respect to the $z$-axis (see Eq.~\ref{eq:halo}). Radial and angular dynamics are coupled for stars with smaller ejection speed ($v_{\rm ej} \lesssim 800$~\kms); these stars start  acquiring significant $v_\varphi$ values after the outer turnaround. However, stars with larger ejection speed die before this increase of $v_\varphi$ can happen. The gray dashed line corresponds to $v_r = 0$. 
	\label{fig:vphi_vr_different_vej}}
\end{figure}

\begin{figure}[ht]
	\includegraphics[width=9.7cm,height=8.4cm]{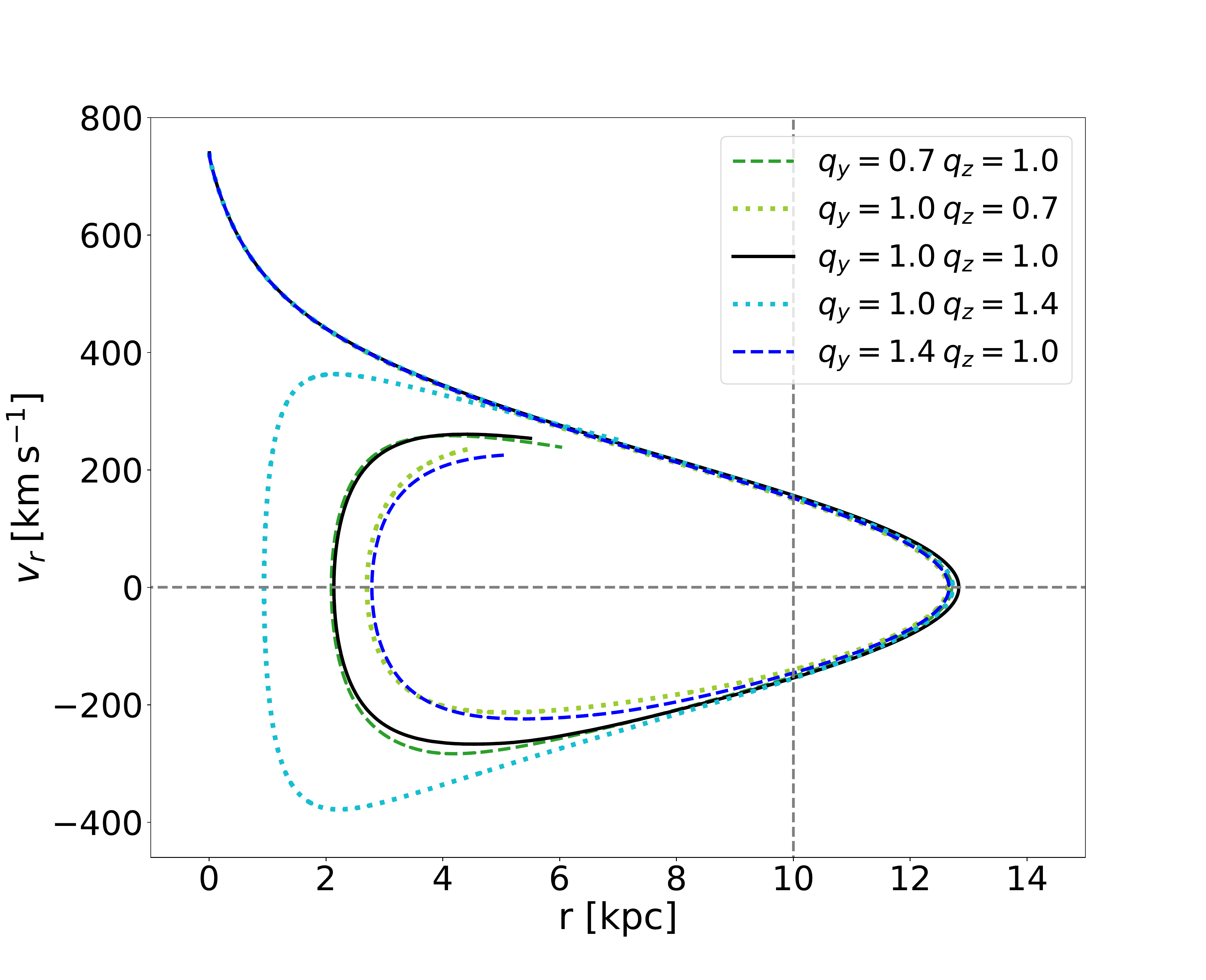}
	\caption{Galactocentric radial velocity $v_r$ as a function of the galactocentric distance $r$ for HVSs ejected radially outward with an ejection velocity of $v_{\rm ej} = 740$~\kms~ in the direction $\hat n (\vartheta, \varphi) = \hat n (45^\circ , 45^\circ)$, for different shapes of the DM halo. The triaxiality parameters ($q_y,q_z$) of the DM halos are listed in the inset. The travel time is $160$ Myr. The vertical and the horizontal dashed lines correspond to $r = 10$~kpc and $v_r = 0$, respectively.
	\label{fig:vr_r_different_shapes}}
\end{figure}

The case of a non-spherical DM halo is slightly more complicated than the case of the spherical halo explored so far. Here, both the disk and the DM halo generate a non-null latitudinal velocity, $v_\vartheta$; furthermore, fully triaxial or spheroidal DM halos with a symmetry axis misaligned with respect to the $z$-axis are also sources of a non-null azimuthal velocity, $v_\varphi$, because the axial symmetry of the Galactic potential is broken. 

However, we again find that both $v_\vartheta$ and $v_\varphi$ are strongly coupled with $v_{\rm r}$ for the stars with $v_{\rm ej} \lesssim 800$~\kms, as it happened for $v_\vartheta$ in the case of a spherical DM halo.
As an example, for the case of a $(q_y,q_z) = (0.8,1)$ DM halo, Fig.~\ref{fig:vphi_vr_different_vej} shows the relation between the azimuthal  and  radial velocity components. The stars' travel time is again 160 Myr, namely the largest possible travel time for 4~$M_\sun$ stars.
The figure shows that for stars with $v_{\rm ej} \lesssim 800$~\kms, $|v_\varphi|$ starts increasing after the outer turnaround, and reaches very large values close to the inner turnaround, irrespective of the shape of the DM halo that generated the non-null $v_\varphi$'s. 
To exclude all the stars that had acquired large $v_\varphi$ and correspondingly large $v_\theta$ after the outer turnaround, we again selected stars with $v_r > 0$ and $r>10$~kpc, as we did in the case of a spherical DM halo.

We note that these two selection criteria, illustrated for both a spherical DM halo (see Fig.~\ref{fig:vtheta_vr_vr_r_different_vej}) and a non-spherical DM halo with $(q_y,q_z) = (0.8,1)$ (see Fig.~\ref{fig:vphi_vr_different_vej}), actually hold for a DM halo with any combination of axis ratios among those investigated in this work, as illustrated in Fig.~\ref{fig:vr_r_different_shapes}. This figure shows the radial velocity, $v_r$, as a function of the galactocentric distance, $r$, for a star ejected at $v_{\rm ej} = 740$~\kms~ and traveling in a Galaxy with DM halos of different shapes: a spherical DM halo, an extremely oblate or prolate DM halo which is axisymmetric about the $z$-axis, and an extremely oblate or prolate DM halo which is axisymmetric about the $y$-axis.
Regardless of the axis ratios, stars ejected at $v_{\rm ej} = 740$~\kms~ can never reach a galactocentric distance $r > 10$~kpc, after having experienced the inner turnaround.

We emphasize that the stars' outer and inner turnaround radii depend on the gravitational potential chosen for the computation of the trajectories \citep[see also][]{chakrabarty2022}. 
Thus, the kinematic selection criteria derived in this work  
would differ in different models of the Galactic gravitational potential.
In addition, the kinematic selection criteria depend on the mass of the HVSs. In particular, HVSs with $M<4\, M_{\sun}$ would have longer lifetimes, and thus a higher probability to reach the outer turnaround, undergo the inner turnaround, and go back to large galactocentric distances before dying out; this would result in a threshold radius larger than 10 kpc.

\subsection{Effect of the initial conditions on the distributions of the shape indicators}
\label{sec:IC_impact}

In Sect.~\ref{sec:halo_impact}, we identified the distribution of the magnitude of the HVS latitudinal velocity, $|v_\vartheta|$, as the only distribution of HVS phase space coordinates to be significantly affected by a change of shape of the DM halo from spherical to spheroidal with axis of symmetry about the $z$-axis. 
Based on this finding, and extending the argument to a more general, non-axisymmetric Galactic potential, we identified the components of $\vec v_{\rm t}$ (i.e., $v_\vartheta$ and $v_\varphi$) as indicators of the triaxiality parameters of a DM halo.
Hereafter, we generally refer to $\omega$ as each of the quantities used as an indicator of the shape of the DM halo; we define the distribution of the shape indicator $\omega$ as $D_\omega$.

In Sect.~\ref{sec:halo_impact} we presented results obtained for simulated HVSs from mock catalogs A, that are unambiguously determined by the shape of the DM halo because the set of stars is ejected with the same initial conditions in halos of different shapes.
If a set of real HVSs were ejected with the initial conditions used to generate our mock catalogs A, recovering the shape of the MW DM halo from the distributions $D_\omega$ of real stars would be straightforward.
In reality, however, the situation is more complex.
At any given time $t$, we can observe the phase space distribution of a sample of HVSs that are traveling in a DM halo whose unknown shape we want to recover. Furthermore, the ejection conditions of these stars are also unknown: even assuming an ejection mechanism (e.g., the Hills mechanism, in our case), we do not know the initial conditions of the stars' trajectories, which are subject to statistical fluctuations.
Therefore, recovering the shape of the DM halo requires the comparison of the distributions $D_\omega$ of the real HVS sample with mock $D_\omega$'s that were generated with all possible combinations of initial conditions and shape of the DM halo.  
In the following, we illustrate the impact of the statistical fluctuations of the initial conditions of the ejected stars on the distributions $D_\omega$ and on our ability to detect deviations from spherical of the shape of the DM halo.

When simulating the trajectories of a sample of HVSs, a different set of initial conditions for the trajectories yields a different distribution of the HVS phase space coordinates at $t=t_{\rm obs}$. Consequently, the results of the KS test, that compares the distribution $D_\omega$ of the shape indicators against a reference set of $D_\omega$'s of DM halos with different degrees of triaxiality, depend on the initial conditions.
In Sect.~\ref{sec:halo_impact}, we already explored the effect of using different sets of initial conditions, each of them applied to all mock catalogs A generated for a series of spheroidal DM halos with different shapes.
These different sets are responsible for the fluctuations of the $p$-value of the KS test within the ranges reported in Table \ref{tab:ks} for some of the HVS phase space coordinates. As a result, mild deviations from spherical of the shape of the DM halo (i.e., $0.05<|q_z-1|<0.1$) may not be recognized.

The situation becomes more complex when different sets of initial conditions are applied to each mock catalog.
To investigate this case, we resorted to our HVS mock catalogs B (see Sect.~\ref{sec:mock_catalogs}): these catalogs differ from one another both for the shape of the DM halo, as catalogs A, and for the set of the stars' initial conditions. 

Depending on the combination of initial conditions and triaxiality parameters that characterizes each mock catalog B, the comparison of the resulting $D_\omega$'s for any of these catalogs with a catalog of a spherical halo, based on the KS test, may lead to two different incorrect conclusions: (i) $p < \alpha$ (with $\alpha = 5\%$), suggesting a deviation from the spherical halo,
at the significance level $\alpha$, even though both star samples have traveled in the same, spherical DM halo;
(ii) $p > \alpha$, suggesting that both the DM halos crossed by the two star samples are spherical, even though one of the two halos is not.

Situation (i) would never occur in the comparison of two catalogs A, with the same initial conditions: the KS test for two $D_\omega$'s from different  catalogs A would always yield $p=100\%$ for identical halo shapes.
Conversely, for two mock catalogs B, situation (i) can occur as a consequence of a fundamental property of the $p$-value: when we compare two $D_\omega$'s with the KS test, if the null hypothesis is true (i.e., if the two $D_\omega$'s are drawn from the same parent distribution), the value of the KS test statistics will be at least as large as the observed value in a fraction $p$ of the cases \citep[e.g.,][]{press2007}.
Because the $p$-value of the KS test is a random variable itself, and because for a true null hypothesis it is uniformly distributed in the range $[0;1]$ \citep[e.g.,][]{Hung1997,Donahue1999,bhattacharya2002}, one time out of 20 the $p$-value will be $\le 5\%$ for $D_\omega$'s that are drawn from the same parent distribution. At the significance level $\alpha = 5\%$, this probability implies that a null hypothesis that is actually true (i.e., the shapes of the two DM halos are equal) will be rejected in 5\% of the cases.
\footnote{In statistical hypothesis testing, this incorrect rejection is known as ``type I error'', or ``false positive''.}

Situation (ii) can instead occur also for catalogs A, when the shapes of the DM halos are only mildly different. However, our success in detecting actual deviations from the spherical shape with the KS test applied to the $D_\omega$'s is lower for catalogs B because of the additional effect of the statistical fluctuations of the initial conditions.
Indeed, whereas in catalogs A we cannot distinguish a spherical DM halo from a spheroidal DM halo, symmetric about the $z$-axis, whose deviation from spherical is $|q_z - 1| < 0.1$, with catalogs B the range of non-detectable deviations widens to $|q_z-1| \lesssim 0.2$. 
The failure to detect the different shapes of DM halos is a failure to reject a null hypothesis that is actually false.
\footnote{In statistical hypothesis testing, this failure to reject a false null hypothesis is known as ``type II error'', or ``false negative''.}

We note that false negatives may significantly hamper the recovery of the halo shape, especially for mild deviation from spherical of the shape of a spheroidal DM halo axisymmetric about the $z$-axis.
As an example, we mention that comparing the $D_\omega$'s obtained in a spheroidal DM halo axisymmetric about the $z$-axis and with $q_z = 0.9$ (i.e., $|q_z -1|=0.1$) against those obtained in a spherical DM halo yields a rate of false negatives of $\sim 20\%$: in other words, we are not able to distinguish the shapes of the two DM halos in $\sim 20\%$ of the cases.
Even though the rate of false negatives decreases with an increasing deviation from the spherical shape, and it becomes null for $|q_z-1| \gtrsim 0.2$, it is important to reduce the rate of false negatives to improve the efficiency of the shape recovery of the DM halo. 

Summarizing, when the initial conditions of the trajectories of a set of HVSs are unknown, as it happens for real HVSs, comparing the $D_\omega$'s of real stars with mock $D_\omega$'s with a KS test may yield results on the shape of the DM halo which are significantly affected by the initial conditions of real HVSs, especially for mild deviations of the shape of the DM halo from spherical. 
Therefore, it is important to design a method that optimizes the recovery of the shape of the DM halo, minimizing the effects of the statistical fluctuations of the stars' initial conditions. We propose this optimized method in the following.


\section{A new method to constrain the shape of the DM halo} 
\label{sec:method}
Our final goal is to recover the shape of the DM halo from the distribution $D_\omega$ of the shape indicators $\omega$ of a real sample of HVSs, properly accounting for the effects of the unknown initial conditions of the stars' trajectories.
To this aim, we designed a method based on (i) the use of the KS test to compare the $D_\omega$ of the sample of real HVSs  with the corresponding distributions generated with a series of DM halos of different shape; (ii) the property of the KS test's $p$-value mentioned in Sect.~\ref{sec:IC_impact}: 
its uniform probability density function for a true null hypothesis which, in our case, occurs when two $D_\omega$'s are drawn from the same parent distribution.

To implement the method and evaluate its efficiency, we resorted to our HVS mock catalogs B (see Sect.~\ref{sec:mock_catalogs}), that differ from one another for the shape of the DM halo and/or the set of the stars' initial conditions.
We also constructed a sample of synthetic HVSs, hereafter referred to as the ``observed sample'': the stars of this sample are ejected from the Galactic center, according to the Hills mechanism,  with a statistically random set of initial conditions, and move across a Galaxy whose DM halo has a known shape. 
In our analysis, the observed sample mimics a real sample of HVSs: we used it to test the efficiency of our method in recovering the correct, known shape of the DM halo from its $D_\omega$  at $t=t_{\rm obs}$. 
The distributions of the kinematic properties of one of our observed samples is highlighted in green color in Fig.~\ref{fig:mock_A_triaxial}.
We emphasize that our HVS observed samples are ideal: we include neither observational uncertainties nor observational cuts imposed by the star observability, like its magnitude or position within the Galaxy. The method success rates that we estimate in Sects.~\ref{sec:success_rate_axisymmetric_Galactic_potential} and \ref{sec:success_rate_non-axisymmetric_Galactic_potential} are thus valid for these samples alone and not for the HVS samples that might actually be observed.

We illustrate the basic concepts of our method in Sect.~\ref{sec:method_fundamentals}, and the method implementation in Sect.~\ref{sec:method_implementation}.

\subsection{Fundamentals of the method} 
\label{sec:method_fundamentals}

In our approach, recovering the shape of the DM halo crossed by an  observed sample of HVSs requires the comparison of the  $D_\omega$  of the  observed sample   with a series of corresponding $D_\omega$'s generated in mock catalogs characterized by a different shape of the DM halo and by a different set of initial conditions. 
We considered the shape of the DM halo of the mock sample that best matches the  observed sample  as the actual shape of the DM  halo crossed by the HVS  observed sample. 

In principle, the comparison could be performed by means of a KS test, whose null hypothesis $H_0$ states that the two compared $D_\omega$'s are drawn from the same parent distribution.
At the significance level $\alpha$, we would accept $H_0$ in those cases where the test returns $p>\alpha$, and we would reject it otherwise. Accepting $H_0$ would correspond to considering the $D_\omega$ of the  observed sample  indistinguishable from the mock $D_\omega$  selected for the comparison, thus associating to the  observed sample  a DM halo with the same shape of the mock sample's DM halo.
However, because of the effect of the statistical fluctuations of the initial conditions of the stars' trajectories (see Sect.~\ref{sec:IC_impact}), the $D_\omega$ of the   observed sample  may either turn out to be indistinguishable from a significant number of mock $D_\omega$'s obtained in DM halos with different shapes (false negatives), or turn out to be significantly different from mock $D_\omega$'s obtained in DM halos with identical shapes (false positives). 

To select the ``best match'' between observed and mock sample, we resorted to a property of the $p$-value, already mentioned in Sect.~\ref{sec:IC_impact}.
When the null hypothesis $H_0$ is true, the $p$-value is uniformly distributed in the range $[0;1]$ \citep[e.g.,][]{Hung1997,Donahue1999}; thus, its median value is $p_{\rm med} =  0.5$. 
On the other hand, when the alternative hypothesis, $H_{\rm a}$, is true, the distribution of the $p$-values is markedly skewed towards low $p$-values,
because small values of $p$ are more likely; thus, the median $p$-value under $H_{\rm a}$ will be $p_{\rm med}\ll 0.5$. 

As a consequence, performing the KS test for a number $n_{\rm t}$ of $D_\omega$ pairs that are randomly drawn from the same parent distribution, namely   
from the ensemble of the $D_\omega$ obtained from mock catalogs B that are characterized by the same shape of the DM halo but by different initial conditions, 
yields a uniform distribution of $p$-values. 
Conversely, performing the KS test for a number $n_{\rm t}$ of $D_\omega$ pairs that are randomly drawn from different  parent distributions, namely from different mock catalogs B, each of them characterized by a different shape of the DM halo and different initial conditions, yields a distribution of $p$-values  which is markedly skewed towards small $p$-values. 
The larger is the difference in shape, the more skewed is the distribution. 

\begin{figure}[ht]
	\includegraphics[width=7.9cm,height=5.6cm]{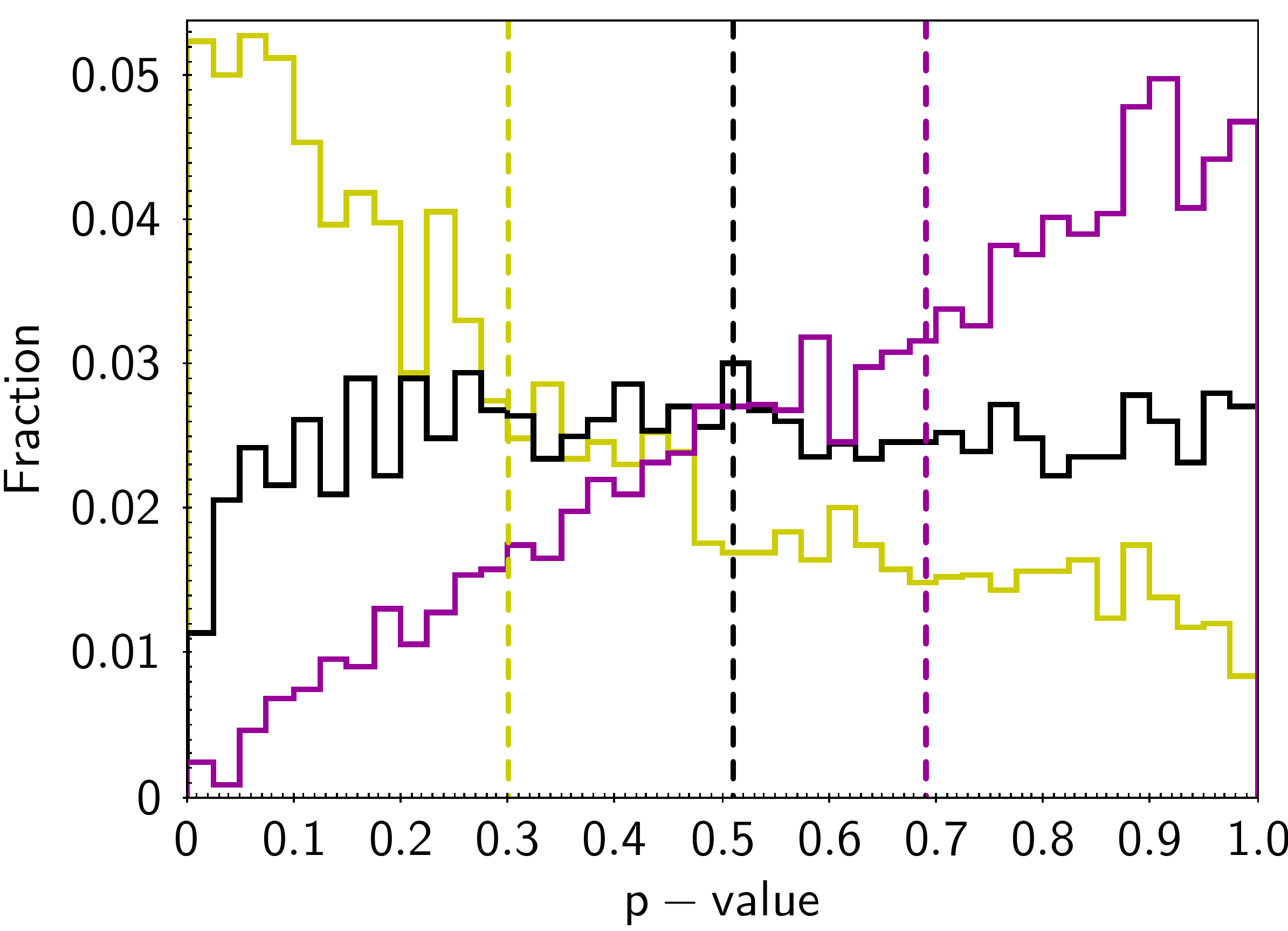}
	\caption{Distributions of the $p$-values obtained by comparing $\tilde D_\omega$ from each of $n=3$ observed samples of HVSs against the $D_\omega$'s of a series of $n_{\rm t}=5,000$ mock samples. The observed samples and the mock samples have all crossed a spherical DM halo. The black histogram shows an approximately uniform distribution of $p$, with a median value $p_{\rm med}\simeq 0.5$; the olive histogram shows the case of a skewed distribution with $p_{\rm med}\simeq 0.3$; the purple histogram shows the case of a skewed distribution with $p_{\rm med}\simeq 0.7$. The median $p$-values of the three distributions are marked by the vertical dashed lines.
	\label{fig:fluctuation}}
\end{figure}

The situation becomes more complex when we pick a specific distribution of $\omega$, say $\tilde D_\omega$, as that of the  observed sample, and we compare it against a series of mock distributions, performing a number $n_{\rm t}$ of KS tests.
Even though both the HVS  observed sample  and the mock samples have traveled in the same DM halo, the distributions of the $p$-values is not necessarily uniform. It may be approximately uniform over the range $[0;1]$, skewed towards low $p$-values, or skewed towards high $p$-values, with a corresponding median $p$-value $p_{\rm med}\simeq 0.5$, $p_{\rm med}< 0.5$, and $p_{\rm med}>0.5$, respectively.
This situation is illustrated in Fig.~\ref{fig:fluctuation}, that shows the distributions of the $p$-values obtained from $n_{\rm t} = 5,000$ KS test comparisons of $\tilde D_\omega$ for each of $n=3$ different observed samples against all the $D_\omega$'s of a mock catalog B generated with a spherical DM halo. 
In each of the three cases, the  observed sample  and all the mock samples of HVSs have traveled in the same, spherical DM halo; however, the $p$-value distributions are markedly different: the black histogram shows the case of an approximately uniform distribution, with $p_{\rm med}\simeq 0.5$; the olive histogram shows the case of a distribution skewed towards low $p$-values, with $p_{\rm med}\simeq 0.3$; the purple histogram shows the case of a distribution skewed towards high $p$-values, with $p_{\rm med} \simeq 0.7$.

\begin{figure}[ht]
	\includegraphics[width=8.3cm,height=6.1cm]{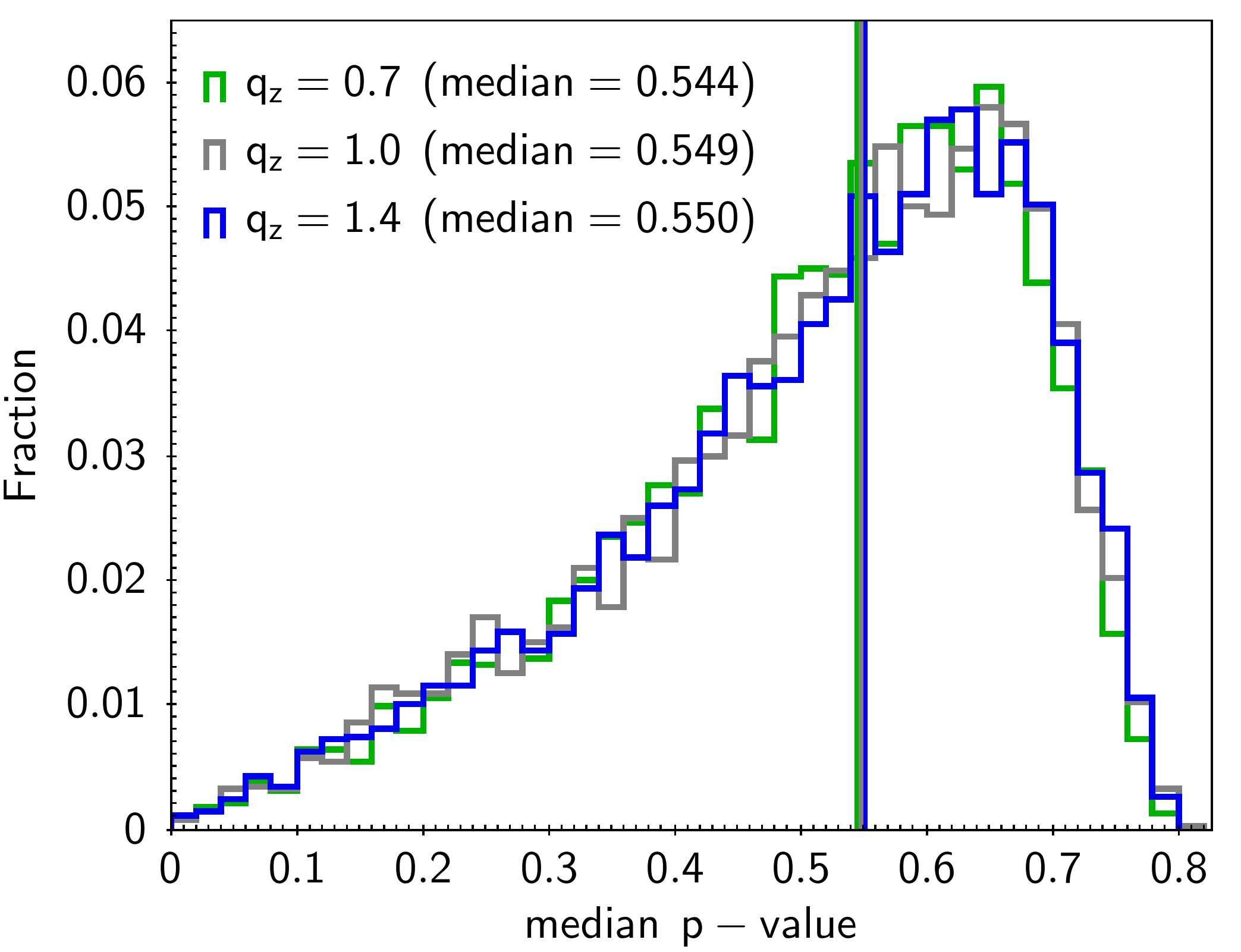}
	\caption{Distribution of the median $p$-values obtained from the KS test comparison of $\tilde D_\omega$ from all possible observed samples against the $D_\omega$'s from mock catalogs B generated in the same DM halo; the DM halo is spherical (gray histogram) or spheroidal with axis of symmetry about the $z$-axis with an oblate ($q_z$=0.7; green histogram) or prolate ($q_z=1.4$; blue histogram) shape. The gray, green, and blue solid lines are the medians of the corresponding distributions. 
	\label{fig:med_distr}}
\end{figure} 

The fact that a uniform distribution of $p$-values is not guaranteed even for the same shapes of the DM halo is a consequence of the random selection of the observed sample's $\tilde D_\omega$ that we compare against all of the mock $D_\omega$'s. 
If we repeat the exercise of Fig.~\ref{fig:fluctuation} with a series of $n > 3$ observed samples, we find $n$ different distributions of $p$, each of them with a different median $p$-value. However, for sufficiently large $n$, the sum of all these distributions yields a uniform distribution.
Moreover, the distribution of the $n$ median $p$-values is markedly skewed towards high $p_{\rm med}$'s, as the one shown in Fig.~\ref{fig:med_distr}: here, the gray  histogram is obtained assuming in turn as observed sample each of the $n_{\rm t}=5,000$ mock catalogs B with a spherical DM halo and comparing its $\tilde  D_\omega$ against all the $D_\omega$'s of the remaining mock catalogs, for a total of $n=n_{\rm t}=5,000$ KS tests per observed sample, and $n_{\rm t}$ corresponding $p$-values whose distribution has a median $p_{\rm med}$. Repeating the exercise $n=n_{\rm t}=5,000$ times yields a distribution of $n=n_{\rm t}=5,000$ median $p$-values.
In this distribution, $\sim 80\%$ of the $p_{\rm med}$'s are in the range $0.3-0.7$, and the median is $M_0 \simeq 0.55$.

We note that the characteristics of the distribution of the median $p$-values primarily depend on the validity of $H_0$; indeed, the distribution is independent of the shape of the DM halo under test. As shown in  Fig.~\ref{fig:med_distr}, the $p_{\rm med}$'s have indistinguishable  distributions for extremely prolate ($q_z=1.4$; blue histogram), spherical ($q_z=1$; gray histogram), and extremely oblate ($q_z=0.7$; green histogram) DM halos. The compatibility of the three distributions is also confirmed by a KS test.
For all of the distributions, the median $p_{\rm med}$ is $M_0 \simeq 0.55$.

Summarizing,
Figs.~\ref{fig:fluctuation} and \ref{fig:med_distr} show 
how the KS test comparison of the $\tilde D_\omega$ of an observed sample of HVSs with a series of $n_{\rm t}$ mock $D_\omega$'s may  yield a $p$-value distribution different from uniform even though the observed sample and the mock sample crossed DM halos with the same shape. This variety of $p$-value distributions corresponds to a variety of $p_{\rm med}$'s. Higher $p_{\rm med}$'s are more likely than lower $p_{\rm med}$'s, implying that a distribution of $p$-values skewed towards higher $p$'s is more likely to occur under the null hypothesis $H_0$. 
Conversely, a distribution of $p$-values skewed towards lower $p$'s, and thus characterized by a lower $p_{\rm med}$, is a more likely outcome under the alternative hypothesis $H_{\rm a}$ that the DM halos have different shapes.

Therefore, to select the ``best match'' between the observed sample and the mock sample, we chose the $p$-value distribution characterized by the largest $p_{\rm med}$.
The DM halo crossed by the HVS observed sample was then assigned the shape of the mock sample that yielded the ``best match''.

\subsection{Implementation of the method} 
\label{sec:method_implementation}

To implement our method, we first explored the case of an axisymmetric Galactic potential, that includes the case of a spherical DM halo and that of a spheroidal DM halo axisymmetric about the $z$-axis; we then explored the more complex case of a non-axisymmetric Galactic potential, that includes the case of a spheroidal DM halo with a symmetry axis misaligned with respect to the $z$-axis, and that of a fully triaxial DM halo. 

In both the axisymmetric and non-axisymmetric case studies, we resorted to mock catalogs B ({Sect.~\ref{sec:mock_catalogs}}), that we constructed as follows.
We defined a series of $n_{\rm s}$ reference shapes for the DM halo, characterized by: 
(i) triaxiality parameters $q_y=1$ and $q_z$ that varied in steps of 0.1 within the range $0.7-1.4$, for the axisymmetric Galactic potential;
(ii) triaxiality parameters $q_y$ and $q_z$ that both varied in steps of 0.1 within the range $0.7-1.4$, and $q_y \ne 1$, for the non-axisymmetric Galactic potential.
For each of the $n_{\rm s}$ shapes of the DM halo, we generated an ensemble of $n_{\rm t}=5,000$ HVS mock catalogs B, each of them including the phase space coordinates of a sample of HVSs characterized by a different random set of initial conditions.
We also generated one observed sample of HVSs, randomly drawn from one of the mock catalogs B: this sample plays the role of a real HVS sample, and was used to evaluate the efficiency of the method in recovering the known shape of the DM halo crossed by the observed sample. 

For each of the $n_{\rm s}$ simulated shapes of the DM halo, we performed a number $n_{\rm t}=5,000$ of KS test comparisons of the $\tilde D_\omega$ of the observed sample against the $D_\omega$'s of the mock samples.
From these $n_{\rm t}$ KS tests, we got a distribution of $n_{\rm t}$ $p$-values, whose median is $p_{\rm med}$.
We repeated the procedure for all the $n_{\rm s}$ simulated shapes, and eventually obtained a set of $n_{\rm s}$ $p_{\rm med}$'s.
We selected the mock sample corresponding to the largest value of $p_{\rm med}$ as the sample that ``best matched'' the observed sample, and we associated the shape of its DM halo to the DM halo of the observed sample.

To evaluate the success rate of our method in recovering the shape of the DM halo crossed by the observed sample, we generated a series of $n=n_{\rm t}$ observed samples corresponding to the same shape of the DM halo, by randomly varying the set of the HVS initial conditions, and we computed the fraction of cases where the method correctly recovers the known shape of the DM halo crossed by the observed sample.
Finally, to study the dependence of the success rate of our method on the shape of the DM halo, we repeated our analysis for the observed samples of HVSs drawn from DM halos of different shapes.

In Sect.~\ref{sec:axisymmetric_Galactic_potential} we illustrate the application of our method to a series of observed samples generated in an axisymmetric Galactic potential, where the shape indicator $\omega$ is $|v_\vartheta|$. The test used for our analysis is the two-sample, one-dimensional KS test that we also used in the previous sections. 
In Sect.~\ref{sec:non-axisymmetric_Galactic_potential} we illustrate the application of our method to a series of observed samples generated in a non-axisymmetric Galactic potential, where the shape indicators $\omega$ are $|v_\vartheta|$ and a function of $v_\varphi$.
The test used for our analysis is the two-sample, two-dimensional KS test.

A possible alternative to the KS test chosen for our method is represented by the Anderson-Darling (AD) test \citep{AD_1952,AD_1954}. 
However, the AD test currently exists only in its one-dimensional formulation; the use of this test would thus be limited to the case of axisymmetric Galactic potentials, when the shape of the DM halo can be recovered from the distribution of one shape indicator only, $|v_{\vartheta}|$. 
In the comparison of two one-dimensional distributions of $|v_{\vartheta}|$, the AD test is more sensitive to the tails of the distributions than the KS test. These tails are sensibly affected by the shape of the DM halo, as can be seen from Fig.~\ref{fig:v_theta_axisymmetric}.
Therefore, the use of the AD test for our method is expected to improve our method success rate (see Sect.~\ref{sec:success_rate_axisymmetric_Galactic_potential}) in recovering the correct shape of an axisymmetric DM halo.
Consequently, our choice of the KS test as the statistical test of our method represents a conservative choice in terms of method success rate.

Unlike the AD test, the KS test is available also in its two-dimensional formulation. We can thus also apply the method to non-axisymmetric Galactic potentials, where the comparison of the distributions of the two shape indicators, $|v_{\vartheta}|$ and a function of $v_{\varphi}$, is required.
Adopting the same statistical test guarantees a consistent comparison of the success rate in axisymmetric and non-axisymmetric scenarios.

We remind that, from mock catalogs B and for the observed sample, we always selected subsamples of stars that fulfill the criteria defined in Sect.~\ref{sec:sample_selection}, namely subsamples composed of $N \simeq N_{\rm obs}/10 \simeq 800$ stars located at $r>10$~kpc and with $v_{r}>0$.

\section{Constraining the shape of the DM halo in an axisymmetric Galactic potential} 
\label{sec:axisymmetric_Galactic_potential} 

If the DM halo of our Galaxy is either spherical (i.e., with triaxiality parameters $q_y=q_z=1$) or spheroidal with axial symmetry about the $z$-axis (i.e., with triaxiality parameters $q_z \ne q_y = 1$),  the total gravitational potential of the Galaxy (Eq.~\ref{eq:phi_grav}) is axisymmetric about the $z$-axis. 
For an axisymmetric Galactic potential, we show that the method presented in Sect.~\ref{sec:method} can effectively recover the axial ratio $q_z$ of the DM halo from the distribution $\tilde D_\omega$ of the shape indicator $\omega$ of an observed sample of HVSs (Sect.~\ref{sec:shape_recovery_axisymmetric_Galactic_potential}). We also present the evaluation of the success rate of the method (Sect.~\ref{sec:success_rate_axisymmetric_Galactic_potential}).

As demonstrated in Sect.~\ref{sec:halo_impact}, if the Galactic potential is axisymmetric, there is only one shape indicator $\omega$ of the DM halo: the magnitude of the latitudinal velocity of the HVSs, $|v_\vartheta|$. We refer to its distribution as $D_{|v_\vartheta|}$. A few examples of $D_{|v_\vartheta|}$ for DM halos with different shapes were shown in Fig.~\ref{fig:v_theta_axisymmetric}.

\subsection{Shape recovery} 
\label{sec:shape_recovery_axisymmetric_Galactic_potential}

We defined a series of $n_{\rm s}=8$ reference shapes for the DM halo by varying $q_z$ in steps of 0.1 in the range $0.7-1.4$,
For each shape, we generated an ensemble of $n_{\rm t}=5,000$ mock catalogs B, one per different set of initial conditions of the star trajectories. We thus got $n_{\rm t}$ mock samples of stars, and $n_{\rm t}$ corresponding distributions  $D_{|v_\vartheta|}$ for each of the $n_{\rm s}$ shapes of the DM halo.

\begin{figure*}[ht]
	\includegraphics[width=8.4cm,height=6.1cm]{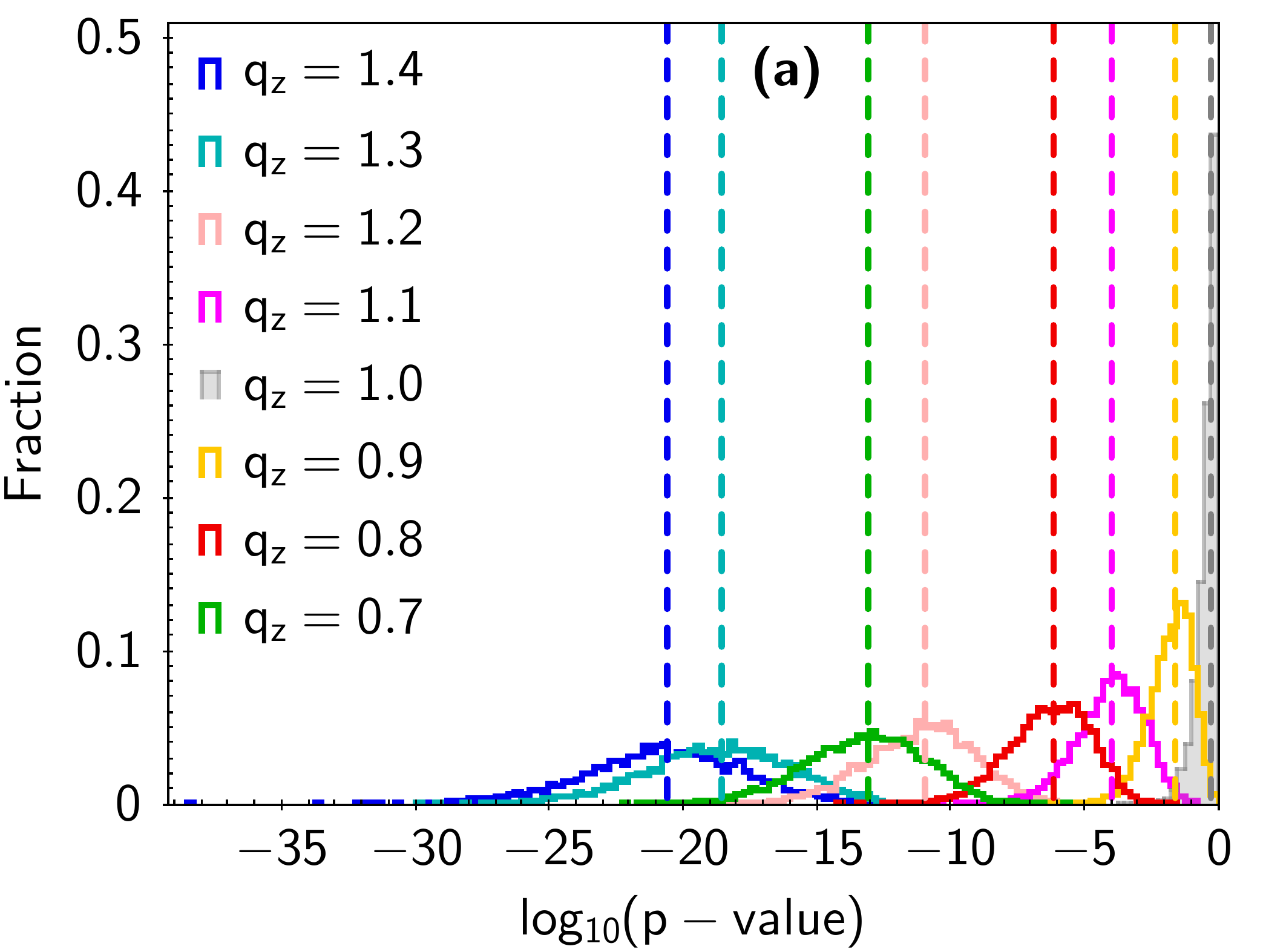}
	\includegraphics[width=8.5cm,height=6.1cm]{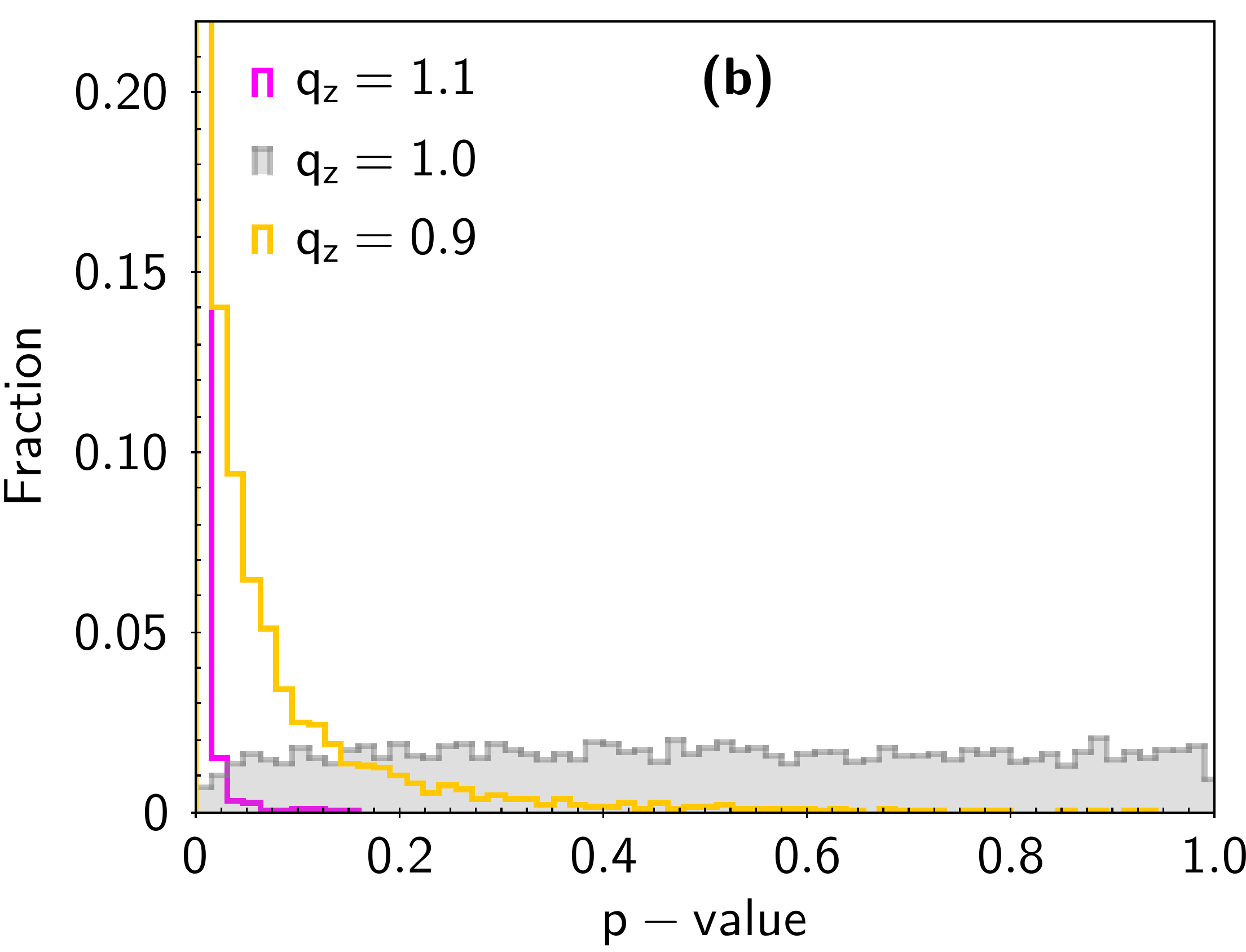}
	\caption{{\it Panel a:} Distributions of the $p$-values (in logarithmic scale) obtained from the KS test comparison of the distribution $\tilde D_{|v_\vartheta|}$ of one HVS observed sample generated in a spherical DM halo, against each of the $n_{\rm s} = 8$ ensembles of $n_{\rm t}=5,000$ mock distributions $D_{|v_\vartheta|}$ generated in DM halos with different shapes, as listed in the panel. The vertical, dashed lines mark the median $p$-value of each distribution.
	{\it Panel b}: Enlargement of the right-most part of panel $a$ with the $p_{\rm med}$ axis in linear scale.
	The difference in shape of the distributions in panels $a$ and $b$ are due to the different size of the histogram bins in the logarithmic and linear scales. 
	\label{fig:axisymmetric_example}}
\end{figure*}

As a first test, we chose as the HVS observed sample one random mock sample of HVSs that have crossed a spherical DM halo (with $q_z=q_y=1$). We now show that our method successfully recovers the axis ratio $q_z=1$.

For each of the $n_{\rm s}=8$ reference shapes, we performed a set of $n_{\rm t}=5,000$ KS test comparisons of the observed sample's $\tilde D_{|v_\vartheta|}$ against each of the $n_{\rm t}$ $D_{|v_\vartheta|}$'s of the mock samples associated to that reference shape.
Figure~\ref{fig:axisymmetric_example} shows the outcome of this test. Specifically, panel $b$ of Fig.~\ref{fig:axisymmetric_example} shows the test result for three of these $n_{\rm s}$ sets of comparisons: the $n_{\rm t}$ KS test comparisons of the $\tilde D_{|v_\vartheta|}$ of the observed sample against the $D_{|v_\vartheta|}$'s of the mock star samples that traveled in the spherical DM halo yields a uniform distribution of $p$-values (filled, gray histogram) whose median is $p_{\rm med}\simeq 0.5$; the comparison of $\tilde D_{|v_\vartheta|}$ against a mock sample generated either in a slightly oblate ($q_z=0.9$; yellow histogram) or in a slightly prolate ($q_z=1.1$; magenta histogram) DM halo yields a distribution of $p$-values which is markedly skewed towards very low $p$'s; the median $p$-values are $p_{\rm med}=2\times 10^{-2}$ for the observed vs. oblate comparison, and $p_{\rm med}=10^{-4}$ for the observed vs. prolate comparison.

As shown in panel $a$ of Fig.~\ref{fig:axisymmetric_example}, the median $p$-value becomes smaller and smaller for increasing departure from spherical of the shape of the DM halo of the mock catalogs used for the comparison: it reaches $p_{\rm med} \simeq 10^{-13}$ for the comparison against the most oblate DM halo case ($q_z=0.7$; green histogram) considered in this work, and $p_{\rm med} \simeq 10^{-21}$ for the comparison against the most prolate DM halo case ($q_z=1.4$; blue histogram).
We thus confirm that the largest $p_{\rm med}$, that we obtained in the observed vs. spherical comparison, identifies the correct shape of the DM halo crossed by the observed sample (i.e., the spherical shape, with $q_z=q_y=1$).

\subsection{Success rate $S$ of the method} 
\label{sec:success_rate_axisymmetric_Galactic_potential}

Even though the result shown in the above test is a likely result, it is not guaranteed.
Indeed, as pointed out in Sect.~\ref{sec:method}, the $p$-value distributions of Fig.~\ref{fig:axisymmetric_example} are not unique, and depend on the specific observed sample that we pick for the comparison (i.e., from the set of random initial conditions of the simulated HVSs). Therefore, it may happen that only in a fraction of cases the largest $p_{\rm med}$ is the correct indicator of the shape of the DM halo.
We defined this fraction as the success rate $S$ of our method in recovering the correct axis ratio of the DM halo.
We now illustrate the evaluation of the method success rate, $S$, for the case of a spherical DM halo (Sect.~\ref{sec:success_rate_axisymmetric_Galactic_potential_1shape}), and the investigation of the dependence of $S$ on the shape of the DM halo (Sect.~\ref{sec:success_rate_axisymmetric_Galactic_potential_shapes}).

\begin{figure*}[ht]
	\includegraphics[width=8.2cm,height=6.1cm]{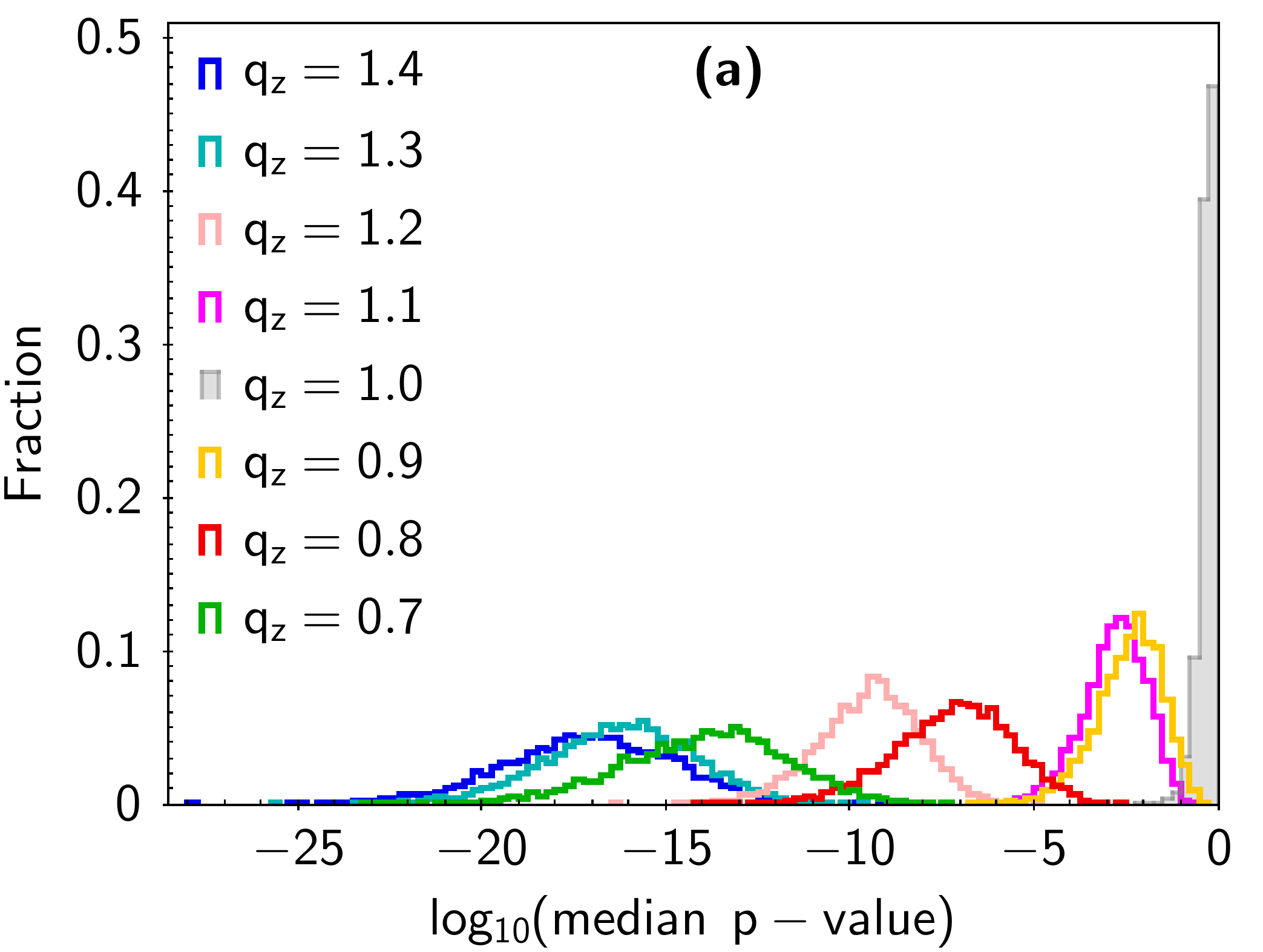}
	\includegraphics[width=8.6cm,height=6.1cm]{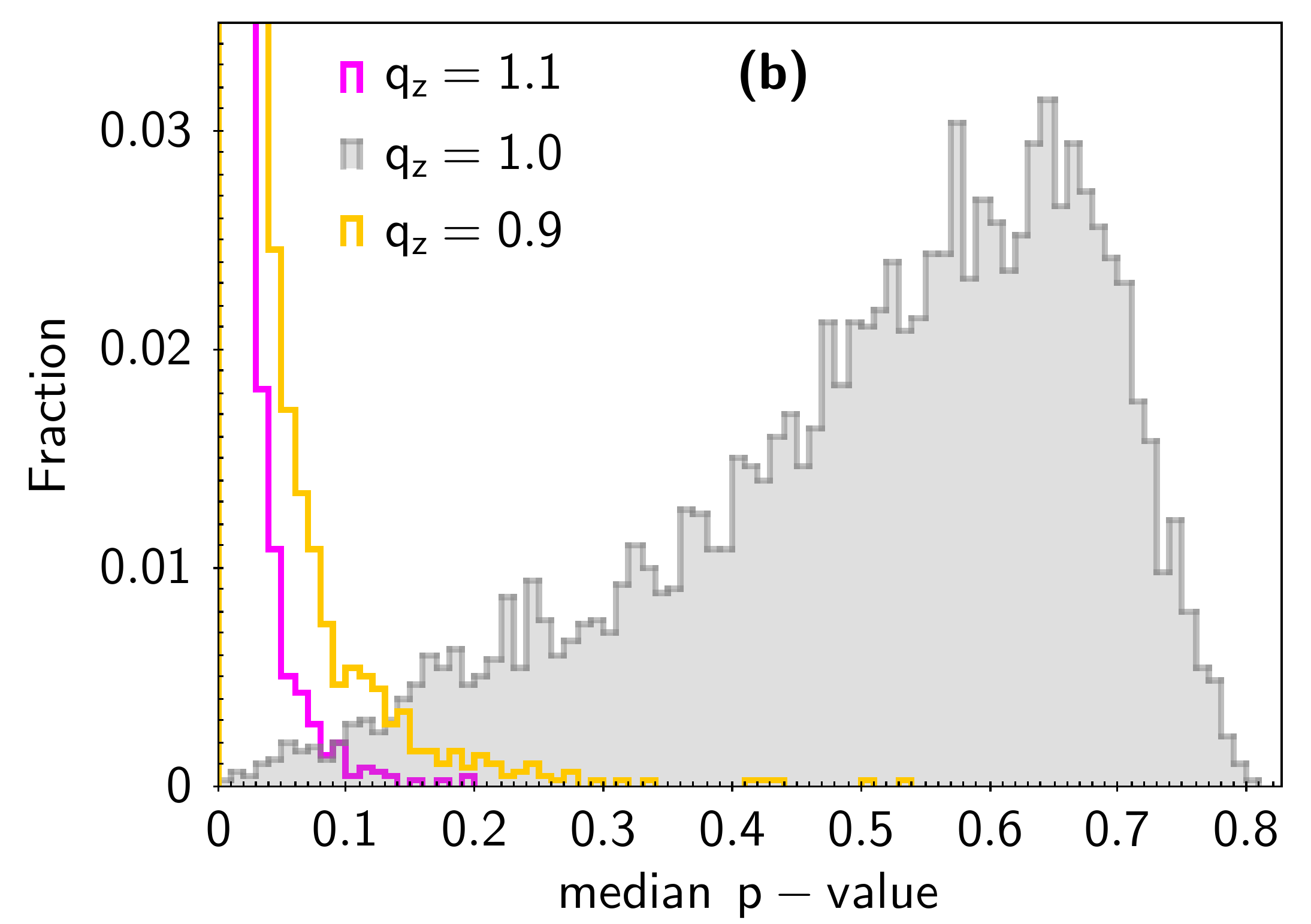}
	\caption{{\it Panel a:} Distributions of the median $p$-values, $p_{\rm med}$ (in logarithmic scale), obtained from the KS test comparison of the HVS observed sample generated in a spherical halo against each of the $n_{\rm s}=8$ ensembles of mock HVS samples generated in a DM halo with different shape.
	Each distribution is the result of the KS test comparison of the $\tilde D_{|v_\vartheta|}$'s of the $n=n_{\rm t}$ observed samples against the $n_{\rm t}=5,000$ mock samples generated in a DM halo with different shape as listed in the panel.
	{\it Panel b}: Enlargement of the right-most portion of panel $a$, with the $p_{\rm med}$ axis in linear scale. 
	The yellow and pink distributions are the only distributions with non-null overlap with the gray distribution.
	The different shape of the distributions in panels $a$ and $b$ are due to the different size of the histogram bins in the logarithmic and linear scales.
	\label{fig:axisymmetric_example_pmed}}
\end{figure*}

\subsubsection{The case of a spherical DM halo} \label{sec:success_rate_axisymmetric_Galactic_potential_1shape}

To evaluate the success rate $S$ of our method in recovering the correct axis ratio $q_z=1$, we constructed the distributions of all possible $p_{\rm med}$'s that could be obtained by comparing the observed sample against the mock samples corresponding to the reference shapes of the DM halo.
To generate these distributions, we simulated a series of $n=n_{\rm t}$ HVS observed samples in a spherical DM halo, by randomly varying the set of the stars' initial conditions. For each observed sample, we performed the $n_{\rm t}$ KS test comparisons against all the mock samples of a given shape, and we obtained a $p_{\rm med}$; performing the procedure for $n=n_{\rm t}$ observed samples yields a distribution of $n=n_{\rm t}$ values of $p_{\rm med}$ for each comparison of the observed samples against the mock samples of a given shape of DM halo.
Repeating this exercise for each of the $n_{\rm s}=8$ shapes of DM halo yields the $n_{\rm s}$ distributions of $p_{\rm med}$'s that we show in Fig.~\ref{fig:axisymmetric_example_pmed}.

Panel $a$ of Fig.~\ref{fig:axisymmetric_example_pmed} shows that the comparison of the observed samples (generated in a spherical DM halo) against the samples generated in a spherical DM halo returns a distribution (the gray histogram) where most of the $p_{\rm med}$'s are larger than the $p_{\rm med}$'s of the other distributions, confirming that the method correctly recovers the shape of the DM halo crossed by the observed sample in most of the cases. 
However, panel $a$ also shows that the distribution corresponding to the comparison of the observed samples against the samples generated in the spherical DM halo displays a non-null overlap with the two distributions of $p_{\rm med}$ that correspond to the comparison of  
(i) the observed samples against the samples generated in a slightly oblate DM halo ($q_z=0.9$; yellow histogram), and (ii) the observed samples against the samples generated in a slightly prolate DM halo ($q_z=1.1$; magenta histogram).
This non-null overlap, that can be better appreciated in panel $b$, implies that, for an observed sample generated in a spherical DM halo, the shapes that mildly deviate from spherical ($|\Delta q_z|=0.1$, in our mock catalogs) might be erroneously associated to the observed sample of HVSs, based on our method. 
The moderate, non-null overlap, however, does not necessarily imply that erroneous associations do occur with a rate proportional to the overlapping areas. 

To illustrate this concept, let us consider two overlapping distributions of $p_{\rm med}$, one which is right-skewed and the other which is left-skewed, with a non-null overlap (as, e.g., the gray and yellow distributions of Fig.~\ref{fig:axisymmetric_example_pmed}).
Each value of $p_{\rm med}$ of the distributions is contributed by a specific HVS observed sample, characterized by a specific random set of initial conditions and a specific $\tilde D_{|v_\vartheta|}$ (see Sect.~\ref{sec:shape_recovery_axisymmetric_Galactic_potential} for details). Only when the $p_{\rm med}$ that contributes to the left-skewed distribution ($p_{\rm med, ls}$) is higher than the $p_{\rm med}$ that contributes to the right-skewed distribution ($p_{\rm med, rs}$) for the same random observed sample, the erroneous association does occur. Other similar pairs $(p_{\rm med, ls}, p_{\rm med, rs})$, that can be randomly drawn from the overlapping portion of the two distributions, may never occur in reality.
Therefore, whereas a null overlap ensures a null rate of erroneous shape associations, the existence of a non-null overlap of two distributions is only an indication that some erroneous associations may occur, without quantifying their rate of occurrence. However, the larger is the overlap, the higher is the probability of erroneous associations.

To evaluate the rate of success of our method in recovering the shape of a spherical DM halo, for each distribution of $p_{\rm med}$'s (as the yellow and magenta distributions in Fig.~\ref{fig:axisymmetric_example_pmed}) that has a non-null overlap with the $p_{\rm med}$ distribution corresponding to the observed vs. spherical comparison (i.e., the gray distribution in Fig.~\ref{fig:axisymmetric_example_pmed}), we computed the rate of occurrence of pairs $(p_{\rm med, ls}, p_{\rm med, rs})$ where $p_{\rm med, ls} > p_{\rm med, rs}$, thus providing an erroneous shape recovery. 
We found that in $1.16\%$ of the cases (i.e., in 58 cases out of $5,000$) the KS test comparison of the $\tilde D_{|v_\vartheta|}$ of the observed sample against the $n_{\rm t}=5,000$ $D_{|v_\vartheta|}$'s of a mildly oblate ($q_z=0.9$) DM halo yields a $p_{\rm med}$ value larger than that obtained in the comparison against the $n_{\rm t}=5,000$ $D_{|v_\vartheta|}$'s of a spherical DM halo (i.e., $p_{\rm med, ls} > p_{\rm med, rs}$), leading to erroneously classify as mildly oblate a spherical halo; this result is equivalent to a success rate $S=98.84\%$.
The fraction of erroneous associations drops to 0.4\% for the comparison against a mildly prolate ($q_z=1.1$) DM halo, yielding $S=99.6\%$.
For $|\Delta q_z| \ge 0.2$, the overlap of the $p_{\rm med}$ distributions with the distribution corresponding to the spherical DM halo is null, implying a null rate of erroneous associations, and a success rate of our method $S=100\%$.

Overall, for an observed sample generated in a spherical DM halo, our method enables to recover the correct axis ratio $q_z=1$ of the DM halo in more than 98\% of the cases; in other words, the method has a success rate $S \gtrsim 98\%$.
We stress that, in the rare cases of an erroneous shape association, the recovered axis ratio $q_z$ is off by only $|\Delta q_z|=0.1$.

\subsubsection{Dependence of the success rate $S$ on the shape of the DM halo of the observed sample} \label{sec:success_rate_axisymmetric_Galactic_potential_shapes}

\begin{figure*}[ht]
	\includegraphics[width=8.6cm,height=6cm]{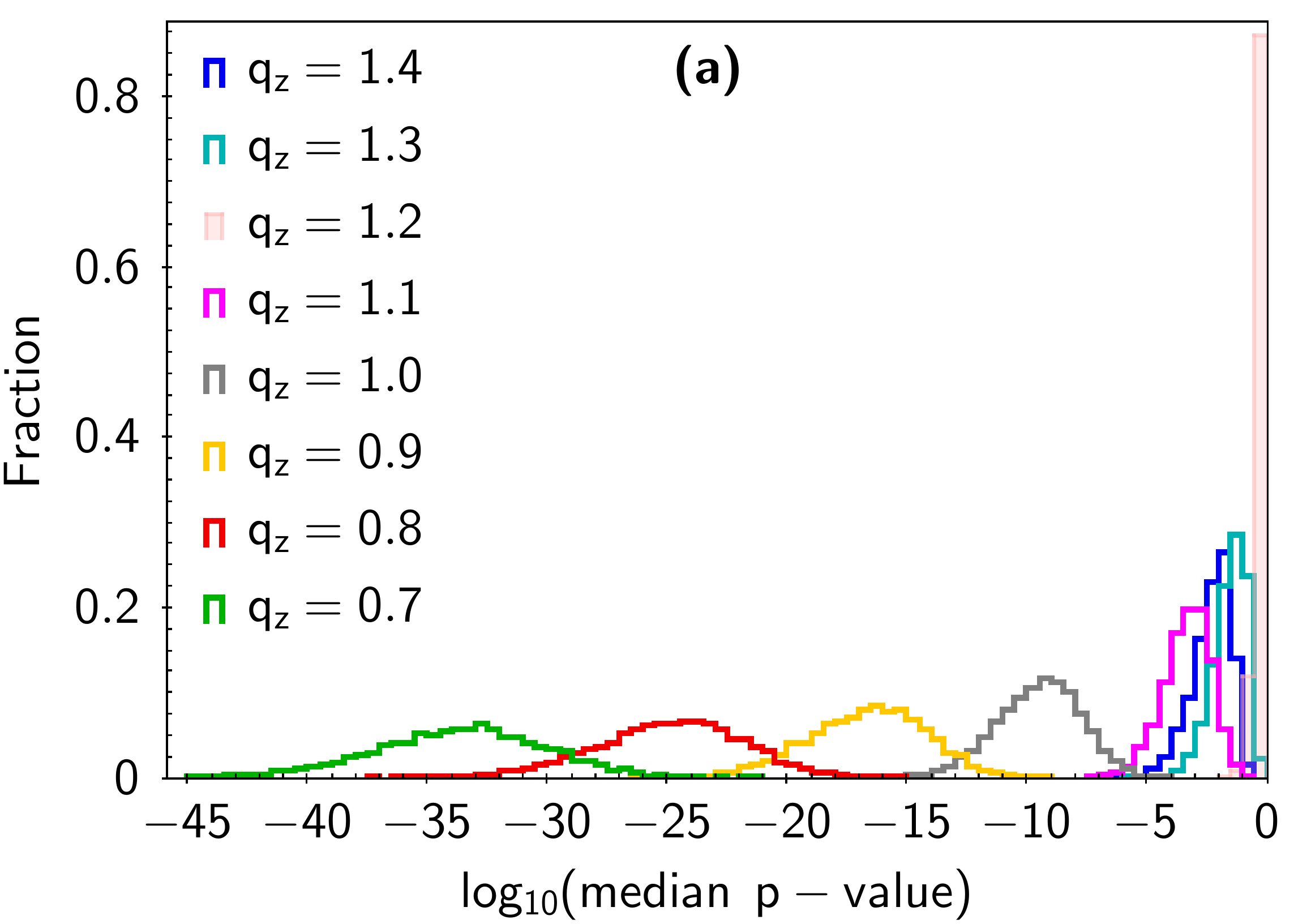}
	\includegraphics[width=8.cm,height=6cm]{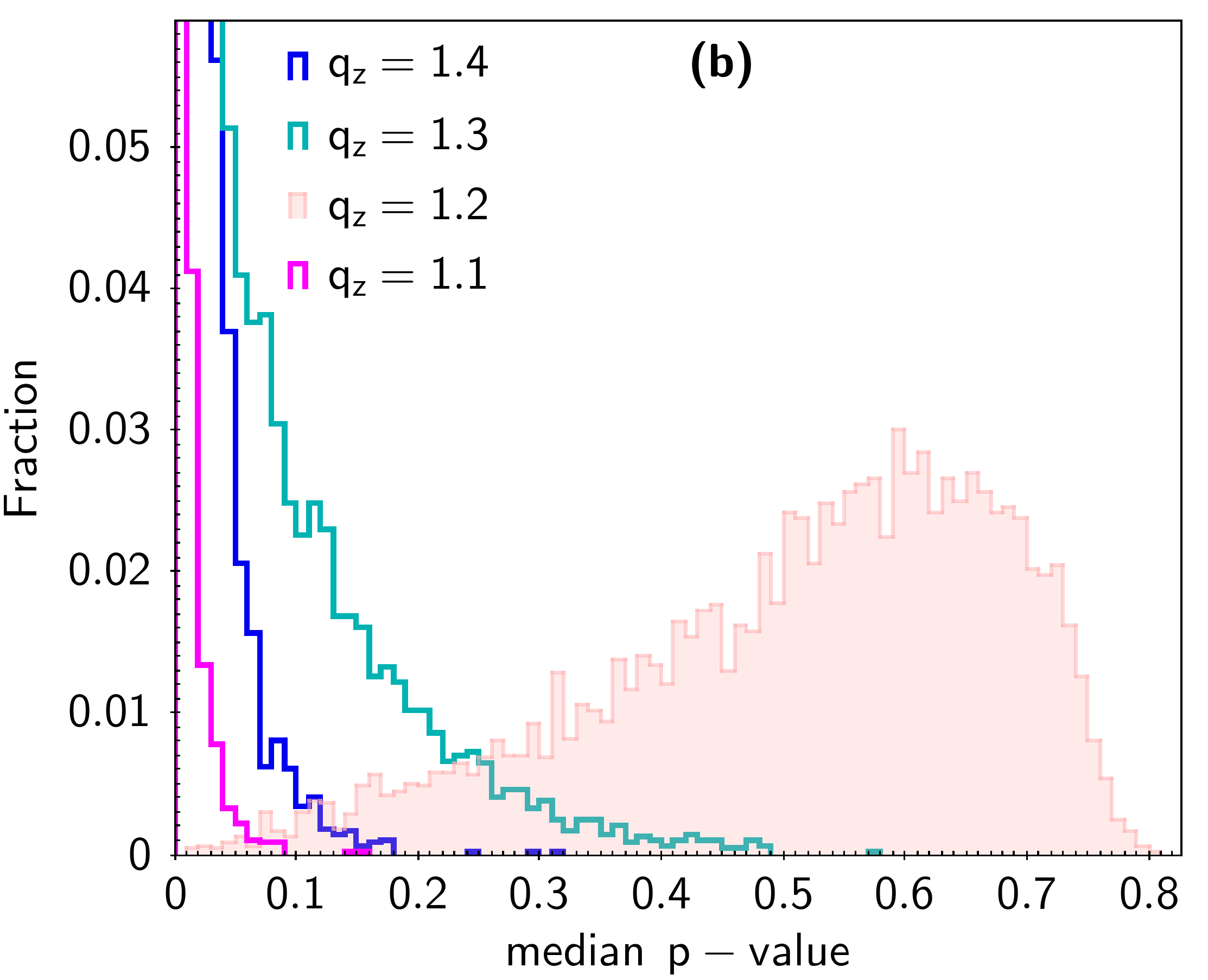}
	\caption{ {\it Panel a:} Distributions of the median $p$-values, $p_{\rm med}$ (in logarithmic scale), obtained from the KS test comparison of the HVS observed sample against each of the $n_{\rm s}=8$ ensembles of mock HVS samples generated in a DM halo with different shape.
	Each distribution is the result of the KS test comparison of the $\tilde D_{|v_\vartheta|}$'s of the $n=n_{\rm t}$ observed samples obtained in a prolate DM halo with $q_z=1.2$ against the $n_{\rm t}=5,000$ mock samples generated in a DM halo with different shape as listed in the panel.
	{\it Panel b}: Enlargement of the right-most part of panel $a$ with the $p_{\rm med}$ axis in linear scale. The cyan, blue, and magenta distributions are the only distributions with non-null overlap with the pink distribution.
	The difference in shape of the distributions in panels $a$ and $b$ are due to the different size of the histogram bins in the logarithmic and linear scales.
	\label{fig:shape_dependence_axisymmetric}}
\end{figure*}

\begin{table*}[ht]
	\caption{Success rate $S$ of our method in recovering the axis ratio $q_z$ of the DM halo of an axisymmetric Galactic potential from the distribution of the magnitudes of the azimuthal velocities, $D_{|v_{\vartheta}|}$, of an observed sample of HVSs.}
    \centering
	\begin{tabular}{ccccccccc}
		\hline \hline
		$q_z$ & $1.4$\tablefootmark{a} & $1.3$ &$1.2$ & $1.1$ &$1.0$ &$0.9$ &$0.8$ &$0.7$\tablefootmark{a} \\
		\hline
		$S$ &88.9\% & 92.6\% & 96.6\% & 99.2\% & 98.4\% & 97.4\% & 97.2\% & 96.6\% \\
		\hline
	\end{tabular}
	\tablefoot{
	\tablefoottext{a}{To properly compute the success rate in recovering the shape of DM halos with $q_z = 1.4$ and $q_z = 0.7$ we generated also reference sets of $n_{\rm t} = 5,000$ HVS mock catalogs for DM halos of $q_z = 1.5$ and $q_z = 0.6$, respectively.}}
	\label{tab:S}
\end{table*}

The success rate $S$ of our method displays a weak, non obvious dependence on the actual shape of the DM halo crossed by the HVS observed sample.
Indeed, repeating the exercise of
Sect.~\ref{sec:success_rate_axisymmetric_Galactic_potential_1shape} for an HVS observed sample that traveled in each of our $n_s=8$ reference DM halos (with $q_z=0.7-1.4$), returned a success rate that varies from $89\%$ to $99\%$, as listed in Table~\ref{tab:S}. 
The success rate is slightly higher for spherical ($q_z=1.0$) or mildly prolate ($q_z=1.1$) DM halos than for oblate and markedly prolate DM halos; in other words, our method recovers more easily the shape of a DM halo crossed by an observed sample of HVSs when this DM halo is not too different from spherical. In addition, $S$ is slightly lower for markedly prolate than for markedly oblate DM halos; this result indicates that the method recovers more easily the shape of an oblate DM halo than the shape of a prolate DM halo.

To illustrate this effect,  Fig.~\ref{fig:shape_dependence_axisymmetric} shows the analogs of the distributions of Fig.~\ref{fig:axisymmetric_example_pmed} for the case of an HVS observed sample that traveled in a prolate DM halo with $q_z=1.2$. 
Panel $a$ shows that the comparison of the HVS observed sample against the mock sample that traveled in a DM halo with the same shape returns a distribution (the pink histogram) where most of the $p_{\rm med}$'s are larger than the $p_{\rm med}$'s of the other distributions. This result confirms that the correct shape is recovered also for non-spherical DM halos. 

However, panel $a$ and the corresponding enlargement in panel $b$ also show that three distributions of $p_{\rm med}$ display a non-null overlap with the distribution corresponding to the correct shape: two of these distributions correspond to the comparison of the observed samples against the samples generated in more prolate DM halos ($q_z=1.3$ and $q_z=1.4$; cyan and blue histograms, respectively); the third one corresponds to the comparison against a less prolate DM halo ($q_z=1.1$; magenta histogram).
The number of overlapping distributions is thus larger (three instead of two) than in the case of the spherical DM halo illustrated in Fig.~\ref{fig:axisymmetric_example_pmed}; furthermore, the overlapping area of the distributions is also larger than in Fig.~\ref{fig:axisymmetric_example_pmed} for the same $\Delta q_z$. Consequently, when the HVS observed sample traveled in a markedly prolate DM halo with $q_z=1.2$, we get a larger number of erroneous shape associations for this halo than in the case of HVSs traveling in a spherical halo.

We note that, while the larger number of overlapping distributions is only a characteristics of the $p_{\rm med}$ distributions  associated with markedly prolate ($q_z \ge 1.2$) DM halos, the larger overlapping area is a property shared by all the oblate and markedly prolate DM halos. 
The excess in the number of overlapping distributions, where present, determines erroneous associations of the shape of the DM halo with spheroids whose $q_z$ differs by 0.2 from the true $q_z$; however, it negligibly affects the success rate $S$ of our method ($\lesssim 0.02\%$). 
On the other hand, the larger overlapping area of the $p_{\rm med}$ distributions associated to DM halos whose $q_z$ differs by 0.1 from the true $q_z$ is the main responsible for the shape dependence of $S$, because the larger area increases the probability that a DM halo is assigned a shape whose $q_z$ is off by 0.1.
Summarizing, in the infrequent cases of erroneous shape associations, the recovered axis ratio $q_z$ is typically off by $|\Delta q_z|= 0.1$, although  $|\Delta q_z|=0.2$ can seldom occur ($\lesssim  0.04\%$ of the cases) when the DM halo crossed by the HVS sample is markedly prolate.

As illustrated at the beginning of this section, the weak dependence of the success rate $S$ on the shape of the DM halo crossed by the HVS observed sample manifests itself with a slightly higher $S$ for spherical and mildly prolate DM halos, and a slightly lower $S$ for markedly prolate than for markedly oblate DM halos. This weak shape dependence is a direct consequence of the following facts: (i) the difference between the $D_{|v_\vartheta|}$'s of HVSs that traveled in a spherical or in a mildly prolate DM halo ($q_z=1.0-1.1$) and the $D_{|v_\vartheta|}$'s generated in DM halos with $q_z'=q_z\pm 0.1$ is more pronounced than the difference between the $D_{|v_\vartheta|}$'s generated in DM halos that are either oblate ($q_z=0.7-0.9$) or markedly prolate ($q_z=1.2-1.4$) and the $D_{|v_\vartheta|}$'s produced in DM halos with $q_z'=q_z\pm 0.1$; (ii) the differences among the $D_{|v_\vartheta|}$'s of markedly prolate halos are milder than those among oblate halos.
We ascribe these two effects to a combination of (a) the choice of a fixed resolution in $q_z$ of our mock catalogs (i.e., $\Delta q_z=0.1$), and (b) to the superposition of the gravitational actions of the disk and of the DM halo.

Specifically, effect (i) is mostly due to reason (a), that is the resolution in $q_z$ of our mock catalogs.
Indeed, more and more prolate (oblate) DM halos, obtained from the spherical halo ($q_z=1.0$) by progressively increasing (decreasing) $q_z$ by $\Delta q_z=0.1$, generate a response in the  $D_{|v_\vartheta|}$'s that is stronger when $q_z$ is closer to 1. This is true independently of the presence of the axisymmetric disk potential.

On the other hand, effect (ii) is due to a combination of reasons (a) and (b). Reason (a) causes the markedly prolate DM halos considered in our mock catalogs to have $q_z$'s (1.2, 1.3, and 1.4) that render their shapes more similar to one another than the $q_z$'s of the oblate DM halos considered in this work (0.7, 0.8, and 0.9); indeed, the DM halos whose axis ratios are the reciprocals of the axis ratios  of these oblate halos would be characterized by $q_z$'s equal to $\simeq 1.1, 1.25$ and $1.4$, respectively.
Thus, the $D_{|v_\vartheta|}$'s generated by the markedly prolate DM halos are expected to be less different from one another than the corresponding distributions obtained for the oblate DM halos.

On top of this effect, the gravitational pull of the disk makes the $D_{|v_\vartheta|}$'s generated by the markedly prolate DM halos even less different from one another. Indeed, in a prolate DM halo the gravitational pull of the halo is opposed to the pull of the disk. For $q_z \le 1.2$, the presence of the DM halo increases the fraction of HVSs with low $|v_\vartheta|$'s, rendering the distributions $D_{|v_\vartheta|}$'s more left-skewed than that of a spherical DM halo.
However, for $q_z \ge 1.3$, the action of the DM halo not only compensates the gravitational pull of the disk, but it overcomes this pull, by bending the HVS trajectories towards the $z$-axis: consequently the fraction of low $|v_\vartheta|$'s drop - because $v_\vartheta$ increases in magnitudes and changes sign - and renders the $D_{|v_\vartheta|}$'s less left-skewed and more similar to those of the less prolate DM halos. 

The above effects are responsible for a slightly higher success rate $S$ for oblate DM halos. 
While the effects of the choice of a fixed $\Delta q_z$ to build the mock catalogs can easily be overcome, the effect of the combined gravitational actions of the disk and of the DM halo is inherent to the problem. 

Overall, in an axisymmetric Galactic potential, our method recovers the axis ratio $q_z$ of the DM halo crossed by an observed sample of HVSs with a success rate $S \gtrsim 89\%$, and the erroneous shape association imply $q_z$ that is off by $\pm 0.1$ in the overwhelming majority ($\gtrsim 99.96\%$) of the cases. In a negligible fraction ($\lesssim  0.04\%$ \footnote{The reported fraction of erroneous shape association with $q_z$ off by 0.2 was computed without taking into account the cases corresponding to $q_z =0.7$ and $q_z = 1.4$: indeed, computing this fraction also for these extreme cases would have required the use of mock samples generated in a DM halo with $q_z =0.5$ and $q_z = 1.6$.}) of the cases, and for markedly prolate DM halos only, $q_z$ can be off by 0.2.


\section{Constraining the shape of the DM halo in a non-axisymmetric Galactic potential}  
\label{sec:non-axisymmetric_Galactic_potential}

\begin{figure*}[ht]
	\centering
	\includegraphics[width=0.49\textwidth]{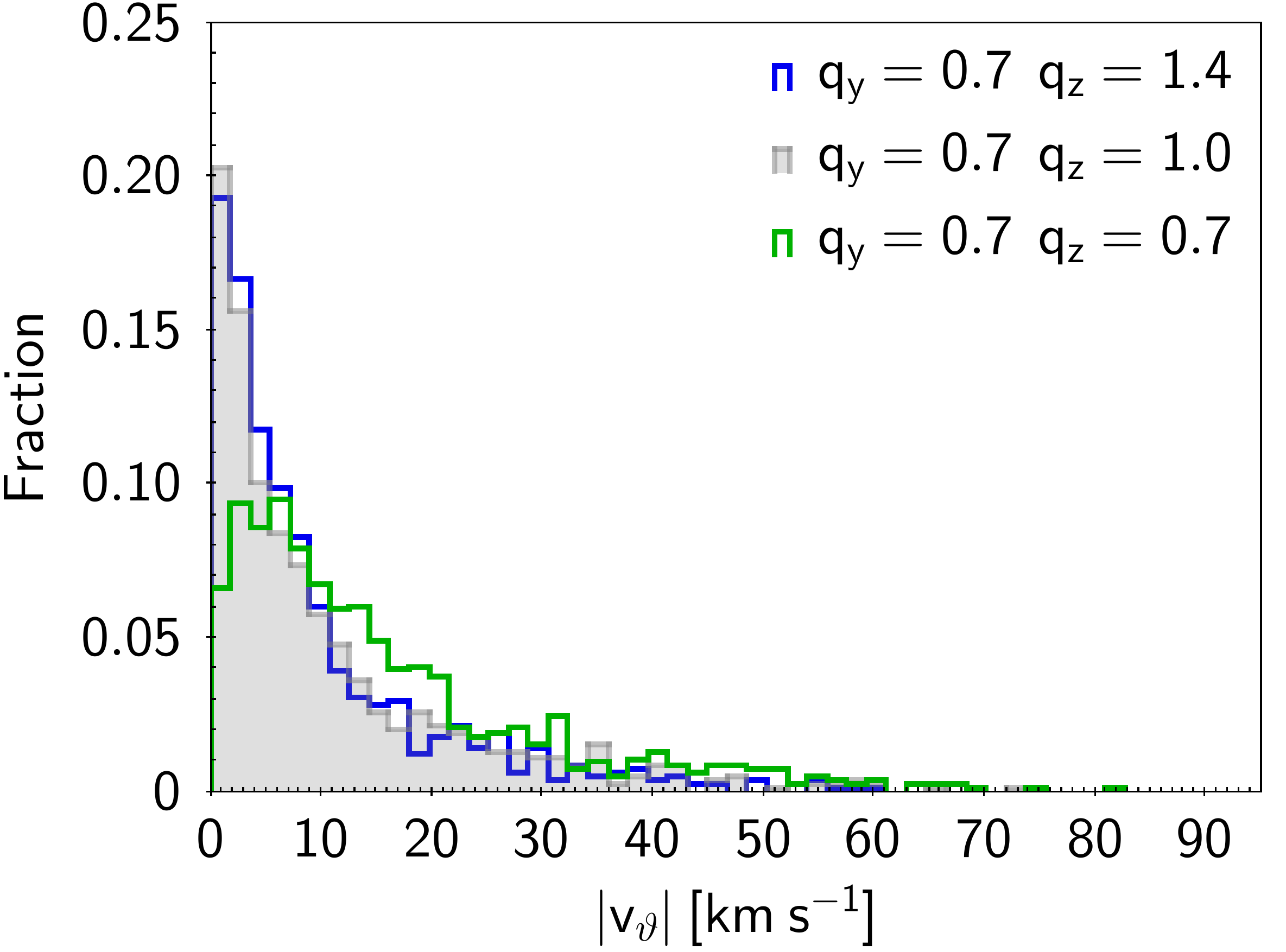} 
	\includegraphics[width=0.49\textwidth]{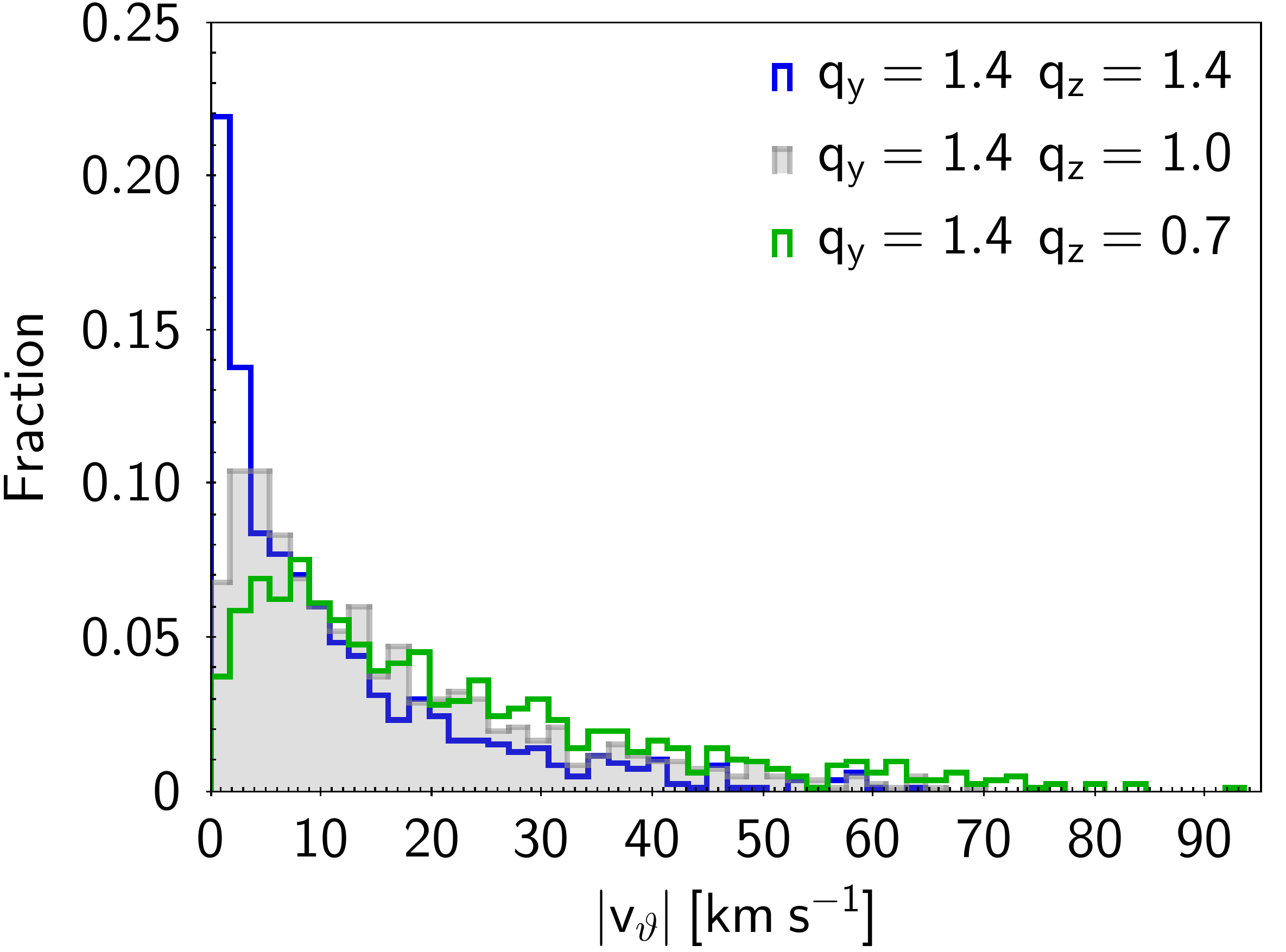}
	\caption{Distributions of the magnitude of the  galactocentric tangential velocity $|v_\vartheta|$ for HVSs that have traveled in DM halos with $q_z = \{1.4 , 1.0 , 0.7\}$, and with $q_y=0.7$ (left panel) and $q_y=1.4$ (right panel). The distributions were generated with the same initial conditions (mock catalogs A), to highlight the effect of the different geometries of the DM halo.
	}
	\label{fig:vtheta_nonaxisymmetric}
\end{figure*}

If our Galaxy has either a fully triaxial DM halo (i.e., triaxiality parameters $q_y \ne q_z$, with $q_y \ne 1$ and  $q_z \ne 1$ ) or a spheroidal DM halo with a symmetry axis misaligned with respect to the $z$-axis (i.e. triaxiality parameters $q_y \ne q_z=1$, or $q_y=q_z \ne 1$), the total gravitational potential of the Galaxy (Eq.~\ref{eq:phi_grav}) is non-axisymmetric. 
For a non-axisymmetric Galactic potential, we show that the method presented in Sect.~\ref{sec:method} can effectively recover the axial ratios $q_y$ and $q_z$ from the distribution $\tilde D_\omega$ of the shape indicators $\omega$ of an observed sample of HVSs (Sect.~\ref{sec:shape_recovery_non-axisymmetric_Galactic_potential}). We also present the evaluation of the success rate of the method (Sect.~\ref{sec:success_rate_non-axisymmetric_Galactic_potential}).

As anticipated in Sect.~\ref{sec:halo_impact}, in a non-axisymmetric Galactic potential both components of the tangential velocity $\vec v_{\rm t}=(v_\vartheta,v_\varphi)$ are affected by the halo triaxiality, and can thus be used as indicators of the shape of the DM halo. 
Specifically, we identify two shape indicators $\omega$: the magnitude of the latitudinal velocity of the HVSs, $|v_\vartheta|$, and a function $\bar v_\varphi$ of the azimuthal velocity $v_\varphi$, that we define in  Sect.~\ref{sec:two_indicators}.
Hereafter, the distributions of the two shape indicators $|v_\vartheta|$ and $\bar v_\varphi$ will be referred to as $D_{|v_\vartheta|}$ and $D_{\bar v_\varphi}$, respectively. The corresponding two-dimensional distribution will be referred to as $D_{|v_\vartheta|, \bar v_\varphi}$.

\subsection{$|v_\vartheta|$ and $\bar v_{\varphi}$: two indicators of the shape of the DM halo} 
\label{sec:two_indicators}

In a Galaxy with a non-axisymmetric  gravitational potential, $D_{|v_\vartheta|}$ and $D_{\bar v_\varphi}$ are both affected by each of the two triaxiality parameters, $q_y$ and $q_z$.

The behavior of $D_{|v_\vartheta|}$ in the case of a non-axisymmetric Galactic potential is similar to the behavior of $D_{|v_\vartheta|}$ in an axisymmetric Galactic potential  (Sect.~\ref{sec:halo_impact}). 

Figure~\ref{fig:vtheta_nonaxisymmetric} shows $D_{|v_\vartheta|}$ for a sample of HVSs that traveled in a Galaxy with DM halos of different shapes: the comparison of different distributions in each of the two panels shows that increasing $q_z$ at fixed $q_y$ leads to $D_{|v_\vartheta|}$'s that are generally more skewed towards low values of $|v_\theta|$.
This effect was already shown in Fig.~\ref{fig:v_theta_axisymmetric} for the case of spheroidal DM halos axisymmetric about the $z$ axis (i.e., with $q_y=1$): the gravitational pull of the disk, that drives the HVSs towards the $x$-$y$ plane, is more and more compensated by a DM distribution which is more and more elongated in the direction of the $z$-axis. We note that the increase of skewness with increasing $q_z$ depends on the value of $q_y$ and stops when the action of the DM halo overcomes that of the disk, making $v_\vartheta$ change sign and increase in magnitude. 
This effect can be noticed in the left panel of Fig.~\ref{fig:vtheta_nonaxisymmetric}, where the $D_{|v_\vartheta|}$'s are comparable for $q_z=1.0$ and $q_z=1.4$.

On the other hand, a comparison of the distributions corresponding to the same value of $q_z$ (i.e., the distributions drawn with the same color) in the two panels in Fig.~\ref{fig:vtheta_nonaxisymmetric} shows the dependence of $D_{|v_\vartheta|}$ on $q_y$, at fixed $q_z$: DM halos with lower $q_y$ at fixed $q_z$ imply larger concentration of dark matter away from the $x$-$y$ plane, and thus generate (i) $D_{|v_\vartheta|}$'s that are more skewed towards lower $|v_\vartheta|$'s, as long as $q_z$ is not too high and the DM halo only compensates the $|v_\vartheta|$'s induced by the disk, (see, e.g., the green and gray histograms, corresponding to $q_z=0.7$ and $q_z=1.0$); (ii) $D_{|v_\vartheta|}$'s that are less skewed towards lower $|v_\vartheta|$'s, for large values of $q_z$, when the pull of the DM halo overcomes the pull of the disk, and the increase in $|v_\vartheta|$ previously discussed is enhanced by a dark matter distribution more concentrated about the $z$-axis (see, e.g., the blue histograms, corresponding to $q_z=1.4$).

A non-null distribution of $v_\varphi$ is a distinctive characteristic of non-axisymmetric Galactic potentials: because in our model the gravitational potential is axially symmetric for the Galactic disk and spherically symmetric for the bulge, the only source of non-zero $v_\varphi$ is the triaxial DM halo with $q_y \neq 1$. This DM halo can be either a fully triaxial DM halo or a spheroidal DM halo with a symmetry axis misaligned with respect to the $z$-axis (see Sect.~\ref{sec:MW_potential}).\footnote{An additional baryonic component of the MW that could in principle contribute to $v_\varphi$ is the MW hot gaseous halo \citep[e.g.,][]{Fang2013,Gatto2013}. However, we recently showed that its effect on the HVS azimuthal velocities is negligible with respect to that of a triaxial DM halo with $q_y \neq 1$ \citep{chakrabarty2022}. Therefore, we do not consider the contribution of the hot gaseous halo in this work.}

\begin{figure*}
	\centering
	\includegraphics[width=0.49\textwidth]{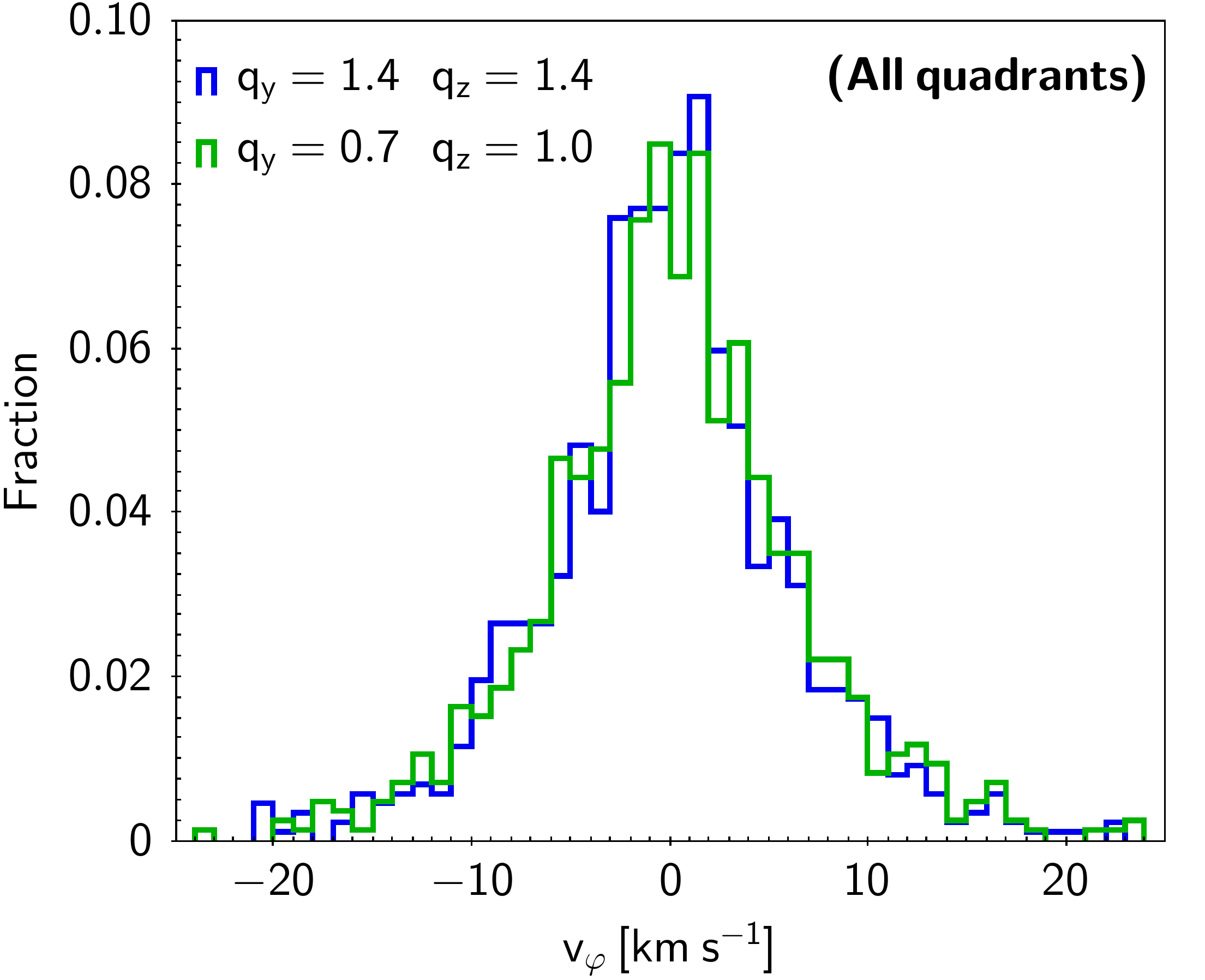}
	\includegraphics[width=0.49\textwidth]{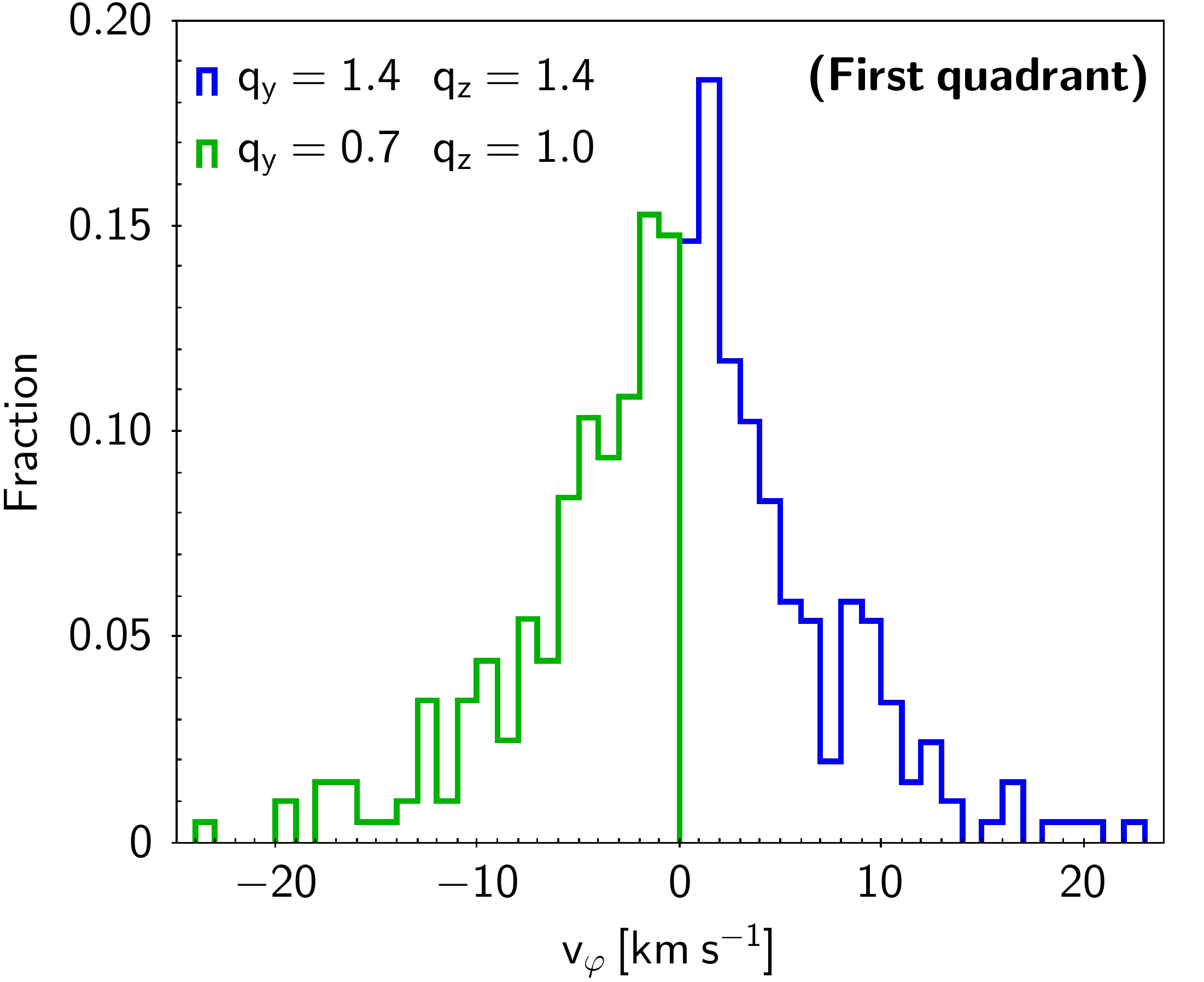}
	\caption{Distributions of the azimuthal velocity $v_\varphi$ of two samples of HVSs that have traveled in gravitational potentials whose DM halos have the same geometrical shape but differ by $90^{\circ}$ in azimuthal orientation.
	{\it Left panel:} The distributions of $v_\varphi$ are indistinguishable from one another,  when the stars from all the quadrants are considered.
	{\it Right panel: } The distributions of $v_\varphi$ become manifestly different when we consider the HVSs located in the first quadrant of the $x$-$y$ plane ($0^\circ < \varphi < 90^\circ$) only. The HVSs in the first quadrant have positive (negative) $v_\varphi$ when $q_y = 1.4$ ($0.7$). The distributions were generated with the same initial conditions (mock catalogs A), to highlight the effect of the different geometries of the DM halo.
}
	\label{fig:vphi_nonaxisymmetric}
\end{figure*}

Even though the distribution of $v_\varphi$ is extremely sensitive to the triaxiality parameters of the DM halo, using the very value of $v_\varphi$ as a shape indicator leads to a degeneracy problem: a triaxial ellipsoid with a given $q_y=q_{y,1}$ and $q_z = q_{z,1}$ is equivalent, in terms of geometric shape, to a triaxial ellipsoid characterized by $q_y = q_{y,2} = 1/q_{y,1}$ and $q_z = q_{z,2} = q_{z,1}/q_{y,1}$; indeed, the role of the semi-major axes $a$ and $b$ is swapped in these two ellipsoids, and a rotation of $90^{\circ}$ about the $z$-axis would make one of the two ellipsoids coincide with the other. 
As a consequence, the resulting distributions of $v_\varphi$ are statistically indistinguishable.
This effect can be seen in the left panel of Fig.~\ref{fig:vphi_nonaxisymmetric}, where we show the case of two samples of HVSs that traveled in two triaxial gravitational potentials whose semi-major axes $a$ and $b$ are swapped: one potential has  $(q_{y,1},q_{z,1}) = (0.7,1.0)$ and the other has $(q_{y,2},q_{z,2}) = (1.4,1.4) \simeq (1/q_{y,1},q_{z,1}/q_{y,1})$.

We note that two DM halos which are degenerate in $v_\varphi$ are also degenerate in $|v_{\vartheta}|$.
This effect can be seen in Fig.~\ref{fig:vtheta_nonaxisymmetric}, where the  $D_{|v_{\vartheta}|}$'s corresponding to a DM halo with $(q_y, q_z) = (0.7, 1.0)$ (left panel, gray histogram) and to a DM halo with $(q_y,q_z) = (1.4,1.4)$ (right panel, blue histogram) are indistinguishable.

This degeneracy problem does not limit our understanding of the halo geometric shape in a strict sense; it rather hampers our ability of  discriminating, for a DM halo with a given geometry, between two halo orientations that differ by $90^{\circ}$ within the adopted reference frame. Breaking this degeneracy would thus enable to constrain not only the degree of triaxiality of the DM halo, but also the orientation of the halo.

The degeneracy might be broken by considering only the HVSs that are traveling in one of the four quadrants of the $x$-$y$ plane. 
Indeed, when $q_y > 1$, the mass distribution is elongated in the direction of the $y$-axis; thus, the stars acquire a $v_\varphi$ that drives them  towards the $y$-axis; conversely, when $q_y <1$, the stars are attracted  towards the $x$-axis. Therefore, when the stars are located in the first quadrant (i.e. they have azimuthal coordinate $0^{\circ}<\varphi<90^{\circ}$) they have positive $v_\varphi$ when $q_y > 1$ and negative $v_\varphi$ when $q_y < 1$, in our sign convention. As shown in the right panel of Fig.~\ref{fig:vphi_nonaxisymmetric}, the distributions of $v_\varphi$ for the HVSs located in the first quadrant are manifestly different and not overlapping: for $q_y =0.7$ the $v_\varphi$'s are all negative, while they are all positive for $q_y=1.4$.

Choosing only the HVSs located in one quadrant is however not the best solution to break the above mentioned degeneracy, because the size of the sample is reduced  by a factor $\sim 4$, lowering the success rate of our method (see Sect.~\ref{sec:sample_size}). 
To recover the original sample size, we defined as shape indicator the variable $\bar v_{\varphi} \equiv  v_{\varphi}\frac{{\rm tan} \varphi}{|{\rm tan} \varphi|}$, where the factor $\frac{{\rm tan} \varphi}{|{\rm tan} \varphi|}$ is a sign plus or minus that renders the value of $v_{\varphi}$ positive when the star is moving towards the y axis ($q_y > 1$), and negative when the star is moving towards the x axis ($q_y < 1$), independently of the quadrant where the star is located. 

\begin{figure}
	\includegraphics[width=9.1cm,height=7.4cm]{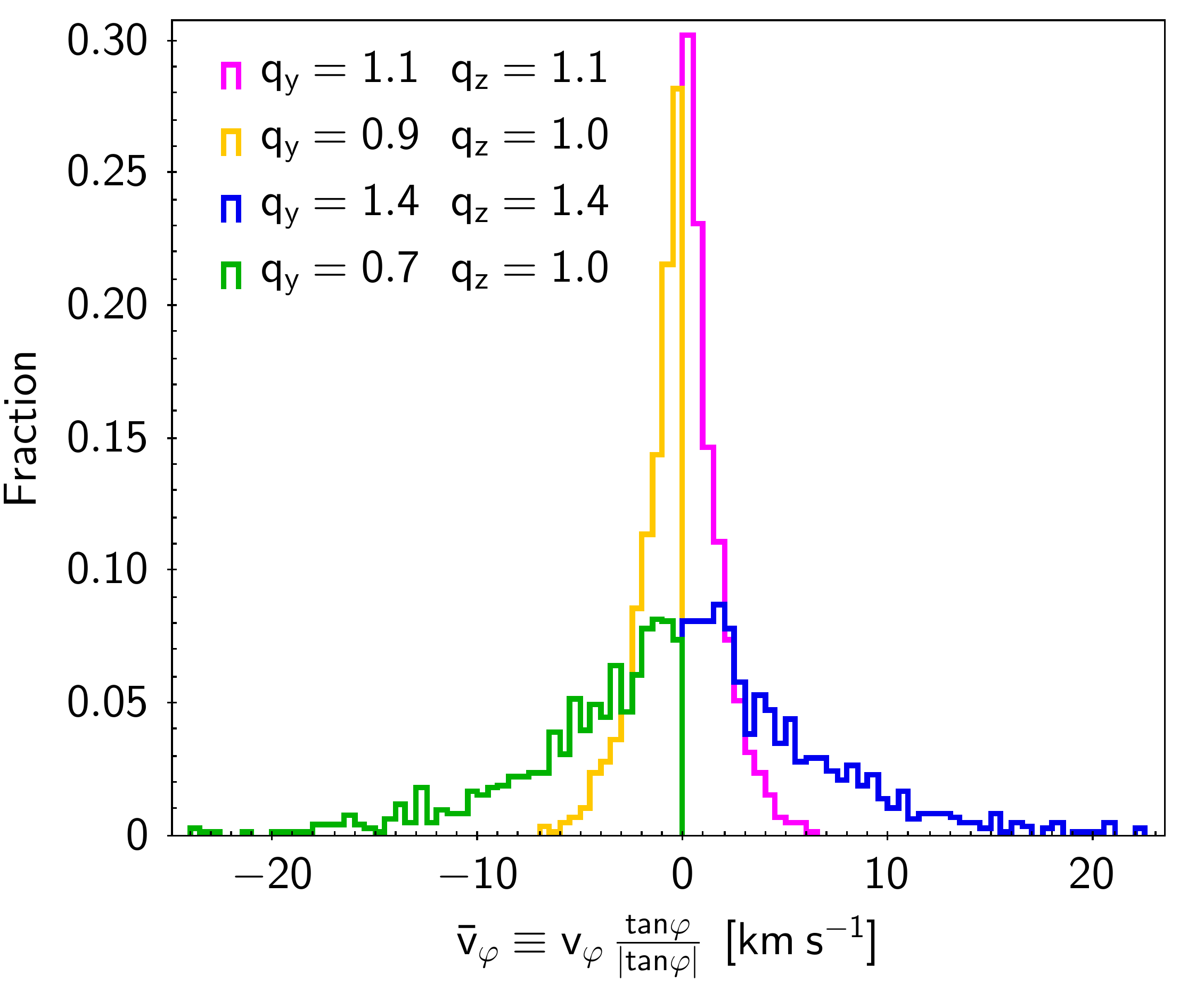}
	\caption{Distributions of $\bar v_{\varphi} \equiv  v_{\varphi}\frac{{\rm tan} \varphi}{|{\rm tan} \varphi|}$ for samples of HVSs that traveled in gravitational potentials whose DM halo has different triaxiality parameters.
	The use of the variable $\bar v_{\varphi}$ instead of $v_\varphi$ enables to easily distinguish distributions of azimuthal velocities generated in DM halos with the same geometry but whose semi-major axes $a$ and $b$ are swapped. The distributions were generated with the same initial conditions (mock catalogs A), to highlight the effect of the different geometries of the DM halo.
	\label{fig:vphi_bar}}
\end{figure}

Figure~\ref{fig:vphi_bar} shows the distributions of $\bar v_\varphi$ for two 
pairs of HVS samples that traveled through DM halos with the same geometric shape but azimuthal orientations that differ by $90^{\circ}$.
One pair is composed of the HVS samples that traveled in DM halos with $(q_y,q_z)= (0.7,1.0)$ (green histogram) and $(q_y,q_z)= (1.4,1.4)$ (blue histogram) (i.e. the same samples investigated in Fig.~\ref{fig:vphi_nonaxisymmetric}); the other pair is composed of the HVS samples that traveled in DM halos with $(q_y,q_z)= (0.9,1.0)$ (yellow histogram) and $(q_y,q_z)= (1.1,1.1)$ (magenta histogram).
The figure shows how the use of the variable $\bar v_{\varphi}$ instead of $v_\varphi$ enables us to easily distinguish distributions of azimuthal velocities generated in pairs of DM halos that have the semi-major axes $a$ and $b$ swapped.
Using the two-dimensional distribution $D_{|v_\vartheta|, \bar v_\varphi}$ thus enables us to overcome the degeneracy problem.

We note that the HVSs represented in Fig.~\ref{fig:vphi_bar} attain values of $\bar v_\varphi$ that are of a few \kms~ for mild $q_y$ deviations (i.e, $|\Delta q_y| = 0.1$) from unity, but may reach $\sim \mathcal{O} (10)$ \kms~ for the most extreme values of $q_y$ considered in our study (i.e., $q_y= 0.7$ and $q_y= 1.4$). This result is a consequence of the extreme flattening of the DM halo along either the $x$-axis or the $y$-axis for values of $q_y$ that are extremely large or extremely small, respectively.

The distributions of $|v_\vartheta|$ and $\bar v_\varphi$ are both very sensitive to the triaxiality parameters of the DM halo. Therefore, their combination can provide a powerful tool to detect non-sphericity of the DM halo and constrain the triaxiality parameters of the corresponding gravitational potential, as we show below.

We compared the two-dimensional distributions $D_{|v_\vartheta|, \bar v_\varphi}$'s of the shape indicators obtained for HVSs that traveled in DM halos of different shapes.
We generated the distributions by using mock catalogs A, which were produced with HVS samples characterized by the same random set of ejection conditions, as we did in Sect.~\ref{sec:halo_impact} for the case of spheroidal DM halos. 
We defined a series of $n_{\rm s}= 56$ total reference shapes by varying both $q_y$ and $q_z$ in steps of 0.1 in the range $[0.7;1.4]$ and imposing $q_y \ne 1$. These reference DM halos have either a fully triaxial shape or a spheroidal shape which is symmetric about the $x$-axis or the $y$-axis, and lead to a non-axisymmetric Galactic potential.
For all these reference shapes, we generated the corresponding HVS mock catalogs. 

We compared the $D_{|v_\vartheta|, \bar v_\varphi}$'s obtained from each of this $n_{\rm s}$ HVS samples against the $D_{|v_\vartheta|, \bar v_\varphi}$ obtained for a spherical DM halo, where  $q_y=q_z=1$, and $D_{\bar v_\varphi}$ is a distribution of null values. For this comparison, we used the two-sample, two-dimensional KS test \citep[][and references therein]{press2007}, hereafter referred to as ``2D KS test''. For all the 2D KS test comparisons, we found null $p$-values, implying that the $D_{|v_\vartheta|, \bar v_\varphi}$'s are significantly different from each other even for the smallest differences in triaxiality parameters with respect to the spherical DM halo. The null $p$-value results from the fact that, in a spherical DM halo, the HVSs do not acquire any azimuthal velocity; any non null $D_{\bar v_{\varphi}}$ is thus a proof that the HVS sample has traveled in a non-axisymmetric gravitational potential.
This result proves that the distribution of $|v_\vartheta|$ and $\bar v_\varphi$ can effectively detect non-spherical shapes of the DM halo. 

A 2D KS test comparison of pairs of $D_{|v_\vartheta|, \bar v_\varphi}$ obtained for DM halos characterized by the same value of $q_z$ but different values of $q_y$ yielded $p$-values always smaller than 5\% even for differences in $q_y$ as small as $\Delta q_y =0.1$. The high sensitivity of the $D_{|v_\vartheta|, \bar v_\varphi}$ to small differences in $q_y$ at fixed $q_z$ is the result of the high sensitivity of $D_{\bar v_\varphi}$ to $q_y$. In turn, this sensitivity comes from the fact that $q_y \ne 1$ is the only source of non-null $v_\varphi$ in triaxial DM halos.

Conversely, small differences in $q_z$ at fixed $q_y$ can lead to comparable $D_{|v_\vartheta|, \bar v_\varphi}$, according to a 2D KS test, with $p$-values that can exceed 5\%. These larger values of $p$ originates from  the fact that differences in $q_z$ affect $v_\vartheta$, but  $v_\vartheta$ is also affected by the gravitational pull of the disk. This effect was already discussed for the case of an axisymmetric Galactic potential, with a DM halo which is either spherical or spheroidal about the $z$-axis, (Sect.~\ref{sec:halo_impact}).
Overall, $\bar v_\varphi$ is a more powerful shape indicator than $|v_\vartheta|$, and the combination of the two shape indicators is the appropriate tool to constrain  the triaxiality parameters of a DM halo with the HVSs.

We note that, in our model of the gravitational potential of the MW, we assumed the DM halo to have its principal axes aligned with the axes of our Cartesian reference frame, $x$, $y$, and $z$, where $x$ indicates the direction from the Sun to the Galactic center and $z$ is orthogonal to the Galactic plane (see Sect.~\ref{sec:MW_potential}). Releasing this assumption has an effect on the shape indicator $\bar v_{\varphi}$ .
Indeed, let us assume that one of the principal axes of the DM halo still coincide with the $z$ axis, while the remaining two principal axes lie on the Galactic plane, but are misaligned with respect to our $x$ and $y$ axes by an angle $0<\phi_0\le 45^{\circ}$. In this case, the HVS $\bar v_{\varphi}$'s would no longer be all negative or all positive. Specifically, for $0^{\circ}<\phi_0\lesssim 45^{\circ}$, higher $\phi_0$ would correspond to higher degrees of mixing of negative and positive values, while for $\phi_0\simeq 45^{\circ}$, the $\bar v_{\varphi}$ distribution will be about half positive and half negative. 

For very small misalignments, the very low degree of mixing of the $\bar v_{\varphi}$ distribution would not prevent us from breaking the degeneracy in the orientation of the DM halo.
On the other hand, for large misalignments, we would not be able to distinguish two DM halos with the same geometrical shape but with orientations that differs by $90^{\circ}$. 
However, we stress that even in those cases the degree of triaxiality of the DM halo could still be determined from the distribution of $v_\varphi$.
\citet{law2009} find that the axes of the DM halo on the plane of the Galactic disk are aligned with the $x$ and $y$ axes within $15^\circ$. In this case, our method would not encounter degeneracy problems.

\subsection{Shape recovery} 
\label{sec:shape_recovery_non-axisymmetric_Galactic_potential}

\begin{table*}[ht]
	\caption{Median value, $p_{\rm med}$, of the distribution of $n_{\rm t}=5,000$ $p$-values obtained from the $n_{\rm t}$ 2D KS test comparisons of the observed sample's $\tilde D_{|v_\vartheta|,\bar v_\varphi}$ against the $D_{|v_\vartheta|,v_\varphi}$'s of the $n_{\rm t}$ mock samples corresponding to different shapes. The largest $p_{\rm med}$ indicates the ``best match'' between the observed sample and the mock sample.
	The observed sample is generated in a DM halo with $q_y=1.2$ and $q_z=0.9$.}
	\begin{center}
		\label{tab:pmed}
		\begin{tabular}{cccccccccc}
			\hline \hline
			$(q_y,q_z)$ & $(1.1,0.8)$ & $(1.1,0.9)$ & $(1.1,1.0)$ & $(1.2,0.8)$ & $(1.2,0.9)$ & $(1.2,1.0)$ & $(1.3,0.8)$ & $(1.3,0.9)$ & $(1.3,1.0)$ \\
			\hline
			$p_{\rm med}$ & $6 \times 10^{-22}$ & $ 10^{-17}$ & $9 \times 10^{-16}$ & 0.006 & 0.443 & 0.004 & $3 \times 10^{-5}$ & $3 \times 10^{-6}$ & $3 \times 10^{-11}$ \\
			\hline
		\end{tabular}
	\end{center}
\end{table*}

\begin{figure*}[ht]
	\includegraphics[width=8.5cm,height=6cm]{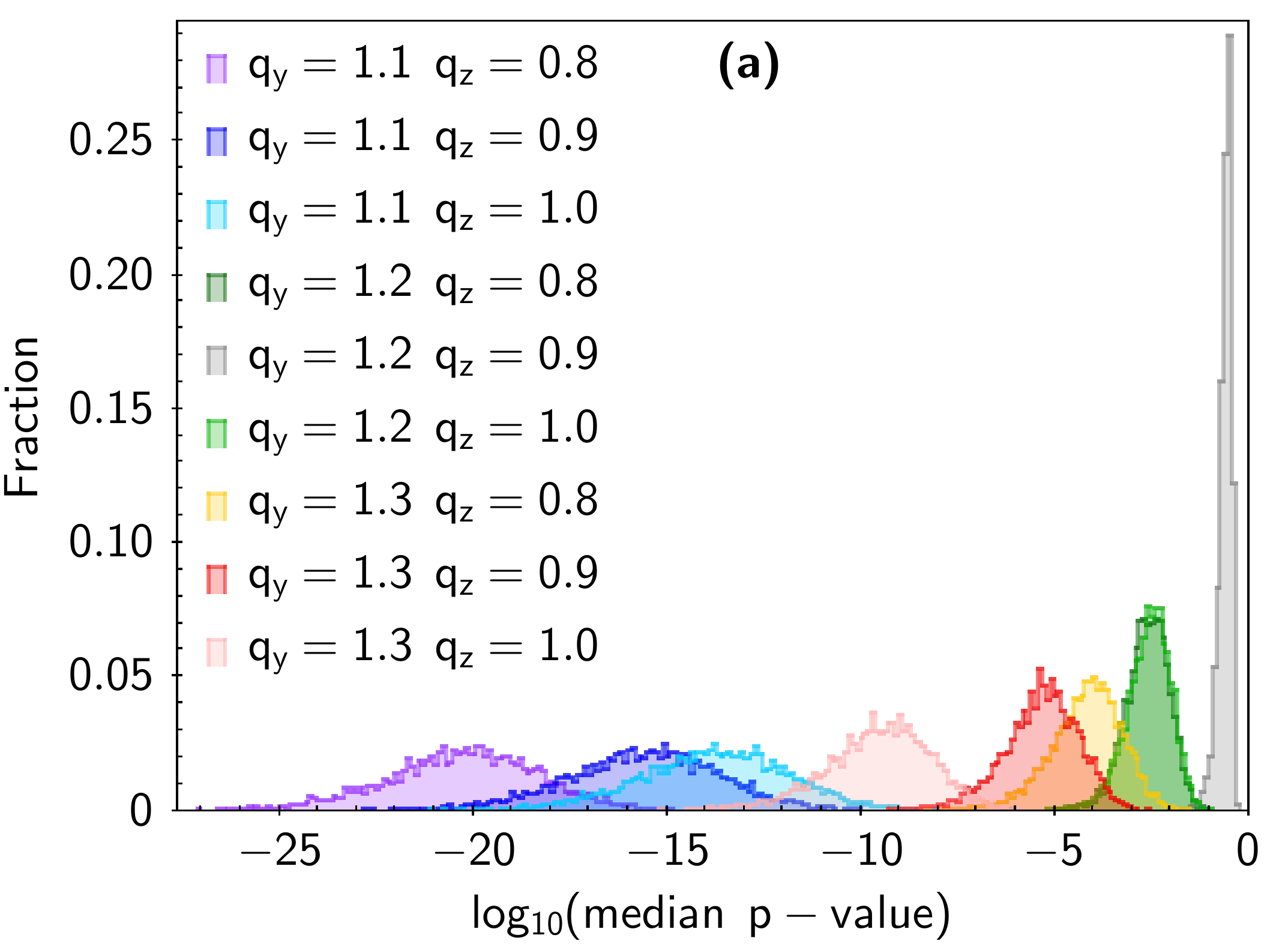}
	\includegraphics[width=8.7cm,height=6cm]{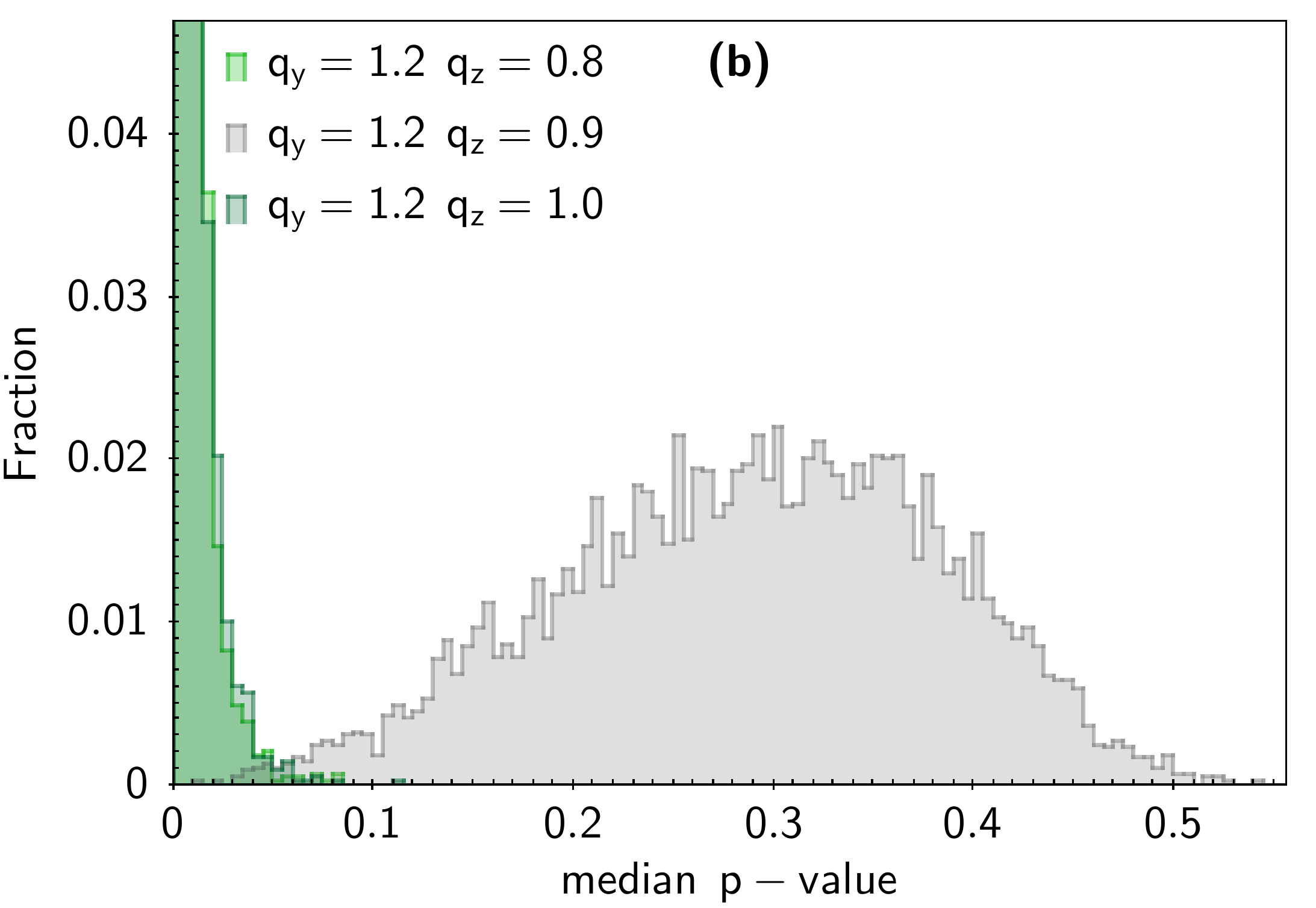}
	\caption{{\it Panel a:} 
		Distributions of the median $p$-values, $p_{\rm med}$ (in logarithmic scale), obtained from the 2D KS test comparison of the HVS observed sample against a selection of $9$ of the $n_{\rm s}=56$ ensembles of mock HVS samples generated in DM halos with different triaxiality parameters.
		Each distribution is the result of the 2D KS test comparisons of the $\tilde D_{|v_\vartheta|,\bar v_\varphi}$'s of the $n=n_{\rm t}=5,000$ observed samples obtained in a DM halo with $q_y = 1.2$ and $q_z = 0.9$ against the $n_{\rm t}$ mock samples generated in a DM halo with a different shape as listed in the panel. 
		{\it Panel b:} 
		Enlargement of the right-most part of panel $a$, with the $p_{\rm med}$ axis in linear scale. 
		The green and dark-green distributions are the only distributions with non-null overlap with the gray distribution.
		The different shapes of the distributions in panels $a$ and $b$ are due to the different size of the histogram bins in the logarithmic and  linear scales.
		\label{fig:nonaxisymmetric_example_pmed}}
\end{figure*}

To recover the shape of the DM halo by means of the distribution $\tilde D_{|v_\vartheta|, \bar v_\varphi}$ of the shape indicators $|v_\vartheta|$ and $\bar v_\varphi$, we made use of the series of $n_{\rm s}=56$ reference shapes for a non-axisymmetric Galactic potential obtained, as described above, by varying both $q_y$ and $q_z$ in steps of 0.1 in the range $0.7-1.4$, and imposing $q_y \ne 1$.
For each shape, we generated an ensemble of $n_{\rm t}=5,000$ mock catalogs B, one per different set of initial conditions of the stars' trajectories. From each mock catalog, we obtained one two-dimensional distribution $D_{|v_\vartheta|, \bar v_\varphi}$ of the shape indicators, for a total of $n_{\rm t}$ distribution $D_{|v_\vartheta|, \bar v_\varphi}$'s per shape of the DM halo.

As an example, we chose as the HVS observed sample one random mock sample of HVSs that traveled in a DM halo with $q_y=1.2$ and $q_z=0.9$. We show here that our method successfully recovers the correct triaxiality parameters $q_y$ and $q_z$.

Following a procedure similar to that described in Sect.~\ref{sec:shape_recovery_axisymmetric_Galactic_potential}, 
for each of the $n_{\rm s}=56$ reference shapes ($q_y,q_z$) of the DM halo we performed the 2D KS test comparisons of the observed sample's $\tilde D_{|v_\vartheta|,\bar v_\varphi}$ against all of the $n_{\rm t}=5,000$ $D_{|v_\vartheta|,v_\varphi}$'s corresponding to each shape.
Thus, for each shape of the DM halo, we obtained a distribution of $n_{\rm t}$ $p$-values and a corresponding $p_{\rm med}$. 

Table \ref{tab:pmed} shows the $p_{\rm med}$'s obtained in this exercise for a selection of DM halos of different triaxiality parameters. We expect that the highest $p_{\rm med}$ points at the shape of the DM halo that ``best matches'' the shape of the DM halo actually crossed by the observed sample (see Sect.~\ref{sec:shape_recovery_axisymmetric_Galactic_potential}). Indeed, the highest $p_{\rm med}=0.443$ occurs for the DM halo halo with triaxiality parameters $q_y = 1.2$ and $q_z = 0.9$, namely for the DM halo actually crossed by the observed sample: this result demonstrates that our method effectively recovers the correct shape of the triaxial DM halo crossed by the observed sample.

\subsection{Success rate}
\label{sec:success_rate_non-axisymmetric_Galactic_potential}

In this subsection, we illustrate the success rate $S$ of our method for the case of a DM halo with a specific shape (Sect.~\ref{sec:success_rate_non-axisymmetric_Galactic_potential_1shape}) and the dependence of $S$ on the shape of the DM halo (Sect.~\ref{sec:success_rate_non-axisymmetric_Galactic_potential_shapes}).

\subsubsection{The case of the triaxial DM halo with $(q_y,q_z) = (1.2,0.9)$} \label{sec:success_rate_non-axisymmetric_Galactic_potential_1shape}

To evaluate the success rate $S$ of our method, we followed a procedure similar to that of Sect.~\ref{sec:success_rate_axisymmetric_Galactic_potential}.
We generated a series of $n=n_{\rm t}$ HVS observed samples in a DM halo with $q_y=1.2$ and $q_z=0.9$, by randomly varying the set of the stars' initial conditions. For each observed sample, we performed the $n_{\rm t}$ 2D KS test comparisons of the observed sample's $\tilde D_{|v_\vartheta|,\bar v_\varphi}$ against the $D_{|v_\vartheta|,\bar v_\varphi}$'s of all the $n_{\rm t}$ mock samples corresponding to a given reference shape of the DM halo, and we obtained a $p_{\rm med}$. Performing the procedure for $n=n_{\rm t}$ observed samples yields a distribution of $n=n_{\rm t}$ values of $p_{\rm med}$ for each given shape of DM halo.
Repeating this procedure for each of the $n_{\rm s}=56$ shapes of DM halo yields $n_{\rm s}$ distributions of $p_{\rm med}$'s. 

We show a representative selection of these distributions in Fig.~\ref{fig:nonaxisymmetric_example_pmed}.
Panels $a$ and $b$ show that the highest $p_{\rm med}$'s correspond to the distribution obtained for the DM halo with triaxiality parameters $q_y = 1.2$ and $q_z = 0.9$ (gray histogram), namely the halo traveled by the observed sample: this proves that our method recovers the correct shape of the triaxial halo in the large majority of the cases.

There are, however, cases where an erroneous shape association can occur. As shown in panel $a$ of Fig.~\ref{fig:nonaxisymmetric_example_pmed}, the distributions of $p_{\rm med}$'s, corresponding to the comparison of the $\tilde D_{|v_\vartheta|,\bar v_\varphi}$'s of the HVS observed sample against the $D_{|v_\vartheta|,\bar v_\varphi}$ of the HVS samples generated in DM halos with the same $q_y=1.2$ and with $q_z=0.8$ or $q_z=0.9$, display a non-null overlap with the distribution corresponding to the comparison of the observed sample against the samples from DM halos with $q_y=1.2$ and $q_z=0.9$.
This overlap can be better appreciated in the enlargement of panel $b$. The overlap implies that, for an observed sample generated in a DM halo with $q_y=1.2$ and $q_z=0.9$, our method can erroneously associate to the observed sample only the shapes characterized by the same $q_y$ and by a $q_z$ slightly different ($|\Delta q_z| \le 0.1$) from the correct $q_z$. 

Erroneous shape associations are however very rare, because the overlap between the distributions is very modest.
In the above example, the success rate of our method is $99.98\%$; only in one case over $5,000$ the highest $p_{\rm med}$ suggests triaxiality parameters for the DM halo crossed by the observed sample that are not the correct ones, namely $(q_y,q_z) = (1.2, 1.0)$ instead of $(q_y,q_z) = (1.2, 0.9)$.
In the analyzed case, any difference $\Delta q_y\ge 0.1$ leads to distributions of $p_{\rm med}$ that do not overlap, namely the rate of erroneous shape associations is equal to zero; the axis ratio $q_y$ is thus correctly recovered in 100\% of cases.

\subsubsection{Dependence of $S$ on the shape of the DM halo of the observed sample} 
\label{sec:success_rate_non-axisymmetric_Galactic_potential_shapes}

To investigate the effect of the shape of the DM halo of the observed sample on the success rate of our method in the case of a non-axisymmetric Galactic potential, we explored three additional shapes for the DM halo. 
Table~\ref{tab:S_nonaxisymmetric} reports the axis ratios $q_y$ and $q_z$ of these DM halos, and the corresponding success rate, $S$. The axis ratio and success rate of the case explored in Sect.~\ref{sec:success_rate_non-axisymmetric_Galactic_potential_1shape} are reported in the same table for completeness.
	
	\begin{table}[ht]
		\caption{Success rate $S$ of our method in recovering the axis ratios $q_y$ and $q_z$ of the DM halo of a non-axisymmetric Galactic potential from the two-dimensional distribution $D_{|v_\vartheta|,\bar v_\varphi}$ of an observed sample of HVSs.}
		\begin{center}
			\label{tab:S_nonaxisymmetric}
			\begin{tabular}{cc}
				\hline \hline
				$(q_y,q_z)$ & $S$ \\
				\hline
				$(0.8,0.8)$ & 100.00\% \\
				$(1.3,0.8)$ &  99.64\% \\
				$(1.2,0.9)$ &  99.98\% \\
				$(1.2,1.3)$ &  96.24\% \\
				\hline
			\end{tabular}
		\end{center}
	\end{table}
	
	We obtained success rates $S > 96\%$ in all the explored cases. Thus, for a non-axisymmetric Galactic potential, the success rate of our method is less sensitive to the actual shape of the DM halo than for an axisymmetric Galactic potential.
	As in the case of an axisymmetric Galactic potential, illustrated in Sect.~\ref{sec:success_rate_axisymmetric_Galactic_potential_shapes}, for a non-axisymmetric Galactic potential we found that the DM halo with the largest $q_z$, namely the case $(q_y,q_z=(1.2,1.3)$, yields the smallest success rate, albeit still larger than 96\%.
	
Even though we investigated only four shapes of DM halos, we selected them to appropriately cover the axis-ratio space.
We thus expect $S \gtrsim 95\%$ also for different combinations of axis ratios.
In particular, less extreme axis ratios are expected to increase the success rate of the method: this tendency, already shown in Sect.~\ref{sec:success_rate_axisymmetric_Galactic_potential_shapes} for the DM halo yielding an axisymmetric Galactic potential, is enhanced in the case of a DM halo that generates a non-axisymmetric Galactic potential, because of the presence of a non-null distribution of $\bar v_{\varphi}$. Therefore, a 2D distribution of shape indicators makes the constraining power of our method higher and more effective against the actual shape of the DM halo that we want to recover.


\section{Sample size and method success rate} 

\label{sec:sample_size}

\begin{figure*}[ht]
	\includegraphics[width=8.2cm,height=6.1cm]{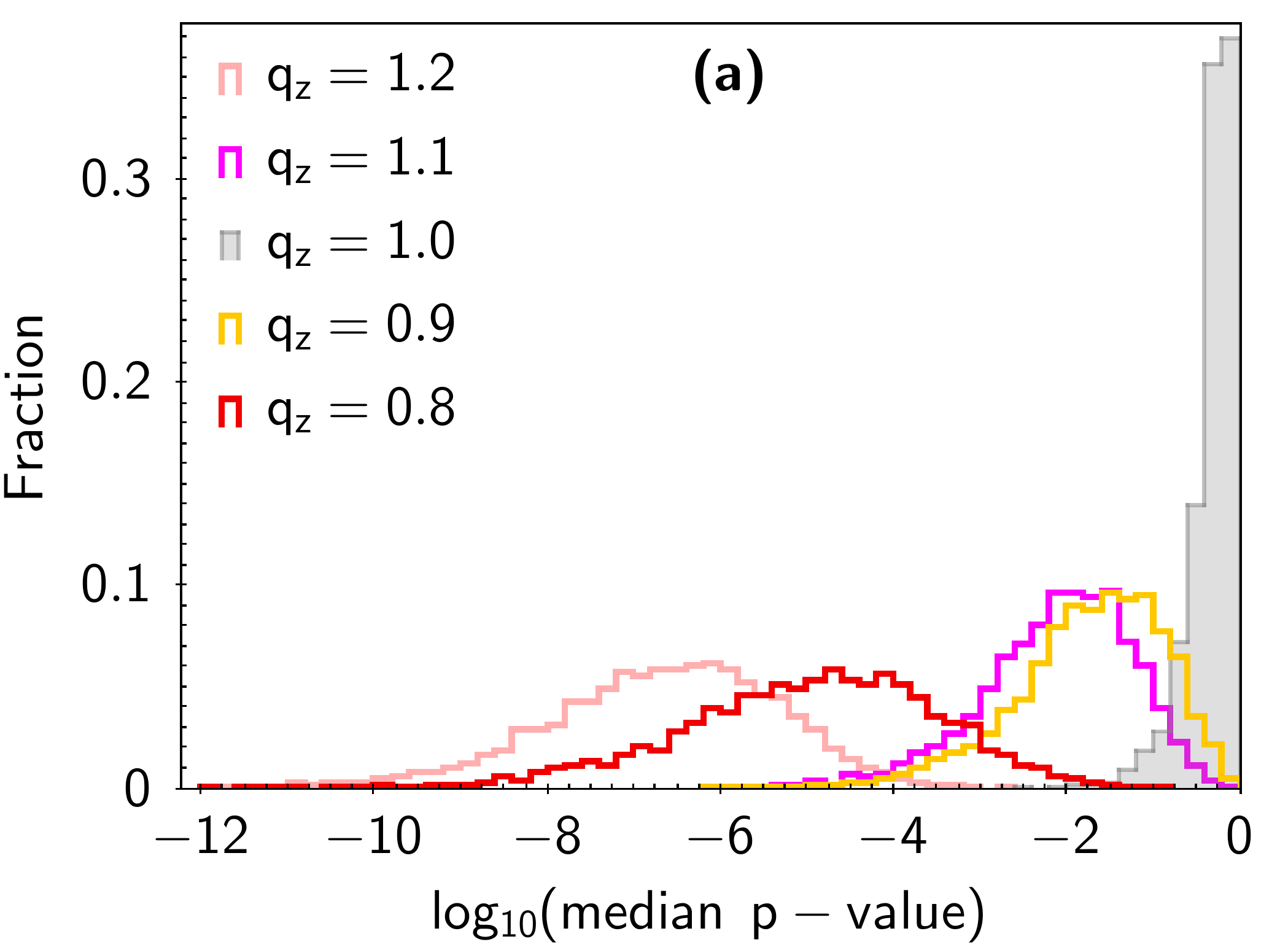}
	\includegraphics[width=8.2cm,height=6.1cm]{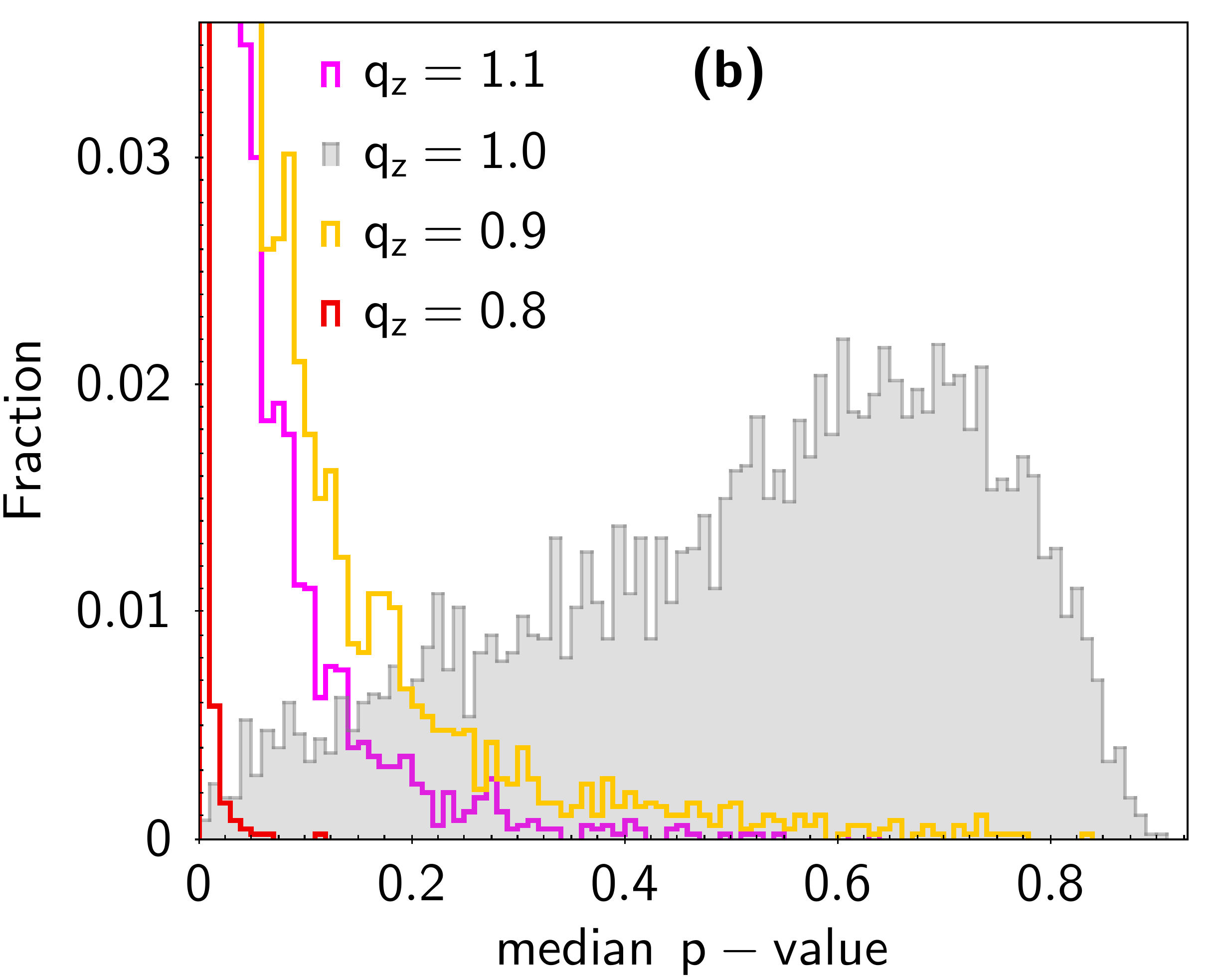}
	\caption{{\it Panel a}: Distributions of the median $p$-values, $p_{\rm med}$ (in logarithmic scale), obtained by comparing each of the $5,000$ observed samples with $\simeq 400$ HVS generated in a spherical DM halo, with the $5,000$ mock samples generated in spheroidal DM halos axisymmetric about the $z$-axis with different $q_z$, as listed in the panel.
	{\it Panel b}: Enlargement of the right-most part of panel $a$ with the $p_{\rm med}$ axis in linear scale. The different shape of the distributions in panels $a$ and $b$ are due to the different size of the histogram bins in the logarithmic and  linear scales.}
	\label{fig:sample_size_impact_example}	
\end{figure*}

\begin{figure}[ht]
	\includegraphics[width=7.8cm,height=5.8cm]{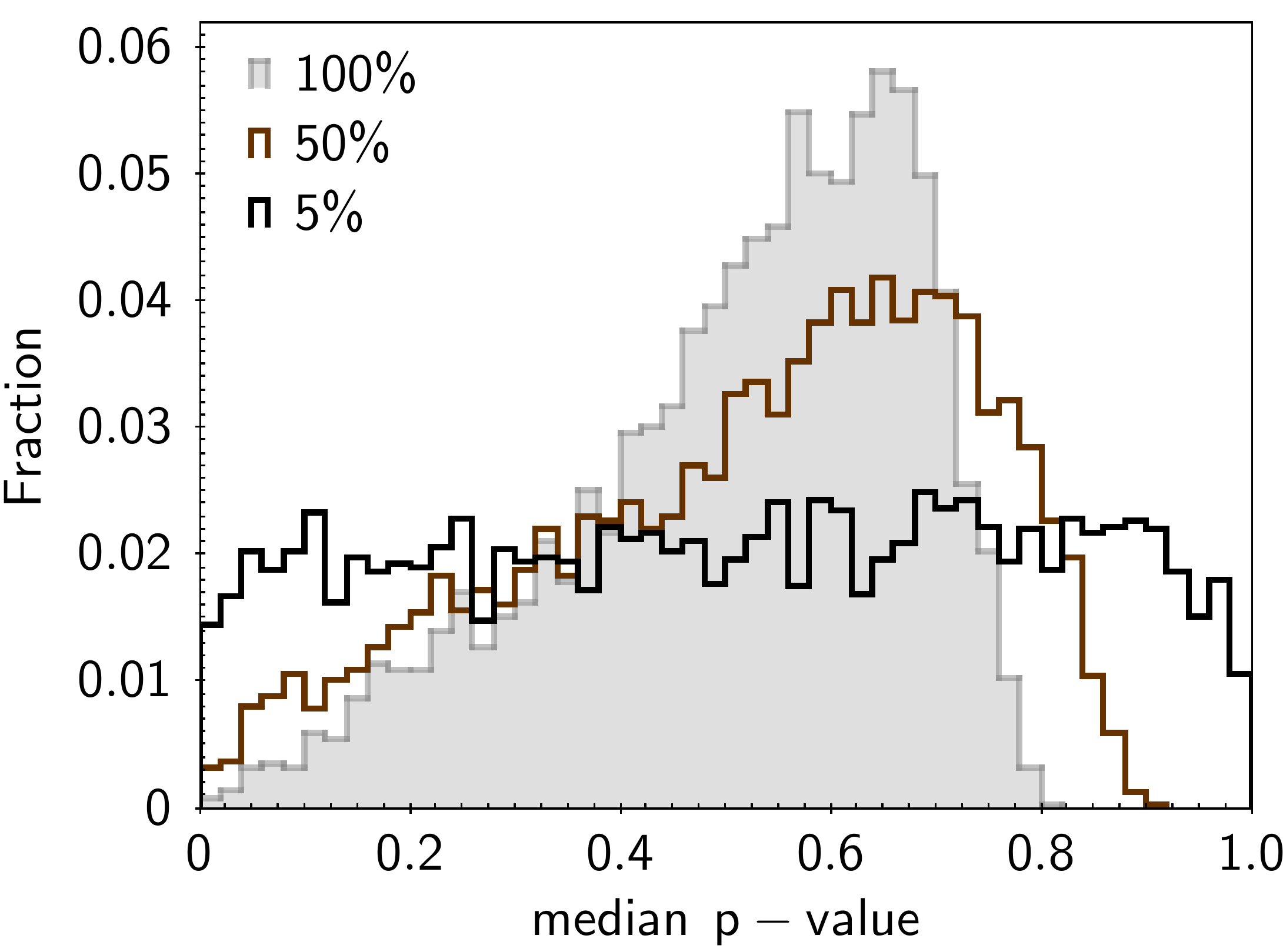}
	\caption{Distributions of the median $p$-values obtained from the KS test comparison of the $\tilde D_{|v_{\vartheta}|}$ of each observed sample, generated in a spherical DM halo, with the $5,000$ mock samples generated in a spherical DM halo.
	The size of the compared samples is 100\% (gray histogram; $N\simeq 800$ stars), 50\% (brown histogram; $N\simeq 400$ stars), and 5\% (black histogram; $N\simeq 40$ stars), respectively, of the original HVS sample size.}
	\label{fig:100_50_5}
\end{figure}

\begin{figure}[ht]
	\includegraphics[width=8.5cm,height=6.8cm]{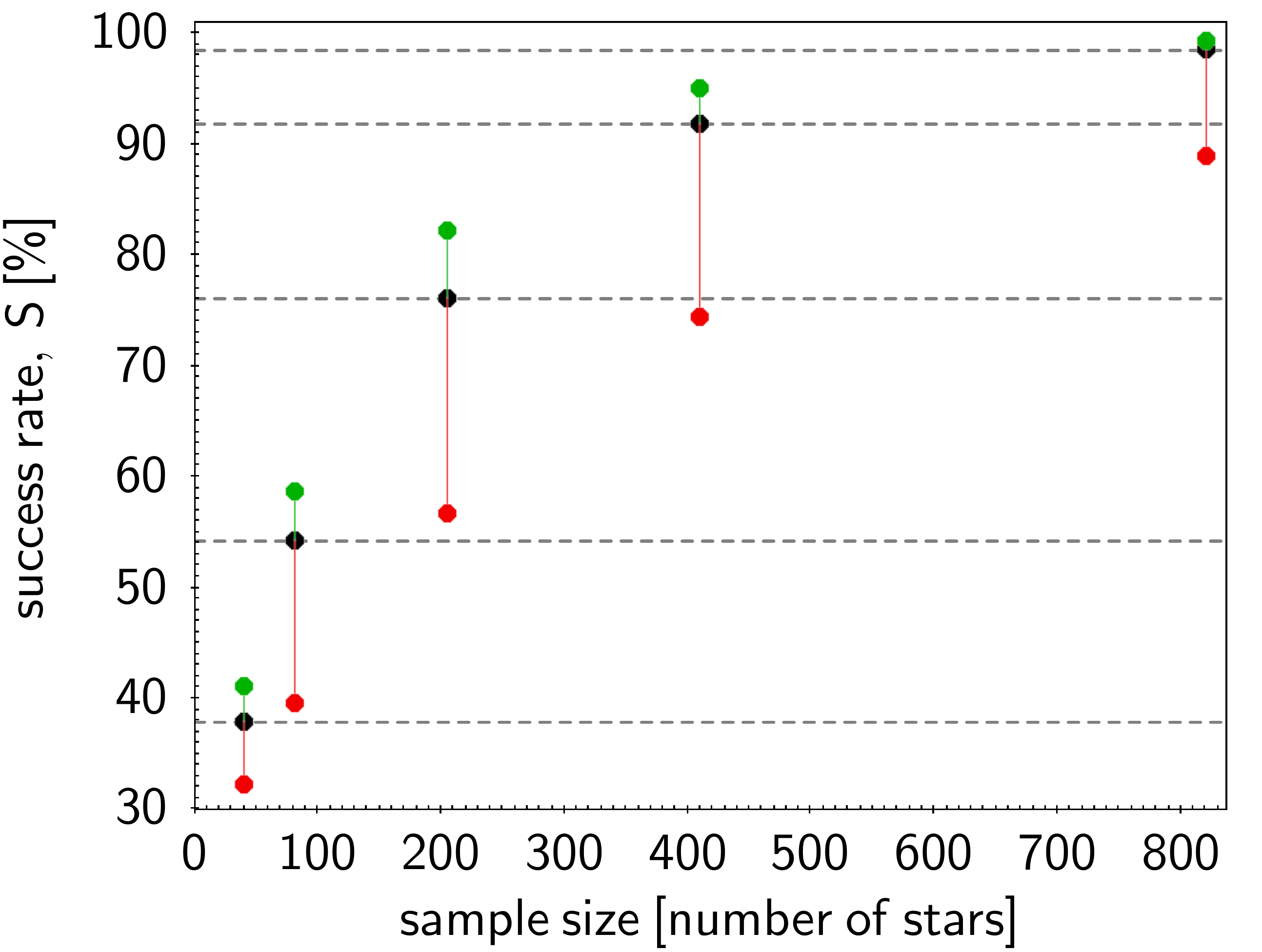}
	\caption{Success rate $S$ of our method in recovering the correct shape of a DM halo as a function of the size of the HVS observed sample, for a spherical DM halo ($q_z= 1.0$; black dots) and for two spheroidal DM halos axisymmetric about the $z$-axis with $q_z= 1.1$ (green dots) and $q_z= 1.4$ (red dots). 
	The sample size reported on the $x$-axis is the average size of the $5,000$ HVS observed samples used to estimate the success rate in the spherical DM halo scenario. $S$ is shown for sample sizes of 100\% ($N\simeq 800$ stars), 50\% ($N\simeq 400$ stars), 25\% ($N\simeq 200$ stars), 10\% ($N\simeq 80$ stars), and 5\% ($N\simeq 40$ stars) of the original sample size.} \label{fig:sample_size_impact_success_rate}
\end{figure}

In Sects.~\ref{sec:axisymmetric_Galactic_potential} and \ref{sec:non-axisymmetric_Galactic_potential} we investigated the efficiency of our method in recovering the shape of the DM halo, in both an axisymmetric and a non-axisymmetric Galactic potential, from a mock observed sample of HVSs composed of $\sim 800$ 4~$M_\sun$ stars. The size of the mock sample, dictated by the combination of the HVS ejection rate and the lifetime of the stars (see Sect.~\ref{sec:orbit_integration}),
as well as by our selection criteria (see Sect.~\ref{sec:halo_impact}), represents the optimal situation, that would occur if the HVSs were actually ejected according to the Hills mechanism with the assumed rate, and if we were able to observe all of them.

Here, we present our investigation of the dependence of the success rate $S$ of our method on the size of the HVS observed sample.
In Sect.~\ref{sec:sample_size_spherical_DM_halo}, we show the results obtained when repeating the analysis performed in Sect.~\ref{sec:axisymmetric_Galactic_potential} on a series of mock HVS observed samples that traveled in a spherical DM halo and whose size is 50\%, 25\%, 10\%, and 5\% of the original sample size. In Sect.~\ref{sec:sample_size_general}, we illustrate the generalization of our results to different shapes of the DM halo.

\subsection{Spherical DM halos} 
\label{sec:sample_size_spherical_DM_halo}

We applied our method to an HVS observed sample that traveled in a spherical DM halo. For the case of an observed sample whose size is 50\% the original size, panel $a$ of Fig.~\ref{fig:sample_size_impact_example} shows the distributions of the median $p$-values obtained by comparing each observed sample with the $5,000$ mock samples generated in spheroidal DM halos axisymmetric about the $z$-axis. This figure is the analog of Fig.~\ref{fig:axisymmetric_example_pmed}, with the exception of the reduced sample size of the observed sample and of the smaller number of $p_{\rm med}$ distributions.

Panel $b$ is the enlargement of panel $a$. It shows that the distributions of the median $p$-values obtained from the comparison of the mock observed samples with the mock samples generated in the spherical DM halo with $q_z = 1.0$ (the gray histogram) is flatter than the corresponding histogram in panel $b$ of Fig.~\ref{fig:axisymmetric_example_pmed}. Indeed, all the distributions of Fig.~\ref{fig:sample_size_impact_example} are flatter than the corresponding distributions of Fig.~\ref{fig:axisymmetric_example_pmed}. 
As a consequence, there is a larger superposition of the gray distribution with the $p_{\rm med}$ distributions obtained from the KS test comparisons of the observed samples and the samples that crossed the mildly oblate ($q_z = 0.9$) and mildly prolate ($q_z = 1.1$) DM halos.
Unlike Fig.~\ref{fig:axisymmetric_example_pmed}, Fig.~\ref{fig:sample_size_impact_example} also shows a non-null overlap of the gray distribution with the red distribution obtained from the comparison of the observed sample against the sample that crossed an oblate DM halo with $q_z = 0.8$; however, this non-null overlap does not correspond to any actual erroneous associations of the observed sample with an oblate DM halo with $q_z = 0.8$, because no comparison yields a $p_{\rm med}$ of the red distribution higher than the $p_{\rm med}$ of the gray distribution (see Sect.~\ref{sec:success_rate_axisymmetric_Galactic_potential_1shape}).

Summarizing, a 50\% smaller HVS sample implies a higher probability of assigning a DM halo with an incorrect axial ratio $q_z$ to an observed sample of HVSs. This corresponds to a lower success rate  $S$ of the method: while for the original HVS observed sample we obtain $S=98.4\%$, for the 50\% smaller sample we obtain  $S=91.8\%$. However, the axial ratio can be off by $|\Delta q_z| = 0.1$ only, as for the full-size sample.

The flattening of the distributions of the median $p$-values keeps on increasing with decreasing size of the observed sample. Figure \ref{fig:100_50_5} shows the change in the distribution of the median $p$-values obtained from the KS test comparison of the HVS observed samples against the mock samples generated in a spherical DM halo, when the size of the HVS observed samples drops from 100\% (gray histogram; $N\simeq  800$ stars) to 50\% (brown histogram; $N\simeq 400$ stars) and to 5\% (black histogram; $N\simeq 40$ stars).
The increasing flattening of the $p_{\rm med}$ distributions for smaller HVS samples occurs because deviations of the distribution of the $p$-values from the uniform distribution increase with decreasing size of the HVS sample. In turn, the difference between $p_{\rm med}$ and the expected value $0.5$ increases and the distributions of $p_{\rm med}$'s thus have higher tails.

For the case of a spherical DM halo, 
the black dots in Fig.~\ref{fig:sample_size_impact_success_rate} show the success rate of our method as a function of the size of the HVS observed sample.
A success rate $S \gtrsim 90\%$ requires an HVS sample of $N \gtrsim 400$ HVSs, while $S \gtrsim 75\%$ is achieved with $N \gtrsim 200$ HVSs, $S \gtrsim 55\%$ is obtained with a sample of $N \gtrsim 80$ HVSs, and $S \gtrsim 40\%$ is obtained with a sample of $N \gtrsim 40$ HVSs.
When the HVS sample size is $N \simeq 400$, the unsuccessful cases yield axial ratios that can be off by $|\Delta q_z| = 0.1$ at most, as it is for the case of the optimal sample size,  $N \simeq 800$. On the other hand, when the sample size is $N \simeq 200$, the axial ratio can be off by $|\Delta q_z| \le 0.2$.
For the smallest sample sizes considered here (i.e., for $N \simeq 80$ and $N \simeq 40$), the axial ratio can be off by $|\Delta q_z| \le 0.4$.
For any sample size, the probability to associate a given  erroneous shape with the DM halo crossed by the observed sample
decreases with increasing difference $|\Delta q_z|$ between the erroneous and the actual axis ratio.

\subsection{Non-spherical DM halos} 
\label{sec:sample_size_general}
For spheroidal DM halos that are axisymmetric about the $z$-axis, the dependence of the success rate $S$ of our method on the sample size is comparable to the dependence found for spherical DM halos,  illustrated in Sect.~\ref{sec:sample_size_spherical_DM_halo}. 
However, the shape dependence of $S$ discussed in Sects.~\ref{sec:success_rate_axisymmetric_Galactic_potential_shapes} and \ref{sec:success_rate_non-axisymmetric_Galactic_potential_shapes} is responsible for fluctuations of the value of $S$ for any given HVS sample size.

Figure~\ref{fig:sample_size_impact_success_rate} shows the dependence of the success rate $S$ on the HVS sample size for the spheroidal DM halos axisymmetric about the $z$-axis that yield the most extreme success rates, namely those with $q_z = 1.1$ (green dots) and $q_z = 1.4$ (red dots), superimposed to the same dependence for spherical DM halos (black dots).
The different $S$'s between each of the two spheroidal cases and the spherical case are represented by the vertical green and red bars superimposed to the black dots. 
The spheroidal DM halo with $q_z = 1.4$ yields the lowest success rate for each sample size, and the difference in $S$ between this case and the spherical case can be as high as $\sim 25\%$. On the other hand, the spheroidal DM halo with $q_z = 1.1$ yields the largest success rate for each sample size, with differences in $S$ from the spherical case always smaller than $10\%$. 

For fully triaxial DM halos or for spheroidal DM halos non-axisymmetric about the $z$-axis, we refrain from performing a detailed analysis: for the optimal sample size of $\sim 800$ HVSs, our method recovers the correct shape of the DM halo with a success rate $S = 96\%-100\%$ for the explored scenarios, depending on the actual shape of the DM halo (see Sect.~\ref{sec:success_rate_non-axisymmetric_Galactic_potential_shapes}).
This success rate is always higher than the success rate obtained, for the optimal sample size, for a spheroidal DM halo axisymmetric about the $z$-axis and with $q_z = 1.4$; for some shapes of the non-axisymmetric DM halo, $S$ can also be higher than that obtained for a spheroidal DM halo axisymmetric about the $z$-axis and with $q_z = 1.1$.
We thus expect that, for  the case of non-axisymmetric DM halos,  $S$ will be generally larger than for the axisymmetric cases with the same sample size.


\section{Discussion}
\label{sec:Discussion} 
The HVSs are ejected from the Galactic center and cross a large range of distances during their journey across the Galaxy. As shown in Sect.~\ref{sec:halo_impact}, their use as tracers of the DM halo of the MW is appropriate in any region where the DM halo is expected to dominate their kinematics, namely at galactocentric distances $r\gtrsim 10$~kpc. However, the phase space coordinates of the HVSs at any radius $r$ stores the information on the triaxiality parameters of the dark halo within $r$. Therefore, the HVSs appear to be a promising  probe of the triaxiality of the DM halo of the MW over a large range of spatial scales.

Conversely, other observational probes can constrain the shape of the DM halo over more limited spatial scales (see Sect.~\ref{sec:introduction}). Our HVS-based method is thus a powerful tool to complement the currently available constraints on the shape of the Galactic DM halo provided by different tracers on different, limited scales.

Clearly, the applicability of the presented method depends on a few working hypotheses, that we discuss below. In addition, the size of the available HVS samples with measured velocity is still an issue both for the success rate of the method and for the magnitude of the offset between the recovered and the actual axis ratio.
However, even though our method was developed in the framework of a specific model of HVS production and with a set of assumptions, our working hypotheses do not limit the validity of the method, as we illustrate in the following.

As for the HVS production, we considered here the Hills mechanism, because of its unique ability in generating both a large number of unbound main-sequence HVSs and B-type stars in close orbit about SgrA$^\star$ (see Sect.~\ref{sec:introduction}).
However, our method can be applied to any mechanism that can eject stars from the Galactic center on a purely radial direction with a velocity $v_{\rm ej}\gtrsim 730$ km~s$^{-1}$, thus enabling them to reach $r>10$ kpc with non-null outward radial velocity.

All our simulated HVSs have a mass of 4~$M_{\sun}$. The large majority of the HVS candidates are B-type stars with mass in the range  $\sim 2-4 \, M_{\sun}$, while other candidates are classified as A, F, G, or K type stars. 
The generation of a realistic sample of ejected stars would require sampling the masses of the binary stars that encounter the SMBH from a mass distribution. 
However, under the simplifying assumption that the population of binary stars were entirely composed of equal-mass binaries, the distribution of the ejection velocities of a sample of HVSs with different masses would be the sum of the independent distributions of the ejection velocities of equal-mass HVSs.
Thus, our simulated mock HVS sample of $4\, M_{\sun}$ stars has the same fractional distribution of ejection velocities that would be obtained for the subsample of $4\, M_{\sun}$ stars in a simulated extended spectrum of  masses for equal-mass binaries.
As a consequence, our mock observed sample would mirror a real sample limited to $4\, M_{\sun}$ HVSs.

In fact, in a realistic scenario, unequal mass binaries are likely. 
For hyperbolic encounters, as those simulated in this work, the primary member of each binary is usually ejected as an HVS; its ejection velocity depends on the mass $m_2$ of the secondary star.
Thus, in a distribution of binaries whose primary member is a $4\:M_{\sun}$ star, the velocity of the ejected 4~$M_{\sun}$ HVS depends on the mass of the companion star, $m_2 \le 4\:M_{\sun}$.
Nevertheless, \citet{bromley2006} show that the fractional velocity distribution of 4~$M_{\sun}$ HVSs located at $r=10-120$~kpc is insensitive to $m_2$, when $m_2=$ 0.5, 1, 2, and 4 $M_{\sun}$.
Our simulations confirm this result for binaries with $m_2$ uniformly sampled in the range $0.1-4 \:M_{\sun}$. 
The insensitivity of the velocity distribution to $m_2$ follows from the fact that the stars that reach $r>10$~kpc populate the high velocity tail of the distribution of ejection velocities ($v_{\rm ej} \gtrsim 730$ \kms); the high speed tail of the ejection velocity distribution obtained in the case of 4+4~$M_{\sun}$ binaries is statistically indistinguishable from the high speed tail of the corresponding distribution obtained when $m_2$ is sampled in the range $0.1-4\:M_{\sun}$.
Thus, limiting our mock HVS sample to $4\, M_{\sun}$ stars ejected from equal mass binaries returns, for stars in the region of interest ($r>10$~kpc), final velocity distributions that are statistically consistent with those we would obtain by fixing the mass of the primary star to $4\, M_{\sun}$ and by varying the mass of the secondary star in the range  $0.1-4\:M_{\sun}$.
However, the normalization of the latter distributions is smaller than the normalization we obtain by simulating $4+4\, M_{\sun}$ binaries. Indeed, when $m_2 = 0.1-4 \:M_{\sun}$, the ejection velocity distribution peaks at lower values, implying that a lower number of HVSs can reach $r>10$~kpc.

The size of the HVS sample also decreases with decreasing ejection rate $R$. 
By adopting $R=10^{-4}$ yr$^{-1}$  (see Sect.~\ref{sec:orbit_integration}), we generated  an optimal  sample of $\sim 800$ $4\, M_\sun$ stars.
We investigated the effect of different sizes of the HVS samples on the method success rate in Sect.~\ref{sec:sample_size}, and showed that the success rate decreases for smaller sample sizes. This analysis is suggestive of the effect of a lower ejection rate on the success rate of our method.
Indeed, the ejection rate depends on a number of assumptions and it is still poorly constrained \citep[see, e.g.,][]{hills1988, yu2003,bromley2012,zhang2013,brown2015}. \citet{bromley2012} assume continuous star formation in the Galactic center and estimate $R\approx 1-2 \times 10^{-3}$ yr$^{-1}$ when integrating over all the mass spectrum of the ejected stars considered. Our adopted $R=10^{-4}$ yr$^{-1}$ for $4\, M_\sun$ stars appears roughly consistent with this analysis of \citet{bromley2012}.  
However, by dropping the assumption of continuous star formation,  \citet{bromley2012} find substantial smaller values: $R\sim 2-8 \times 10^{-5}$ yr$^{-1}$. This range partly overlaps with the range $\sim 10^{-5}-$~a few $10^{-4}$ yr$^{-1}$ found by \citet{zhang2013} who consider different origins of the injected binaries and different models of the Initial Mass Function of the primary stars \citep[see also][]{yu2003}.

The decrease of the success rate with the decreasing size of the HVS sample illustrated in Sect.~\ref{sec:sample_size} also mimics the effect of the reduction of the HVS sample because of observational limitations, like the star magnitude or the star position within the Galaxy.

The dependence of the size of the HVS sample on the mass of the stars is more intricate:
the larger number of stars with $M< 4\, M_\sun$ and lifetime $\tau_{\rm L}>160$ Myr would determine a larger number of HVSs that are alive at the observation time compared to our $4\, M_\odot$ star sample. However, longer-lived, lower-mass stars can reach larger Galactocentric distances after experiencing  the inner turnaround (Sect.~\ref{sec:sample_selection}); therefore, the lower limit $r=10$~kpc we adopted here for the $4\, M_\odot$ stars will be larger for lower mass stars, thus potentially reducing the number of HVSs suitable for our method.  
Opposite effects would be determined by the moderate number of stars with $M> 4\, M_\sun$, and lifetime $\tau_{\rm L}<160$ Myr.
We plan to investigate in future work how these effects combine to determine the optimal sample size and, in turn, the success rate of the method.

Our HVS mock samples include both unbound and bound HVSs, that are HVSs whose ejection velocity does not exceed the escape velocity of the MW. The bound HVSs that satisfy the selection criteria that we defined in Sect.~\ref{sec:sample_selection} are indicators of the shape of the DM halo as good as the unbound HVSs. Therefore, the observation of bound HVSs is of fundamental importance to increase the HVS sample size and the success rate of the method.

As highlighted by \cite{marchetti2021}, bound HVSs may be the best candidate stars ejected from the Galactic center that can be observed in the {\it Gaia} catalogs, while the majority of the unbound HVS population is expected to be too far from the Sun's position \citep{Kenyon2014} for its radial velocity to be measured by the {\it Gaia} Radial Velocity Spectrometer (RVS).
In the current sample of HVS candidates, $\sim 70\%$ of the stars are located at a galactocentric distance $r > 10$~kpc, at the $3\sigma$ level. The possibility to measure the radial velocities of fainter objects in the outer halo will be of fundamental importance to identify new bound and unbound HVS candidates that satisfy our selection criteria. For example, the forthcoming 4-metre Multi-Object Spectroscopic Telescope \citep[4MOST;][]{4most2019} will be able to increase the volume of the spectroscopic sample provided by {\it Gaia}; specifically, it will measure the radial velocities of {\it Gaia} photometric sources with magnitude $G < 20.5$, while the {\it Gaia} RVS will provide radial velocities for sources with $G_{\rm RVS} \le 16.2$~mag, corresponding to $G \le 15.9$ for B0V stars and $G \le 17.4$~mag for K4V stars 
\citep[][]{Jordi2010,Fitzgerald1970}.

If our sample of 4 $M_\sun$ HVSs reached the optimal size of $\sim 800$ stars, in the ideal case of null uncertainties, our method would be able to recover the correct axis ratios of the DM halo in $\gtrsim 90\%$ of cases, while being off by 0.1 in the remaining $\lesssim 10\%$.
The offset of 0.1 found in the minority of unsuccessful cases is set by the resolution in triaxiality parameters that we choose in our simulations, $\Delta q_{y,z} = 0.1$. 

Here, we did not investigate the effect of any observational limitations on our HVS observed sample, nor the effect of the uncertainties on the HVS distances and galactocentric tangential velocities on the success rate of our method. This investigation will be pursued elsewhere.

The main contribution to the uncertainty on the tangential velocity of distant HVSs comes from their proper motion. The proper motion measurements currently available in the {\it Gaia} Early Data Release 3 \citep[EDR3;][]{gaiamission2016,gaiaED3summ2021} are affected by uncertainties that can lead to relative uncertainties on the tangential velocities larger than $100\%$.
On the other hand, for nearby HVS candidates, the largest contribution to the uncertainties on the tangential velocities comes from the star distances inferred from {\it Gaia} parallaxes: in the Gaia EDR3, the uncertainties on the parallaxes range from $0.02-0.03$~mas, for sources with $G < 15$~mag, to $1.3$~mas, for sources with $G = 21$ mag \citep{gaiaED3summ2021}. The relative uncertainties on the parallax-inferred distances of an HVS candidate at $10$~kpc from the Sun can thus vary from $20-30\%$ for the brightest stars to 1300\% for the faintest stars.
To preserve the high success rate of our method, it clearly is of utmost importance to reduce the uncertainties on both the star distances and proper motions.

A future {\it Theia}-like mission \citep{Malbet2016,theiacollaboration2017,malbet2019,theia2021}, designed for unprecedented high precision astrometry, may achieve an end-of-mission uncertainty on proper motions of a few micro-arcseconds per year (i.e., $\sim$100 times smaller than the uncertainty of {\it Gaia}), opening up the possibility for significantly constraining the shape of the DM halo.
The availability of more precise measurements of proper motions will also make it possible to better constrain the birth place of the current HVS candidates.

The confirmation of the galactocentric origin of the trajectories of HVS candidates and their use to constrain the Galactic gravitational potential requires a final cautionary note. This research program is far from obvious, and can suffer from a circularity problem. The galactocentric origin of an HVS candidate can be assessed through a backtracking of the star trajectory \citep[e.g.][]{brown2014,brown2018,marchetti2019,irrgang2018a,koposov2020,kreuzer2020, irrgang2021}; in turn, the backtracking requires an assumption on the gravitational potential that one aims to constrain. 
In addition to the matter distribution of the Galaxy, the gravitational potential is set by the distribution of the satellites of the Galaxy \citep[e.g.,][]{kenyon2018}. Moreover, the gravitational potential of the Galaxy is time dependent because of the interaction of the Galaxy  with its satellites \citep{boubert2020}. On a wider perspective, the galactocentric origin of an HVS candidate also depends on the theory of gravity \citep{chakrabarty2022}. Therefore, a solid confirmation of the galactocentric origin is currently limited to high-speed stars that are close to the Galactic center, where the  effects mentioned above are negligible \citep[e.g.,][]{koposov2020}. An appropriate method for constraining the gravitational potential with slower and more distant HVS candidates would require, for example, an iterative approach. In the absence of a self-consistent procedure, constraints on the Galaxy mass model with HVS candidates remain questionable.


\section{Conclusions} 
\label{sec:conclusions}

We presented a new method to infer the shape of the DM halo of the MW from the distribution of the components of the galactocentric tangential velocities of a sample of HVSs.
We applied our method to an ideal optimal sample of $\sim 800$ 4~$M_{\sun}$ simulated HVSs. We referred to this sample as the observed sample. We illustrated the method for both axisymmetric and non-axisymmetric Galactic potentials.

In the axisymmetric scenario, we recovered the axial ratio of the DM halo from the one-dimensional distribution of one shape indicator only: the magnitude of the latitudinal velocity, $|v_{\vartheta}|$, of the HVSs of the observed sample. In the non-axisymmetric scenario, we recovered the axial ratios from the two-dimensional distribution of two shape indicators: $|v_\vartheta|$ and a function $\bar v_\varphi$ of the azimuthal velocity, $v_\varphi$, of the HVSs of the observed sample. 

The method is based on the use of the one- or two-dimensional KS test to compare the distribution of the shape indicator(s) of the observed sample's,  $D_{|v_{\vartheta}|}$  or $D_{|v_\vartheta|,\bar v_\varphi}$, against the corresponding distributions of HVS mock samples that traveled in DM halos with different shapes. 
The resolution in axial ratios of our ensemble of mock catalogs is $\Delta q_{y,z}=0.1$.

For each shape of the DM halo, we compared the observed sample with a set of $5,000$ $D_{|v_{\vartheta}|}$'s or $D_{|v_\vartheta|,\bar v_\varphi}$'s, which account for the different ejection initial conditions of the HVSs. A distribution of $5,000$ $p$-values was thus obtained for each shape. 
We used the median, $p_{\rm med}$, of these $p$-value distributions to identify the shape of the DM halo that best matched the shape of the DM halo crossed by the observed sample: the highest $p_{\rm med}$ value comes from the $p$-value distribution associated with the correct shape of the DM halo.

In our ideal case of galactocentric velocities with null uncertainties and no observational limitations, the method has a success rate $S \gtrsim 89\%$ in recovering the correct shape of the DM halo of an axisymmetric Galactic potential, and a success rate $S > 96\%$ in recovering the correct shape of the DM halo of the non-axisymmetric Galactic potentials that we explored in this work.

The higher success rate of our method for a non-axisymmetric Galactic potential is due to (i) the availability of two shape indicators, $|v_{\vartheta}|$ and ${\bar v_{\varphi}}$, compared to one shape indicator alone for an axisymmetric Galactic potential; and (ii) the sensitivity of the azimuthal velocity $v_\varphi$ to the shape of the DM halo. 
In the small fraction of unsuccessful cases, an erroneous shape association occurs; however, the discrepancy from  the correct axial ratio is small.
Indeed, in the axisymmetric Galactic potential scenario, the erroneous DM halo axis ratio $q_z$ typically differs from the correct ratio by $|\Delta q_z|= 0.1$; an offset $|\Delta q_z|=0.2$ is very rare ($\lesssim  0.04\%$ of cases). 
In the case of a non-axisymmetric Galactic potential, the incorrect shape associations are expected to be even rarer, because our method has a higher constraining power for a non-axisymmetric than for an axisymmetric Galactic potential.

The success rate of our method depends on the size of the HVS sample, and is higher for larger HVS samples.
In the case of an axisymmetric Galactic potential, a decrease in the sample size corresponds to a decrease of the success rate that depends on the actual shape of the DM halo: for a spherical halo, a decrease of the sample size from 800 to 40 mock observed HVSs implies a drop of $S$ from $\sim 98\%$ to $\sim 38\%$; for a spheroidal halo axisymmetric about the $z$-axis with $q_z = 1.1$, 
which yields the highest $S$, $S$ drops from $\sim 99\%$ to $\sim 41\%$ for the same decrease of the sample size; for a spheroidal halo with $q_z = 1.4$, which yields the lowest $S$, $S$ drops from $\sim 89\%$ to $\sim 32\%$.

On the other hand, for any non-axisymmetric Galactic potential, and for any  given sample size, $S$ is expected to be always larger than the rate obtained in the most unfavorable axisymmetric case ($q_z = 1.4$) and typically also larger than the rate obtained in the most favorable axisymmetric case ($q_z = 1.1$).

In addition to increase the success rate $S$, increasing the size of the HVSs sample decreases the discrepancy between the inferred shape and the correct shape. 

Accurate estimates of the success rate of our method when applied to real data require the generation of more realistic mock observed HVS samples, that account for (i) the uncertainties on the distances and velocities of the HVSs, (ii) the observational limitations of the HVS sample, and (iii) an appropriate mass distribution of the ejected stars.
Nevertheless, our analysis suggests that a robust determination of the shape of the DM potential requires measuring the galactocentric velocities of a few hundred HVSs whose galactocentric origin is robustly confirmed.

We will assess elsewhere the sensitivity that is required for our method and that might be reached with future astrometric space missions \citep{Malbet2016,theiacollaboration2017,malbet2019,theia2021}. Similarly, we will investigate the success rate of our method with expected realistic HVS samples that might come from upcoming spectroscopic surveys \citep{4most2019}.


\begin{acknowledgements}
We sincerely thank Scott Kenyon, Margaret Geller, and Warren Brown for numerous inspiring discussions that stimulated our initial interest in the dynamics of HVSs. We are in debt with the referee whose sharp comments contributed to clarify the presentation of our work. The Departments of Excellence grant L.232/2016 of the Italian Ministry of Education, University and Research (MIUR) fully supported the PhD fellowship of AG and partly supported the fellowship of SSC.
SSC is also supported by the grant ``The Milky Way and Dwarf Weights with Space Scales'' funded by University of Torino and Compagnia di S.~Paolo (UniTO-CSP), by the grant no.~IDROL~70541 IDRF~2020.0756 funded by Fondazione CRT, and by INFN.
We acknowledge partial support from the INFN grant InDark. The work of SE included here was part of his Master Thesis project at the University of Torino.
This research has made use of NASA’s Astrophysics Data System Bibliographic Services, of the routines of the \textsc{Numerical Recipes} \citep{press2007}, of \textsc{Topcat} \citep{Topcat2005} and of the \textsc{NumPy} \citep{numpy} and \textsc{matplotlib} \citep{Matplotlib} \textsc{Python} modules.
\end{acknowledgements}

%
%

\bibliographystyle{aa} 
\bibliography{biblio.bib} 

\begin{appendix}
\section{Mock catalogs and observed sample}
\label{sec:appendixA}

\begin{figure*}[ht!]
    \centering
	\includegraphics[width=9.1cm,height=7cm]{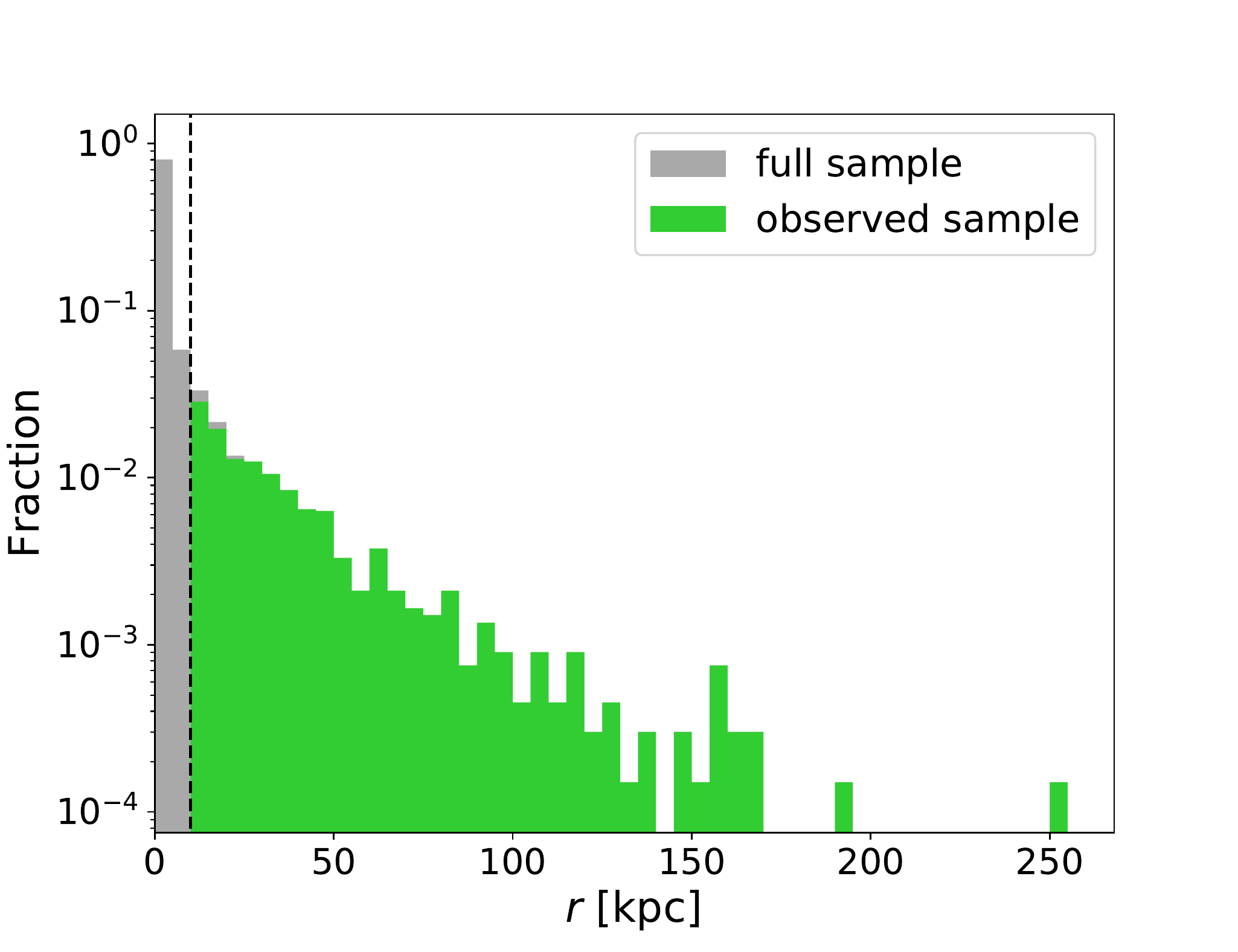}
	\includegraphics[width=9.1cm,height=7cm]{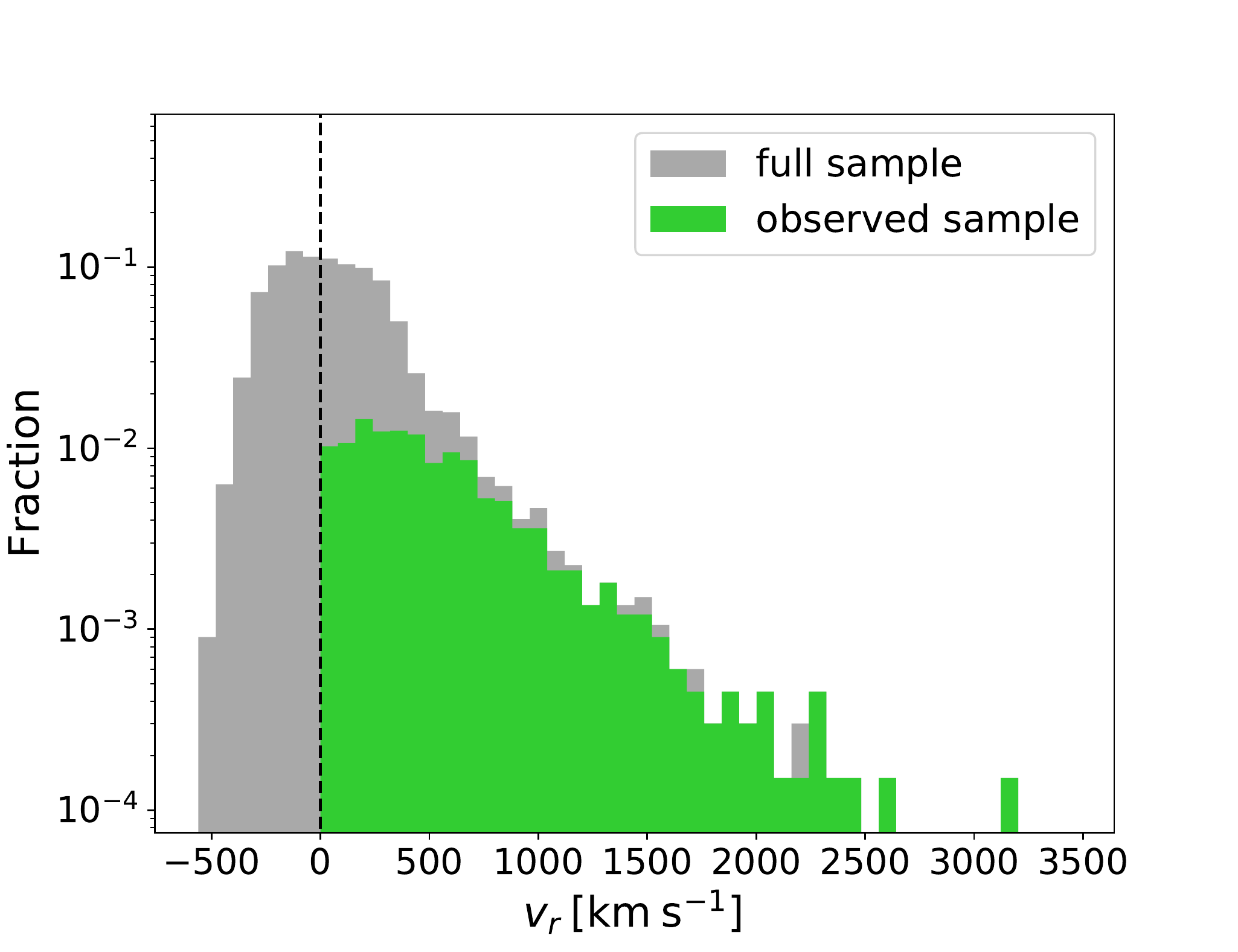}
	\includegraphics[width=9.1cm,height=7cm]{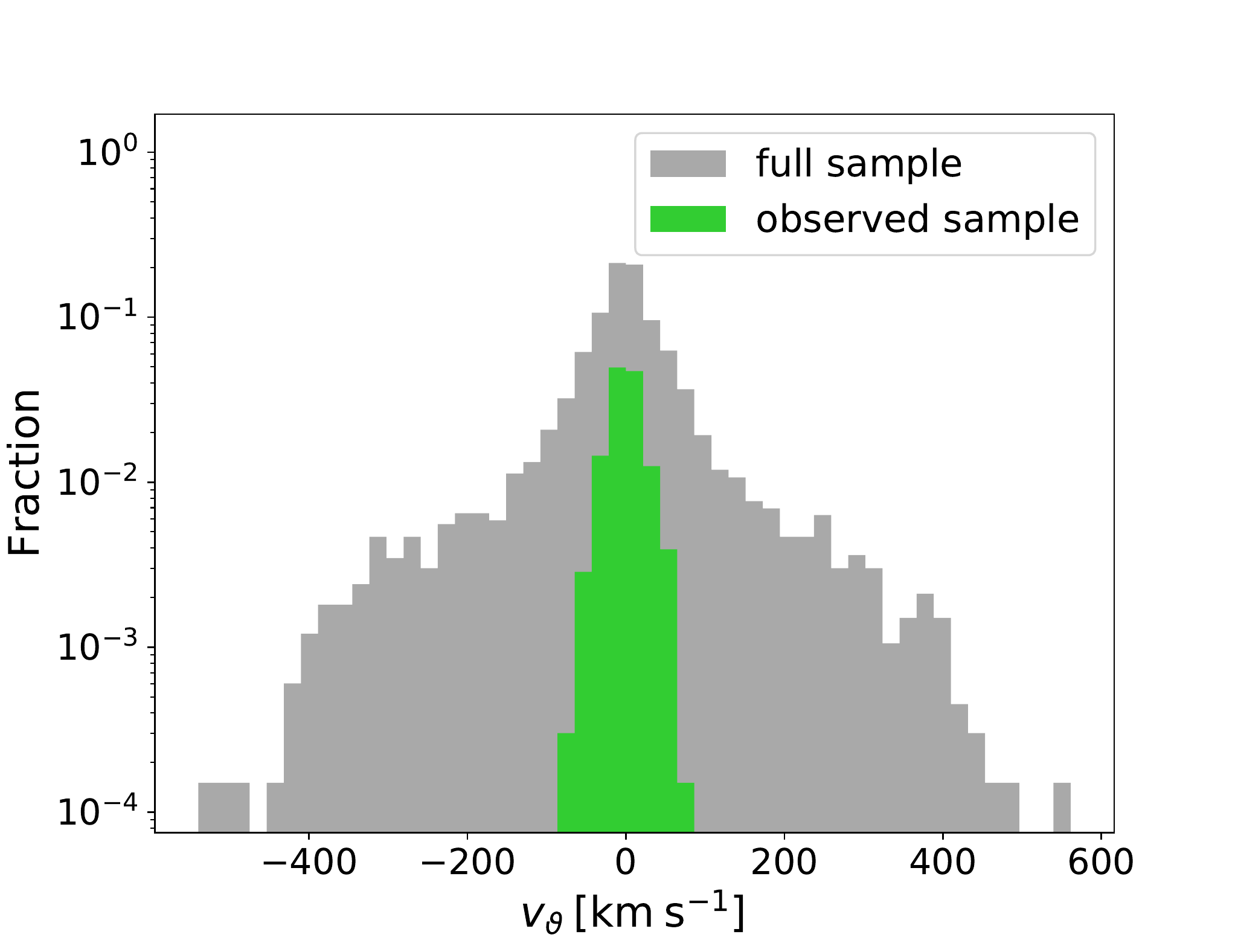}
	\includegraphics[width=9.1cm,height=7cm]{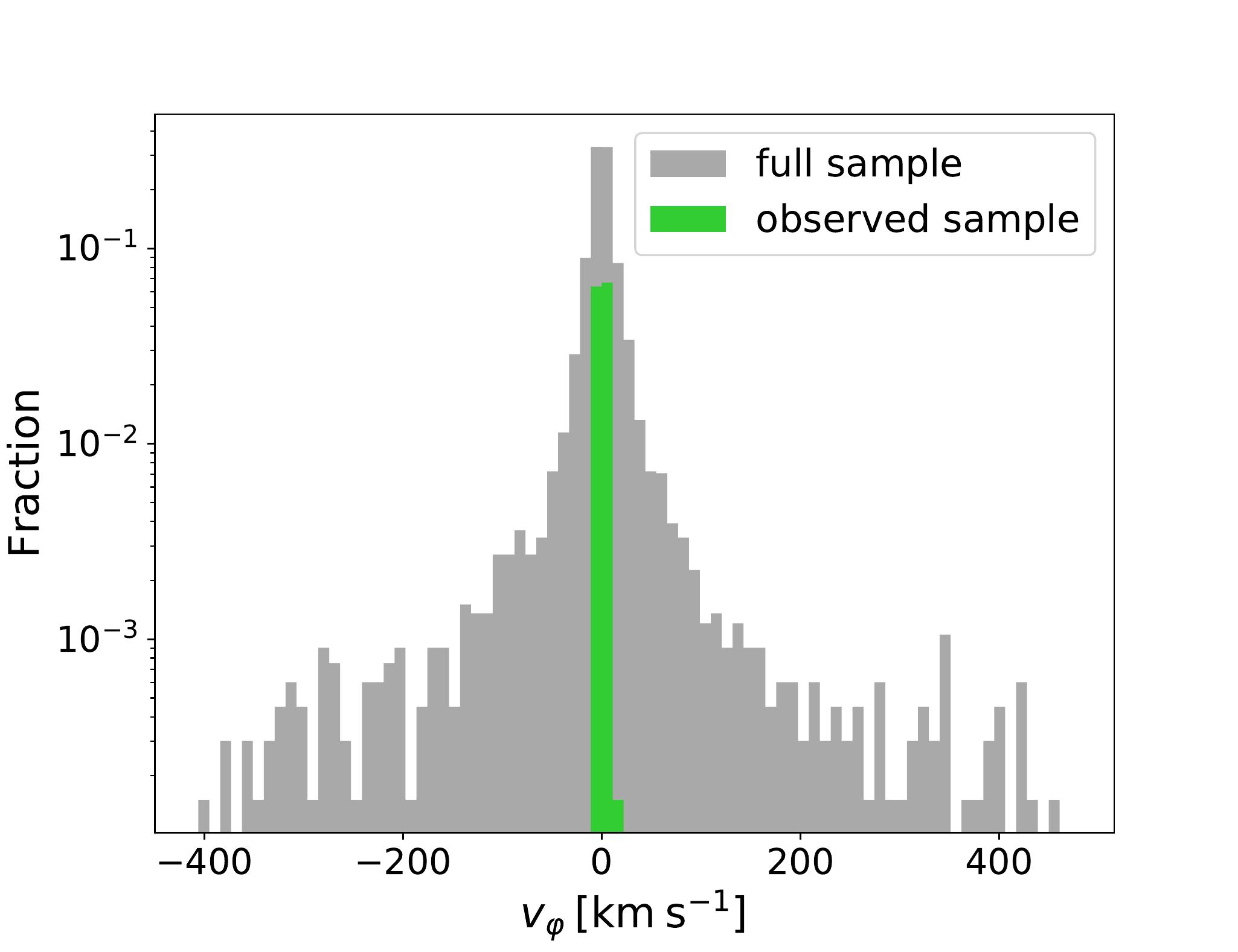}

	\caption{Kinematic properties of the HVSs of one of our mock catalogs, corresponding to a Galaxy  whose DM halo has a triaxial gravitational potential with $(q_y,q_z) = (1.2,0.9)$. The gray histograms represent the full sample of HVSs; the green histograms represent our observed sample, derived from the full sample 
	by requiring $r>10$ kpc and $v_r>0$.
{\it Top left panel}: Distribution of Galactocentric distances; the vertical, dashed line corresponds to our threshold radius $r>10$ kpc. {\it Top right panel}: Distribution of the radial velocities; the vertical, dashed line corresponds to $v_r = 0$.
{\it Bottom left panel}: Distribution of the latitudinal velocities.	{\it Bottom right panel}: Distribution of the azimuthal velocities.}\label{fig:mock_A_triaxial}	
\end{figure*}

Fig.~\ref{fig:mock_A_triaxial} shows the distributions of the distances to the Galactic center, and of the galactocentric radial, latitudinal, and azimuthal velocities of our simulated HVSs at the steady state (i.e., at the observation time $t_{\rm obs}$; see Sect.~\ref{sec:orbit_integration}) for a mock catalog (see Sect.~\ref{sec:mock_catalogs}) generated in a gravitational potential where the  DM halo is triaxial with axis ratios  $(q_y,q_z) = (1.2,0.9)$.  
The gray histograms correspond to the full sample of simulated HVSs, while the green histograms represent our ``observed sample'' of $\sim 800$ HVSs, namely the ideal optimal sample of HVSs that we derived from the full sample by applying the selection criteria $r>10$~kpc and $v_r>0$  (see Sects.~\ref{sec:halo_impact}, \ref{sec:sample_selection}, and \ref{sec:method}).
The distributions of the kinematic properties of the stars in the other Galactic potentials we investigated in this work are qualitatively similar to those shown in Fig.~\ref{fig:mock_A_triaxial}.


\section{Transformation of coordinates and velocities}
\label{sec:appendixB}
The Galactic heliocentric position of a star is $(d,\ell,b)$, with $d$ the heliocentric distance to the star, $\ell$ its Galactic longitude, and $b$ its Galactic latitude; the  Galactic heliocentric velocity is $(v_d,v_\ell,v_b)$, with $v_d$ the radial velocity of the star, and $v_\ell$ and $v_b$ the longitudinal and latitudinal components of the heliocentric tangential velocity.
The Galactic heliocentric position and velocity components are 

\begin{equation}\label{eq:d}
d = \sqrt{(x+R_{\sun})^{2}+ y^2+z^2} \, , 
\end{equation}
\begin{equation}\label{eq:l}
l = \mathrm{arctan}\left( \frac{x~\mathrm{tan}\: \varphi}{x+R_{\sun}}\right)\, ,
\end{equation}
\begin{equation}\label{eq:b}
b = \mathrm{arcsin}\left( \frac{z}{d}\right) \, ,
\end{equation}
and
\begin{eqnarray}
\label{eq:vd}
v_d =& ~(v_x-U_{\sun})\,\mathrm{cos}l\,\mathrm{cos}b
+[v_y-(V_{\sun}+\Theta_0)]\,\mathrm{sin}l\,\mathrm{cos}b \nonumber \\
&+ (v_z-W_{\sun})\,\mathrm{sin}b \, ,
\end{eqnarray}
\begin{equation}
\label{eq:vl}
v_l = ~[v_y-(V_{\sun}+\Theta_0)]\,\mathrm{cos}l-(v_x-U_{\sun})\,\mathrm{sin}l \, ,
\end{equation}
\begin{eqnarray}
\label{eq:vb}
v_b =& ~(U_{\sun}-v_x)\,\mathrm{cos}l\,\mathrm{sin}b-[v_y-(V_{\sun}+\Theta_0)]\,\mathrm{sin}l\,\mathrm{sin}b \nonumber \\
&+ (v_z-W_{\sun})\,\mathrm{cos}b \, . 
\end{eqnarray}
The longitudinal component of the proper motion $\mu$ is 
$\mu_\ell = v_\ell/(d\,\mathrm{cos}b)$, and its latitudinal component is ${\mu_b} = v_b/d$.

We transformed the Galactic heliocentric star position and proper motion to the equatorial system. In this system, the star position is  $(d,\alpha,\delta)$, with $\alpha$ the right ascension and $\delta$ the declination; the star velocity is $(v_d,v_\alpha,v_\delta)$, with $v_d$ still the radial velocity of the star, and $v_\alpha$ and $v_\delta$ the components of the star tangential velocity along the right ascension and declination, respectively; the star proper motion is $\mu=(\mu_\alpha,\mu_\delta)$, with  
$\mu_\alpha = v_\alpha/(d\,\mathrm{cos}\delta)$, and ${\mu_\delta} = v_\delta/d$.

For the star position, we adopted the transformation equations \citep{duffett-smith1979} 
\begin{equation}
\alpha  = \mathrm{arctan}\left[ \frac{\mathrm{cos}b~\mathrm{cos}(\ell-\ell_{\rm asc})}{\mathrm{sin}b~\mathrm{cos}\delta_G-\mathrm{cos}b~\mathrm{sin}\delta_G~\mathrm{sin}(\ell-\ell_{\rm asc})}\right]  + \alpha_G \, ,
\end{equation}
\begin{equation}
\delta = \mathrm{arcsin}[\mathrm{cos}b~\mathrm{cos}\delta_G~\mathrm{sin}(\ell-\ell_{\rm asc}) + \mathrm{sin}b~\mathrm{sin}\delta_G] \, ,
\end{equation}
where $\alpha_G$ and $\delta_G$ are the equatorial coordinates of the North Galactic Pole, and 
$\ell_{\rm asc}$ is the Galactic longitude of the ascending node of the Galactic plane, related to the Galactic longitude of the North Celestial Pole by $\ell_{\rm asc}=\ell_{\rm NCP} - 90^{\circ}$.
We chose J2000.0 as a reference point for the coordinate conversion. This choice yields $\alpha_G = 192.85948123^{\circ}$, $\delta_G = 27.12825120^{\circ}$ \citep{cox2000}, and $\ell_{\rm asc} = 32.93192^{\circ}$ \citep{poleski2013}.

For the components of the proper motion $\mu$, we adopted the  transformation equations \citep{poleski2013}
\begin{equation}
\mu_{\delta} = \left(\frac{C_2\,\mathrm{cos}b}{C_1^2 + C_2^2}\right)\, \left({\mu_\ell}^* +  \frac{C_1}{C_2} {\mu_b} \right)\, , 
\end{equation}
\begin{equation}
\mu_{\alpha}^*  = -\frac{1}{C_1}\left(C_2 \mu_{\delta} - \mu_\ell \mathrm{cos}b \right) \, ,
\end{equation}
with
\begin{eqnarray}
\label{eq:prova}
 C_1 & = & \mathrm{sin}\delta_G~\mathrm{cos}\delta - \mathrm{cos}\delta_G\, \mathrm{sin}\delta\,\mathrm{cos}(\alpha - \alpha_G)  \, ,  \\
C_2 & = &\mathrm{cos}\delta_G~\mathrm{sin}(\alpha - \alpha_G)\, ,
\end{eqnarray}
and $\mu_{\ell}^* =  \mu_{\ell}\,\mathrm{cos}b$,
$\mu_{\alpha}^* =  \mu_{\alpha}\,\mathrm{cos}\delta$,
and $\mu_{\alpha}^{*2} + \mu_{\delta}^2   = \mu_{\ell}^{*2} + \mu_{b}^2=\mu^2.$ 

\end{appendix}

\end{document}